\documentclass[a4paper, aps, pre, preprint, showpacs, showkeys, 10pt, twocolumn]{revtex4-1}

\usepackage{amsmath}
\usepackage{amssymb}
\usepackage{graphicx}
\usepackage[mathscr]{euscript}
\usepackage{natbib}

\newcommand{\opV}{\mathscr{V}}
\newcommand{\opR}{\mathscr{R}}
\newcommand{\opF}{\mathscr{F}}
\newcommand{\opI}{\mathscr{I}}
\newcommand{\opU}{\mathscr{U}}
\newcommand{\opG}{\mathscr{G}}
\newcommand{\opA}{\mathscr{A}}
\newcommand{\mS}{\mathbf{S}}
\newcommand{\fd}{f^\dagger}
\newcommand{\Tr}{\operatorname{Tr}}
\newcommand{\kB}{k_\mathrm{B}}
\newcommand{\Ks}{K_\mathrm{s}}
\newcommand{\partf}{\mathcal{Q}}
\newcommand{\Hamiltonian}{\mathcal{H}}
\newcommand{\spX}{\mathcal{X}}
\newcommand{\iu}{\mathrm{i}}
\newcommand{\ee}{\mathrm{e}}
\newcommand{\Kast}{K^\ast}
\newcommand{\Tw}{T_\mathrm{w}}
\newcommand{\Tc}{T_\mathrm{c}}
\newcommand{\mgn}{\mathfrak{m}}
\newcommand{\flateral}{f_{\mathrm{C}\parallel}}
\newcommand{\fnormal}{f_{\mathrm{C}\perp}}
\newcommand{\ftotal}{\mathbf{f}_{\mathrm{C}}}
\newcommand{\xiAS}{\xi_\parallel^{\left(\mathrm{AS}\right)}}
\newcommand{\xiS}{\xi_\parallel^{\left(\mathrm{S}\right)}}
\newcommand{\xib}{\xi_\mathrm{b}}
\newcommand{\xiw}{\xi_\mathrm{w}}
\newcommand{\xiASp}{\xi_\parallel^{\prime\left(\mathrm{AS}\right)}}
\newcommand{\Last}{L^\ast}
\newcommand{\Laast}{L^{\ast\ast}}
\newcommand{\Pf}{\operatorname{Pf}}
\newcommand{\Fper}{F_\text{per}}
\newcommand{\fper}{f_\text{per}}
\newcommand{\sign}{\operatorname{sign}}
\newcommand{\Fsingle}{F_\text{single}}

\begin{document} 
\title{Lateral critical Casimir force in two--dimensional inhomogeneous Ising strip. Exact results.}
\date{\today}
\author{Piotr Nowakowski}
\email{pionow@is.mpg.de}
\affiliation{Max--Planck--Institut f\"ur Intelligente Systeme, Heisenbergstr.\ 3, 70569 Stuttgart, Germany}
\affiliation{Institut f\"ur Theoretische Physik IV,Universit\"at Stuttgart, Pfaffenwaldring 57, 70569 Stuttgart, Germany}
\author{Marek \surname{Napi\'orkowski}}
\affiliation{Institute of Theoretical Physics, Faculty of Physics, University of Warsaw, ul. Pasteura 5, 02--093 Warszawa, Poland}
\begin{abstract}
We consider two--dimensional Ising strip bounded by two planar, inhomogeneous walls. The inhomogeneity of each wall is modeled by a magnetic field acting on surface spins. It is equal to $+h_1$  except for a group of $N_1$ sites where it is equal to $-h_1$. The inhomogeneities of the upper and lower wall are shifted with respect to each other by a lateral distance $L$. Using exact diagonalization of the transfer matrix, we study both the lateral and normal critical Casimir forces  as well as magnetization profiles for a wide range of temperatures and system parameters. The lateral critical Casimir force tends to reduce the shift between the inhomogeneities, and the excess normal force is attractive. Upon increasing the shift $L$ we observe, depending on the temperature, three different scenarios of breaking of the capillary bridge of negative magnetization connecting the inhomogeneities of the walls across the strip. As long as there exists a capillary bridge in the system, the magnitude of the excess total critical Casimir force is almost constant, with its direction depending on $L$. By investigating the bridge morphologies  we have found a relation between the point at which the  bridge breaks and  the inflection point of the force. We provide a simple argument that some of the properties reported here should also hold for a whole range of different models of the strip with the same type of inhomogeneity.
\end{abstract}
\maketitle
\section{Introduction}\label{sec1}
The fluctuation--induced forces have been known for almost seven decades \cite{Lamoreaux99}. Historically, Hendrik Casimir was the first to point out their existence in the context of quantum electrodynamics \cite{Casimir48}. In this paper we study forces caused by fluctuations of thermodynamic system which are called critical Casimir forces. This type of interaction was first suggested by Fisher and de Gennes \cite{Fisher78}. Since then, the critical Casimir forces have been studied theoretically \cite{Evans94, Krech94, Eisenriegler95,Sprenger06}, numerically \cite{Krech96, Hucht07, Vasilyev11, ParisenToldin13} and experimentally \cite{Garcia02, Ganshin06, Hertlein08, Soyka08, Paladugu15}.

Initial studies were mostly devoted to interactions between symmetric objects like spheres or planar, homogeneous walls immersed in a critical fluid. In such cases the direction of the force follows from the symmetry of the system. When the immersed objects have a chemical or geometrical structure which breaks the symmetry, then the critical Casimir interaction is affected and the determination of the direction of the force becomes nontrivial \cite{Soyka08, Troendle11, ParisenToldin13, Bimonte15}.

In this paper we consider a two--dimensional Ising strip bounded by two planar (one--dimensional) and inhomogeneous walls. In such a system the critical Casimir force has both normal and lateral component. Using the method based on the exact diagonalization of the transfer matrix \cite{Kaufman49, Abraham73, Stecki94} we calculate and analyze both components of the force.  Additionally, we calculate and analyze the magnetization profiles and investigate the relation between the force and morphology of capillary bridges.     

Some properties of the force exerted by a capillary bridge may also prove relevant for the atomic force microscopy \cite{Binnig86, Butt05, Eaton10}. The inhomogeneity on the top wall can be considered as a model of a probe scanning an inhomogeneous bottom surface. Although our model is two--dimensional, it allows to study the effect of condensation of a liquid between the probe and the sample close to the critical point with all thermodynamic fluctuations taken into account.

We have already reported some of our results \cite{Nowakowski14}, inter alia the scaling of the lateral critical Casimir force close to the critical temperature. In this paper we provide both the full discussion of  Casimir forces  and morphology of capillary bridges as well as an extensive description of applied methods of calculations. In particular, we provide a detailed discussion of the Casimir force in the vicinity of wetting transition in the corresponding semi-infinite system.

The paper is organized as follows: In Sec.~\ref{sec2} we define the model, derive the formula for the lateral critical Casimir force and recall some properties of the two--dimensional Ising model. In Sec.~\ref{sec3} we report our results for the lateral force for different values of system parameters, in particular in some limiting cases. Sec.~\ref{sec4} is devoted to the discussion of the excess normal critical Casimir force. In Sec.~\ref{sec4:C} we study the excess total Casimir force. In Sec.~\ref{sec5} we discuss magnetization profiles and in Sec.~\ref{sec6} we study the phenomenon of nucleation and breaking up of capillary bridge. We summarize our results in Sec.~\ref{sec7}.  In App.~\ref{secA} the method of calculation of the partition function for homogeneous Ising strip is recalled and our method of modifying the surface field  to account for inhomogeneities is explained. It also contains the derivation of expressions for the free energy and magnetization. In App.~\ref{secB} an algorithm of numerical evaluation of certain matrix elements (needed to evaluate the formulas in the previous appendix) is described. Finally, in App.~\ref{secC}, we present arguments that some of the properties derived for the two--dimensional Ising strip should also hold for other models and we formulate the necessary conditions thereof.

\section{Model}\label{sec2}
\subsection{Definition of the model}

We consider a two--dimensional Ising strip on a square lattice. The state of each spin is denoted by $s_{m,n}=\pm 1$, where $m=1,2,3,\ldots,M$ and $n=1,2,3,\ldots,N$ enumerate rows and columns, respectively. The nearest--neighbor interactions are parametrized by positive coupling constant $J$. There is no bulk magnetic field. 

\begin{figure}[t]
\begin{center}
\includegraphics[width=0.9\columnwidth]{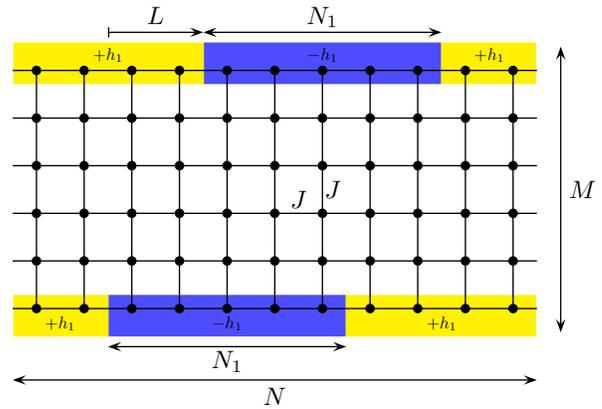}
\end{center}
\caption{\label{sec1:fig1} Schematic plot of the Ising strip. Each spin is represented by a black dot. Neighboring spins interact with a coupling constant $J$. Blue and yellow colors show inhomogeneous surface field equal to $+h_1$ and $-h_1$, respectively, which acts on the first and the last row of spins. $N$ denotes the length of the system and it is taken to be infinite, $M$ is the width of the strip, $N_1$ is the size of both inhomogeneities, and $L$ is the shift between them.}
\end{figure}

The short range interactions of the system with two inhomogeneous surrounding walls are modeled by inhomogeneous surface magnetic fields $h^\prime_n$ and $h^{\prime\prime}_n$ acting on the spins in the bottom ($m=1$) and the top ($m=M$) row, respectively. The Hamiltonian has the standard form 
\begin{multline}\label{sec1:Hamiltonian}
\Hamiltonian\left(\left\{s_{m,n}\right\};h_1,M,L,N_1\right)=-J \sum_{m=1}^M\sum_{n=1}^N s_{m,n-1}s_{m,n}\\
-J\sum_{m=1}^{M-1}\sum_{n=1}^N s_{m,n}s_{m+1,n}-\sum_{n=1}^N\left(h^\prime_n s_{1,n}+h^{\prime\prime}_n s_{M,n}\right),
\end{multline}
where periodic boundary conditions $s_{m,0}\equiv s_{m,N}$ in the lateral direction are imposed. We consider a particular type of each wall inhomogeneity. It is represented by an ``island'' of inverted surface field, i.e., $h^\prime_n=+h_1>0$ for all sites except for a group of $N_1$ subsequent sites at which $h^\prime_n=-h_1$;  similarly for the upper wall. Thus the surface fields $h^\prime_n$ and $h^{\prime\prime}_n$ are given by 
\begin{subequations}
\begin{align}
\label{sec1:hprime}h^\prime_n&=\begin{cases}-h_1 & \text{for } n=1,2,\ldots, N_1,\\ +h_1 & \text{for all other columns $n$,}\end{cases}\\
\label{sec1:hpprime}h^{\prime\prime}_n&=\begin{cases}  -h_1 & \text{for } n=L+1,L+2,\ldots, L+N_1,\\ +h_1 & \text{for all other columns $n$.} \end{cases}
\end{align}
\end{subequations}
Inhomogeneities of the walls are shifted horizontally by a distance $L$. A schematic plot of the system is presented in Fig.~\ref{sec1:fig1}.

All results presented in this paper are calculated in the limit $N\to\infty$ with all other parameters (including the width of the strip $M$) fixed. We assume that the position of the bottom wall is fixed and study the forces acting on the top wall. As our calculations assume that the system is in equilibrium, the critical Casimir force must always be balanced by an external force acting on the top wall. We do not consider any dynamic effects.

In our analysis we shall often refer to two simpler systems --- a symmetric strip (S) described by Hamiltonian \eqref{sec1:Hamiltonian} with $h^\prime_n=h^{\prime\prime}_n=h_1$ and an antisymmetric strip (AS) with $h^\prime_n=-h^{\prime\prime}_n=h_1$ for all $n$. Properties of these systems have been discussed in \cite{Evans94, Maciolek96, Nowakowski09}.

\subsection{Critical Casimir force}

To derive the expression for the free energy of the system, first, we calculate the partition function \linebreak $\partf=\sum_{\left\{s_{m,n}\right\}}\exp\left[-\Hamiltonian/\left(\kB T\right)\right]$ using the transfer matrix in the horizontal direction \cite{Abraham73}. The influence of the surface fields is represented with the help of two additional rows of spins and the wall inhomogeneities are taken into account using ``spin flip'' operators. Full derivation of the corresponding formulas is presented in App.~\ref{secA}. In this paper we consider dimensionless free energy
\begin{equation}
F_{\text{full}}\left(T,h_1,M,L,N_1,N\right)=-\ln \partf 
\end{equation}
from which we subtract the free energy $F^{\left(\mathrm{S}\right)}$ of symmetric strip of the same size and then take the limit $N\to\infty$ to obtain the reduced free energy
\begin{multline}\label{sec2:red_energy}
F\left(T,h_1,M,L,N_1\right)=\\
\lim_{N\to\infty}\left[F_{\text{full}}\left(T,h_1,M,L,N_1,N\right)-F^{\left(\mathrm{S}\right)}\left(T,h_1,M,N\right)\right].
\end{multline}
The resulting reduced free energy $F$ represents the modification of the free energy of the homogeneous system caused by the presence of a chemical pattern on the walls and contains all information about the interaction between the inhomogeneities mediated by the strip.

The critical Casimir force is defined as a minus gradient of the free energy. Here, because $L$ can take only integer values, we use a discrete version of this definition 
\begin{multline}\label{sec2:flateral}
\flateral\left(T,h_1,M,L,N_1\right)=-\Bigg[F\left(T,h_1,M,L+\frac{1}{2},N_1\right)\\
-F\left(T,h_1,M,L-\frac{1}{2},N_1\right)\Bigg],
\end{multline}
which provides the dimensionless lateral critical Casimir force for half--integer $L$.

\subsection{Characteristic temperatures and length--scales}

First, we recall the temperatures and length--scales which characterize bulk and semi--infinite two--dimensional Ising systems on a square lattice. Below the bulk critical temperature $\Tc$, given by \cite{Kramers41, Onsager44}
\begin{equation}
K_\mathrm{c}=J/\left(\kB\Tc\right)=\frac{1}{2}\ln\left(1+\sqrt{2}\right),
\end{equation}
and in the absence of bulk magnetic field, there are two coexisting phases with non--zero magnetization $\pm \mgn_0\left(T\right)$, while above $\Tc$ there is only one phase and $\mgn_0\left(T\right)=0$.

In a semi--infinite system with surface field $h_1$ one observes the wetting transition at temperature $\Tw(h_1) \leqslant \Tc$ which is defined by $W\left(\Tw,h_1\right)=1$ \cite{Abraham80}, where 
\begin{equation}
W\left(T,h_1\right)=\left(\cosh 2 \Kast+1\right)\left(\cosh 2K-\cosh 2H_1\right) 
\end{equation}
with $K=J/\left(\kB T\right)$, $H_1=h_1/\left(\kB T\right)$, and the dual coupling $\Kast$ is defined by  
\begin{equation}\label{sec2:dual}
\sinh 2K \sinh 2K^\ast=1.
\end{equation}
Note that $\Tw(h_1=0)=\Tc$ and it monotonically decreases to zero upon increasing $h_1$. For $h_1\geqslant J$, the wetting temperature is $0$ \cite{Abraham80}.

The spin--spin bulk correlation length is given by \cite{Palmer07}
\begin{equation}
\xib\left(T\right)=\begin{cases}1/\left(4K-4\Kast\right) & \text{for }  T<\Tc,\\ 1/\left(2\Kast-2K\right) & \text{for }  T>\Tc.\end{cases}
\end{equation}
Near the critical point $\xib \approx \xi_0^{\pm}\left|t\right|^{-\nu}$, where $t=\left(T-\Tc\right)/\Tc$ and $\nu=1$, $\xi_0^+=\left[2\ln\left(1+\sqrt{2}\right)\right]^{-1}$ ($\xi_0^-=\xi_0^+/2$) is a supercritical (subcritical) amplitude.

The parallel correlation lengths $\xiS$ and $\xiAS$ characterize the exponential decay of the correlation function of two spins located in the same row in a symmetric and antisymmetric Ising strip, respectively. They are given by
\begin{subequations}
\begin{align}
\label{sec2:xiS}\xiS\left(T,h_1,M\right)&=1/\left(\gamma_2+\gamma_1\right),\\
\label{sec2:xiAS}\xiAS\left(T,h_1,M\right)&=1/\left(\gamma_2-\gamma_1\right),
\end{align}
\end{subequations}
where the coefficients $\gamma_1$ and $\gamma_2$ are evaluated in the process of diagonalization of the transfer matrix, see \eqref{secA:gamma}. We check by direct calculation that in the limit $M\to\infty$%
\begin{subequations}\label{sec2:xi}
\begin{align}
\nonumber&\xiS\left(T,h_1,M\right)=\\
&\hspace{8mm}\begin{cases} A_1\left(T,h_1\right) +\mathrm{O}\left(M\ee^{-2M/\xiw}\right) & \text{for }T<\Tw,\\ \xib\left(T\right)+\mathrm{O}\left(M^{-2}\right) & \text{for } \Tw<T\neq \Tc, \end{cases}\\
\nonumber&\xiAS\left(T,h_1,M\right)=\\
\label{sec2:xiASasymp}&\begin{cases}A_2\left(T,h_1\right) \ee^{M/\xiw}+\mathrm{O}\left(M\ee^{-M/\xiw}\right) & \text{for } T<\Tw,\\ A_3\left(T,h_1\right) M^2+\mathrm{O}\left(M\right) & \text{for } \Tw<T<\Tc,\\ \xib\left(T\right)+\mathrm{O}\left(M^{-2}\right) & \text{for } T>\Tc,\end{cases}
\end{align}
\end{subequations}
where $A_1$, $A_2$, $A_3$ are positive functions and $\xiw=1/\ln W\left(T,h_1\right)$ is a length--scale proportional to the thickness of the adsorbed layer of negative magnetization (wetting layer) in a semi--infinite Ising model with a surface field $h_1 <0 $; $\xiw\sim 1/\left(\Tw-T\right)$. The exponential growth of $\xiAS$ below $\Tw$ is typical for systems with nearly spontaneously broken symmetry \cite{Parry92,Maciolek96}. This length--scale dominates for $T < \Tw$. However, it is also necessary to consider the second largest length--scale characterizing  the correlation function
\begin{equation}
\xiASp\left(T,h_1,M\right)=1/\left(\gamma_3-\gamma_1\right).
\end{equation}
Below $\Tw$ one has $\xiASp\left(T,h_1,M\right)=A_4\left(T,h_1\right)+\mathrm{O}\left(M^{-2}\right)$, where $A_4$ is a positive function, and thus, for $M$ large enough, there is a clear separation of length--scales: $\xiAS\gg\xiASp$. Note that around $\Tw$, for finite $M$, these length--scales are not separated yet. Above $\Tw$ both length--scales $\xiASp$ and $\xiAS$ are of the same order. 

Finally, we introduce the dimensionless surface tension characterizing the horizontal interface between two phases in the strip~\cite{Maciolek96}
\begin{equation}\label{sec2:sigma}
\sigma\left(T,h_1,M\right)=\gamma_1.
\end{equation}
It is defined as a difference between the dimensionless free energy per column of antisymmetric and symmetric strip and --- for finite $M$ --- remains positive even above $\Tc$. Upon increasing $M$, $\sigma\left(T,h_1,M\right)$ approaches the surface tension $\sigma_{\infty}\left(T\right)$ characterizing the coexistence of two bulk  phases 
\begin{equation}
\sigma_{\infty}\left(T\right)=\begin{cases} 2K-2\Kast & \text{for $T<\Tc$,} \\ 0 & \text{for $T\geqslant \Tc$.} \end{cases}
\end{equation}

\section{Lateral force}\label{sec3}
In this section we present the results for the lateral critical Casimir force which has been calculated numerically using \eqref{secA:freeen} and \eqref{sec2:flateral}. We also discuss various limiting cases to obtain better insight into its observed properties.

\subsection{Case: $N_1\gg L$}

\begin{figure*}[t]
\begin{center}
\begin{tabular}{lcl}
(a) & & (b)\\
\includegraphics[width=\columnwidth]{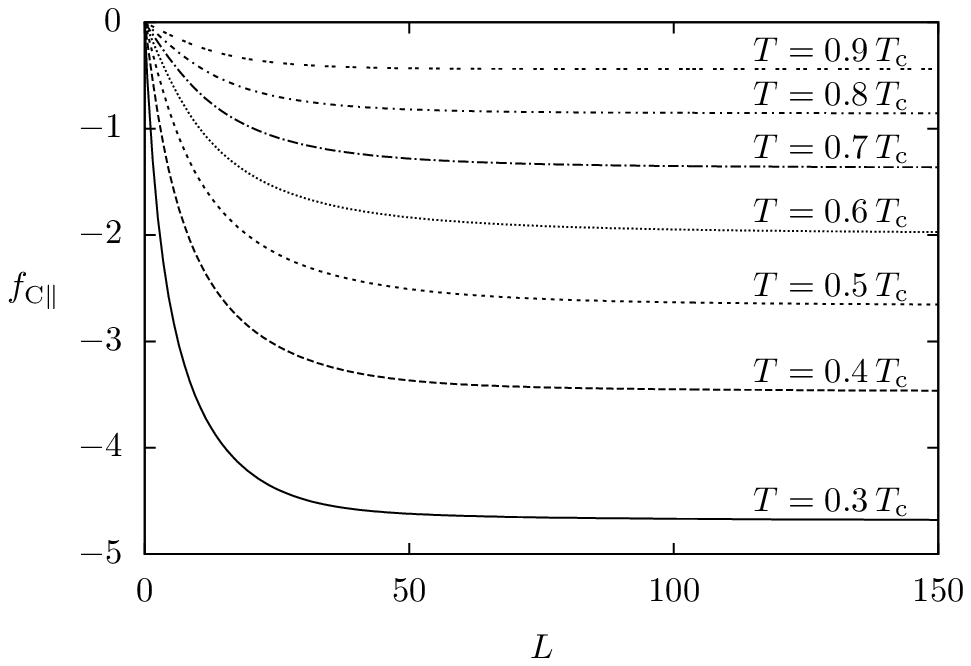} & & \includegraphics[width=\columnwidth]{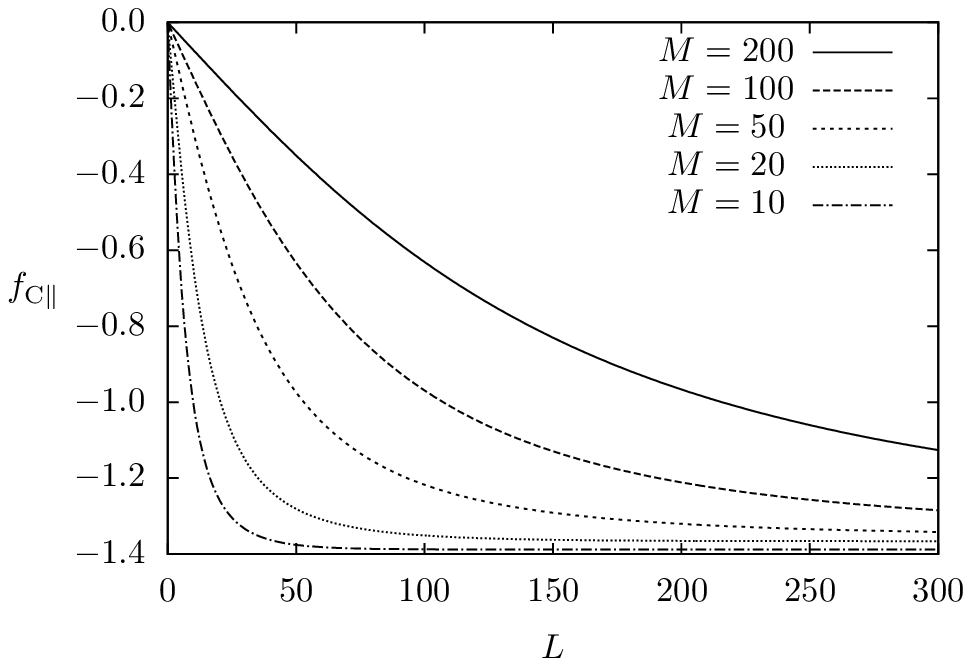}\\
(c) & & (d)\\
\includegraphics[width=\columnwidth]{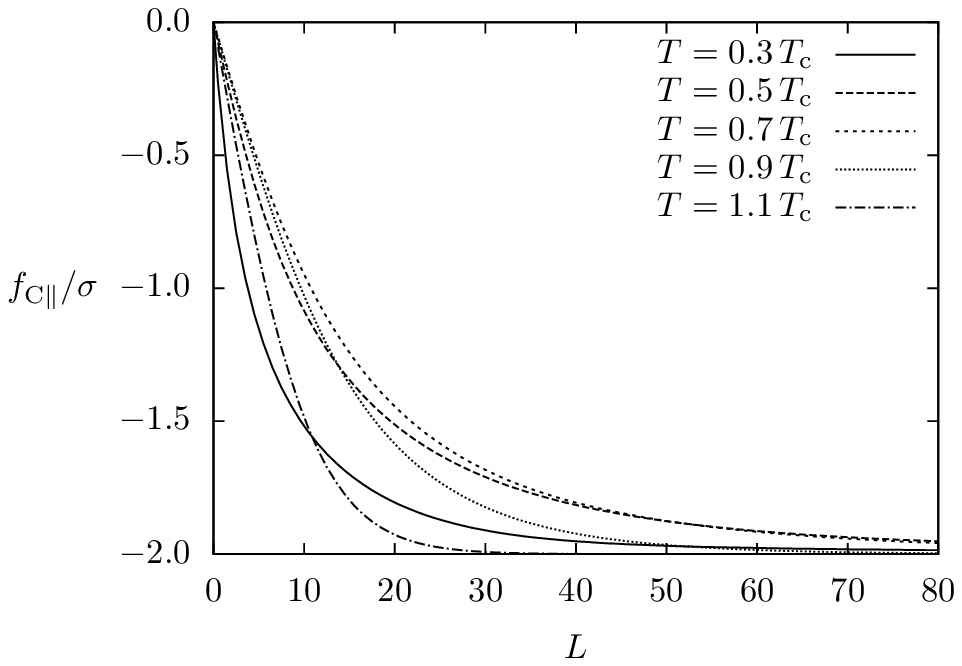} & & \includegraphics[width=\columnwidth]{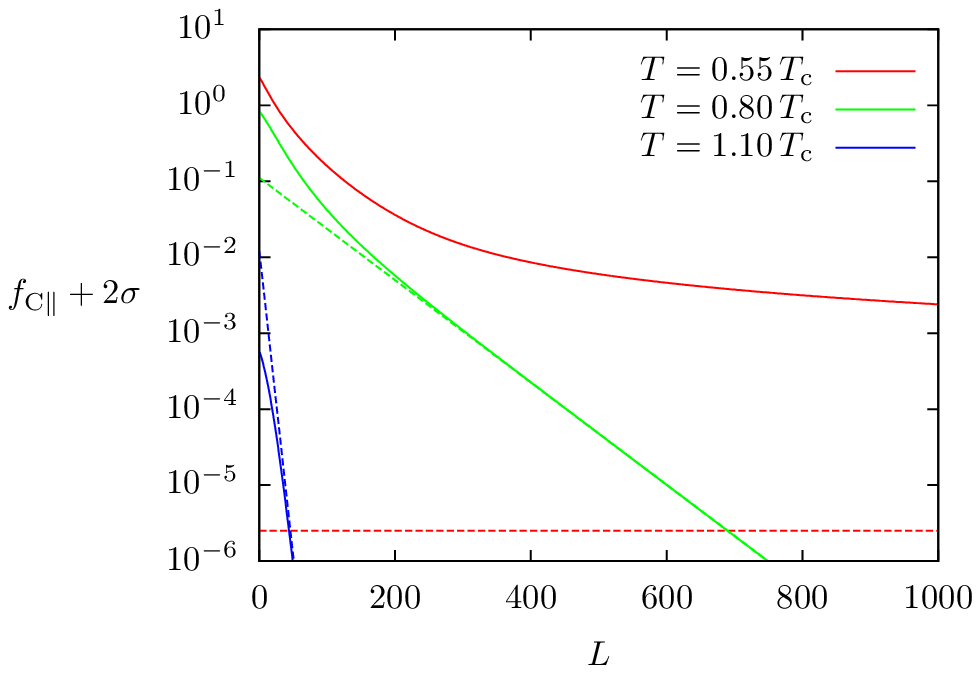}\\
\end{tabular}
\end{center}
\caption{\label{sec3:fig1} The lateral critical Casimir force for infinite size of inhomogeneities ($N_1=\infty$) as function of the shift $L$. Discrete points are connected to guide the eye. For all plots $h_1=0.8\,\Tc$ ($\Tw\approx 0.621\,\Tc$). (a) The force for different temperatures and $M=20$. (b) The force for $T=0.7\,\Tc$ and different widths of the strip. (c) Plot of the force normalized by the surface tension for $M=20$ and different values of temperature. (d) Comparison of the force (solid line) and its approximation \eqref{sec3:forceapprox1} for large $L$ (broken line) for $M=40$ and different temperatures. For $T=0.55\,\Tc$ the parallel correlation length $\xiAS\approx 795000 $ and thus the approximation breaks down for the plotted range of $L$.}
\end{figure*}

We start the analysis from the case of inhomogeneities of an infinite extent, $N_1=\infty$. Since the limit $N_1\to\infty$ is taken after the $N \to \infty$ limit, the relevant regime is $L\ll N_1\ll N$. In this limit, the expression for the lateral force obtained from \eqref{secA:freeen} takes the following form:
\begin{multline}\label{sec3:flatinf}
\flateral\left(T,h_1,M,N_1=\infty,L\right)=\\
-2 \ln \Bigg[\left(\sum_{k=1}^{M+1}t_1^{\left(k\right)}t_2^{\left(k\right)}\ee^{-\left|L-1/2\right|\gamma_k}\right)\\
\times\left(\sum_{k=1}^{M+1}t_1^{\left(k\right)}t_2^{\left(k\right)}\ee^{-\left|L+1/2\right|\gamma_k}\right)^{-1}\Bigg].
\end{multline}
The lateral force is presented in Fig.~\ref{sec3:fig1} for $L>0$. As a function of $L$ it is antisymmetric and negative for $L>0$. This means that the force acts in the direction opposite to the shift and the position $L=0$, where two inhomogeneities are exactly one above another, is (for fixed $M$) a stable mechanical equilibrium point. 

For all temperatures, upon increasing $L$ from 0, the absolute value of the force first grows linearly and then exponentially approaches its limiting value. To study this in more detail, we extract from \eqref{sec3:flatinf} the dominant contributions to the force in the limit  $L\to\infty$ 
\begin{multline}\label{sec3:forceapprox1}
\flateral\left(T,h_1,M,N_1=\infty,L\right)=-2\sigma\left(T,h_1,M\right)\\
-4 t_1^{\left(2\right)}t_2^{\left(2\right)}/\left(t_1^{\left(1\right)}t_2^{\left(1\right)}\right)\sinh\left(2\xiAS\right)^{-1} \ee^{-L/\xiAS}\\
+\mathrm{O}\left(\ee^{-L/\xiASp},\ee^{-L/\left(\xiAS/2\right)}\right).
\end{multline}
Note that the limiting value of the force is $-2\sigma\left(T,h_1,M\right)$ and the characteristic length--scale of the exponential approach to this value is $\xiAS$. Indeed, for large $L$ there are two clusters of columns with opposite surface fields, both of the length $L$ --- one at the beginning and one at the end of inhomogeneities. The contribution to free energy stemming from these two regions is roughly equal to $2 L \sigma$, which gives the limiting value of the lateral force. The leading order correction to this energy comes from the order parameter perturbation around the points at which the sign of the surface field changes. This energy does not depend on $L$ and its presence is not reflected in the expression for the force. The next correction comes from the overlap of the order parameter profiles close to the endpoints of the inhomogeneities. These points are separated by a strip of length $L$ with opposite surface fields, and the contribution to the free energy is of order of $\exp\left(-L/\xiAS\right)$. The form of the leading order correction in \eqref{sec3:forceapprox1} depends on the temperature: the term $\exp\left[-L/\left(\xiAS/2\right)\right]$ dominates for small temperatures while for high temperatures $\exp\left(-L/\xiASp\right)$ dominates. The temperature of the crossover between these two behaviors is located close to $\Tc$ and approaches it upon increasing $M$.

The plots in Fig.~\ref{sec3:fig1} are in agreement with the above picture. In Fig.~\ref{sec3:fig1}(a) we present the force for different temperatures --- the limiting value of the absolute value of the force, similarly to $\sigma$, decreases upon increasing the temperature. In Fig.~\ref{sec3:fig1}(b) the force for different widths of the strip is plotted. The force approaches its limit faster for smaller $M$, which is in an agreement with the growth of $\xiAS$ upon increasing $M$ (see \eqref{sec2:xiASasymp}). Because $\sigma$ depends only weakly on $M$, the limiting value of the force is almost the same for all relevant plots. In Fig.~\ref{sec3:fig1}(c) we have plotted normalized force for different temperatures. The approach to the limiting value is the slowest for temperatures close to $\Tw\approx 0.621\,\Tc$. For $T>\Tw$, upon increasing the temperature, the correlation length $\xiAS$ is decreasing and thus the force approaches its limit faster. For $T$ below $\Tw$ the length--scale $\xiAS$ is much larger than values of $L$ in the plot and thus the limiting value visible on the graph is slightly larger than $-2$. In Fig.~\ref{sec3:fig1}(d) the quality of approximation \eqref{sec3:forceapprox1} is checked. For $T=0.55\,\Tc<\Tw$, where $\xiAS$ is large, the approximation is useful for much larger values of $L$ than presented here.

A similar behavior of the lateral critical Casimir force was observed in the three--dimensional system in the scaling limit analyzed within the mean field theory \cite{Sprenger06}. In this paper, the case of both walls having a periodic pattern of inhomogeneities is considered. If the period of this pattern is much larger than the width of the strip and the shift is small, then --- like in our model --- upon increasing the shift, the amplitude of the force first growths linearly and then saturates at a certain value (see Fig.~26 in \cite{Sprenger06}).

\subsection{Case: $L\approx N_1\gg 1$}

In this Section we study the case when the lateral displacement between the left end of the top inhomogeneity and the right end of the bottom inhomogeneity is small (in the sense described below). We introduce a new variable $P=L-N_1$ and calculate the lateral force in the limit $N_1\to\infty$ with fixed $P$. Using \eqref{secA:freeen}, \eqref{sec2:flateral}, and \eqref{secA:tsymmetry} we obtain
\begin{widetext}
\begin{multline}\label{sec3:flatP}
\flateral\left(T,h_1,M,N_1=\infty,P\right)=\ln\Bigg\{\Bigg[\left(t_1^{\left(1\right)}t_2^{\left(1\right)}\right)^2\left(1+\ee^{-\left(2P+1\right)\gamma_1}\right)-t_1^{\left(1\right)}t_2^{\left(1\right)}\ee^{-\left(P+1/2\right)\gamma_1}\sum_{k=1}^{M+1}t_1^{\left(k\right)}t_2^{\left(k\right)}\ee^{-\left|P+1/2\right|\gamma_k}\Bigg]\\
\times\left[\left(t_1^{\left(1\right)}t_2^{\left(1\right)}\right)^2\left(1+\ee^{-\left(2P-1\right)\gamma_1}\right)-t_1^{\left(1\right)}t_2^{\left(1\right)}\ee^{-\left(P-1/2\right)\gamma_1}\sum_{k=1}^{M+1}t_1^{\left(k\right)}t_2^{\left(k\right)}\ee^{-\left|P-1/2\right|\gamma_k}\right]^{-1}\Bigg\},
\end{multline}
\end{widetext}
and the corrections to the above expression are of the order of $\exp\left(-N_1/\xiAS\right)$.

\begin{figure*}
\begin{center}
\begin{tabular}{lcl}
(a) & & (b)\\
\includegraphics[width=\columnwidth]{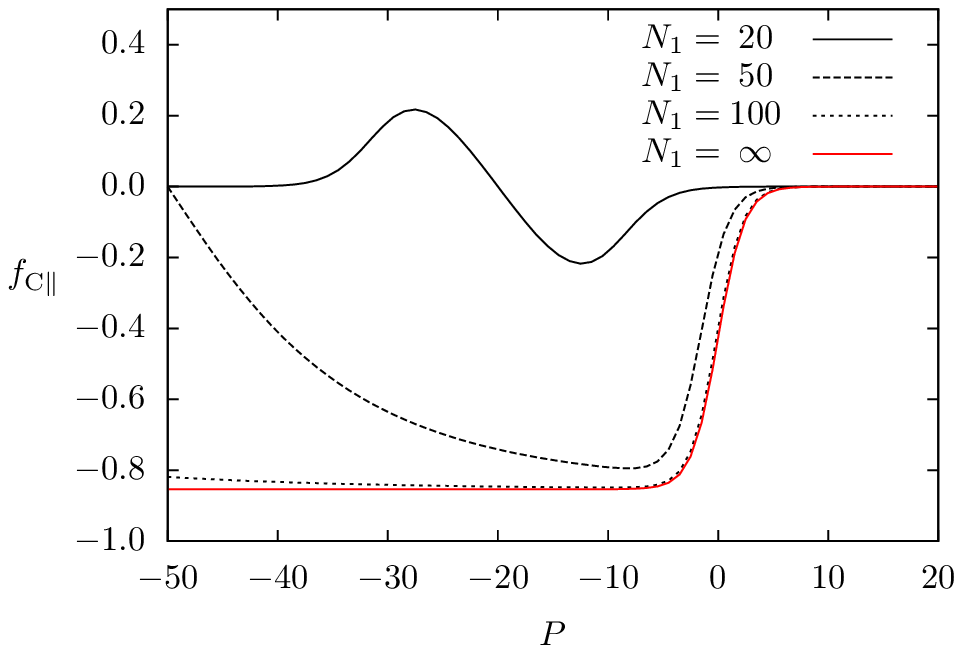} & & \includegraphics[width=\columnwidth]{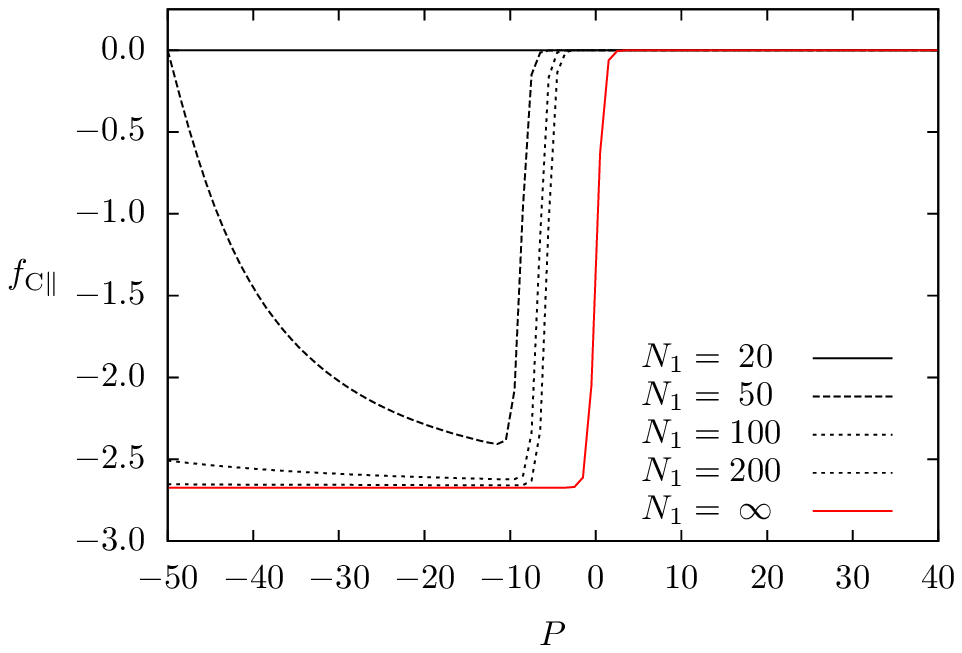}\\
(c) & & (d)\\
\includegraphics[width=\columnwidth]{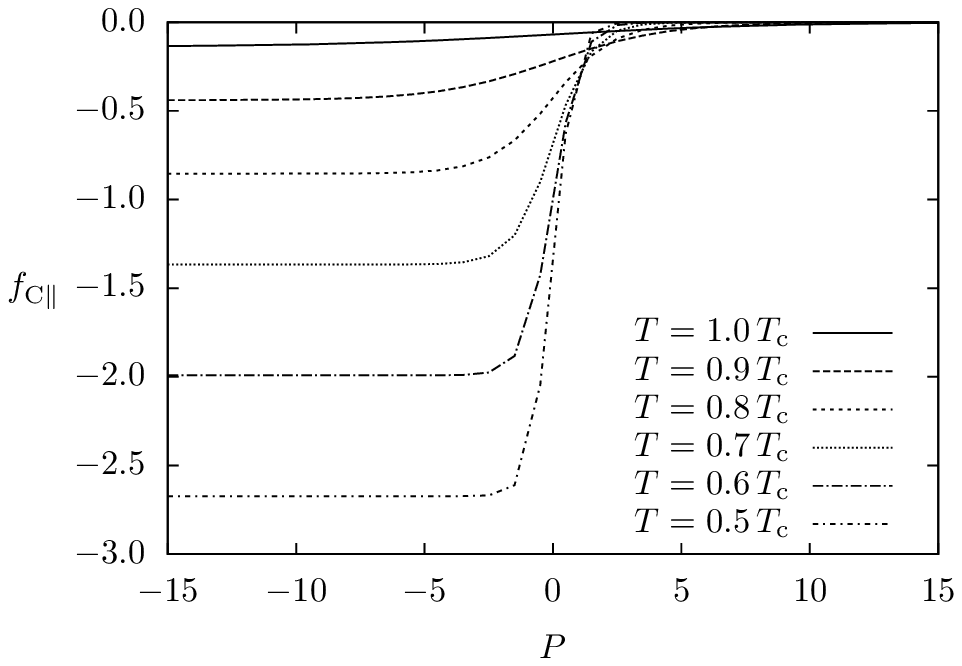} & & \includegraphics[width=\columnwidth]{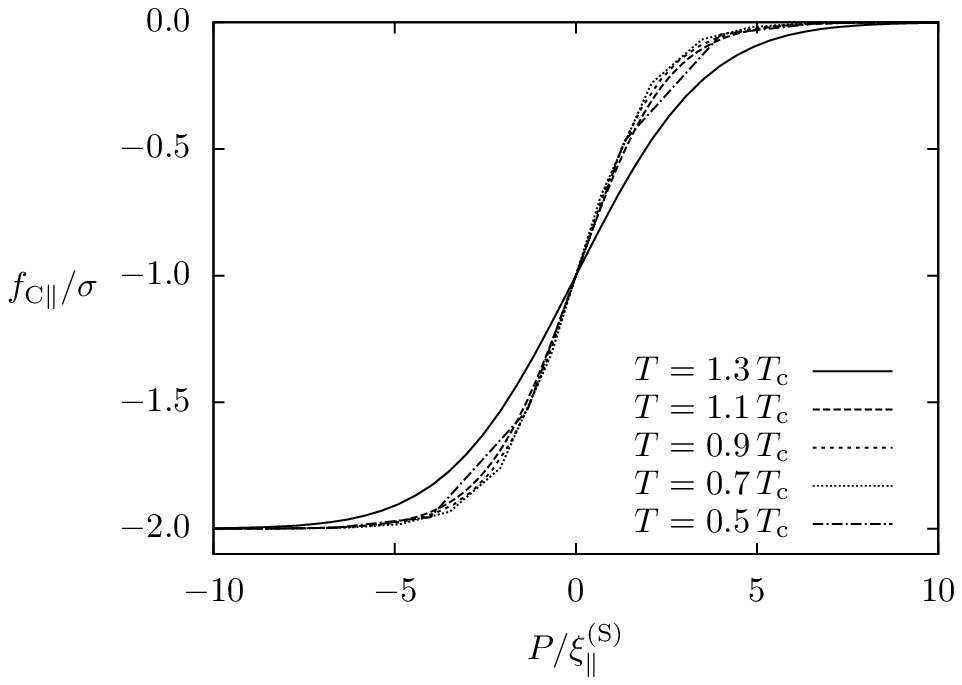}\\
\end{tabular}
\end{center}
\caption{\label{sec3:fig2} Plots of the lateral critical Casimir force as function of $P=L-N_1$ instead of $L$ as the measure of the shift. For all plots $M=20$ and $h_1=0.8\,J$ 
($\Tw \approx 0.621\,\Tc$). The lines connecting discrete points are drawn to guide the eye and the kinks visible on some plots are artifacts of this. (a)~Plots for $T=0.8\,\Tc$ and increasing values of $N_1$ converge to the plot for $N_1=\infty$; (b)~the same plot for $T=0.5\,\Tc<\Tw$ (the convergence is much slower in this case); (c)~plots of the force for $N_1=\infty$ and different temperatures; (d)~the same plot as in (c) in rescaled coordinates.}
\end{figure*}

\begin{figure*}[t]
\begin{center}
\begin{tabular}{lcl}
(a) & & (b)\\
\includegraphics[width=0.45 \textwidth]{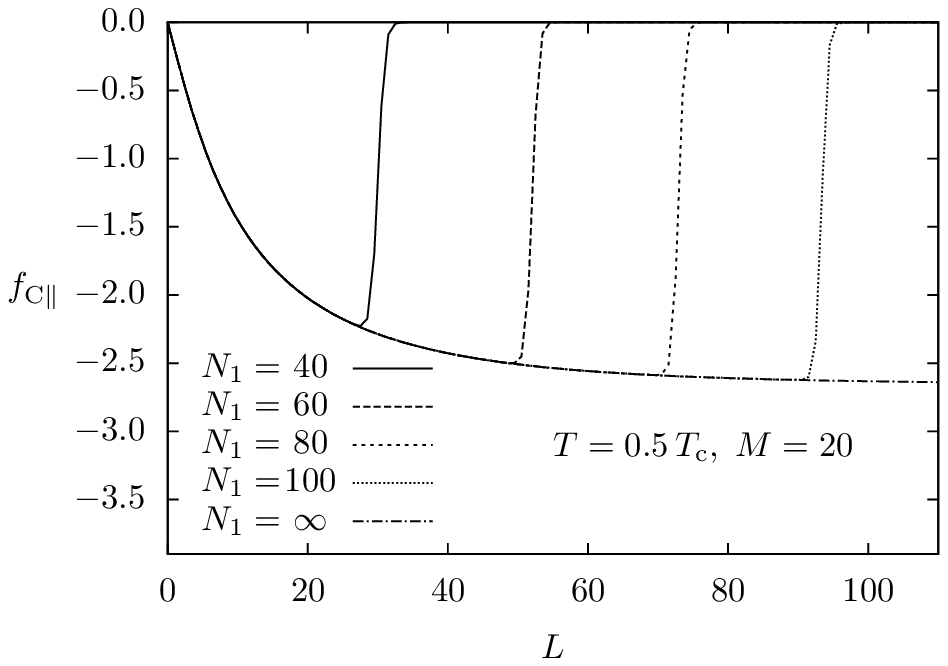} & & \includegraphics[width=0.45 \textwidth]{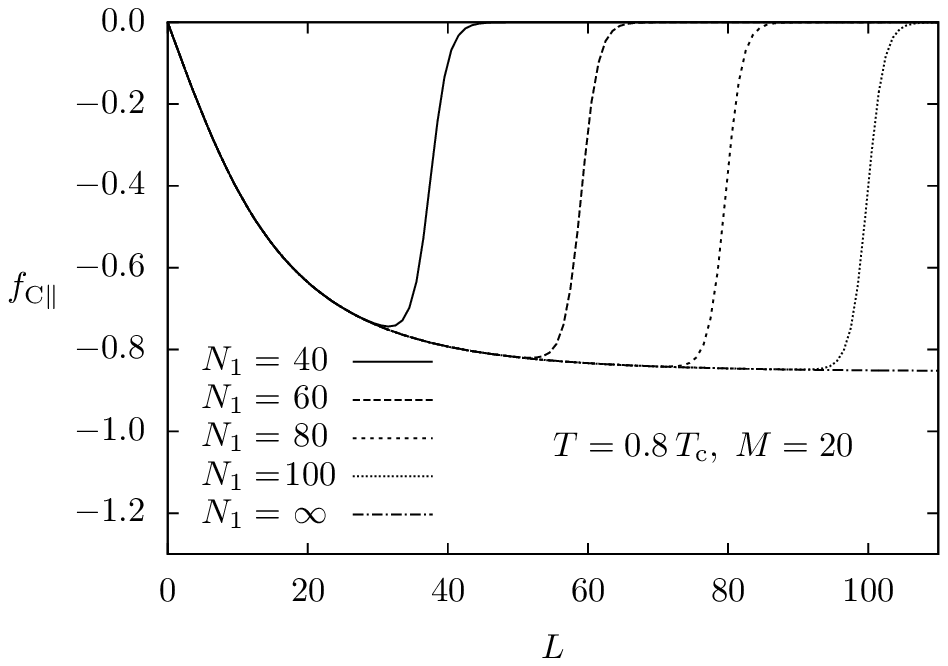}\\
(c) & & (d)\\
\includegraphics[width=0.45 \textwidth]{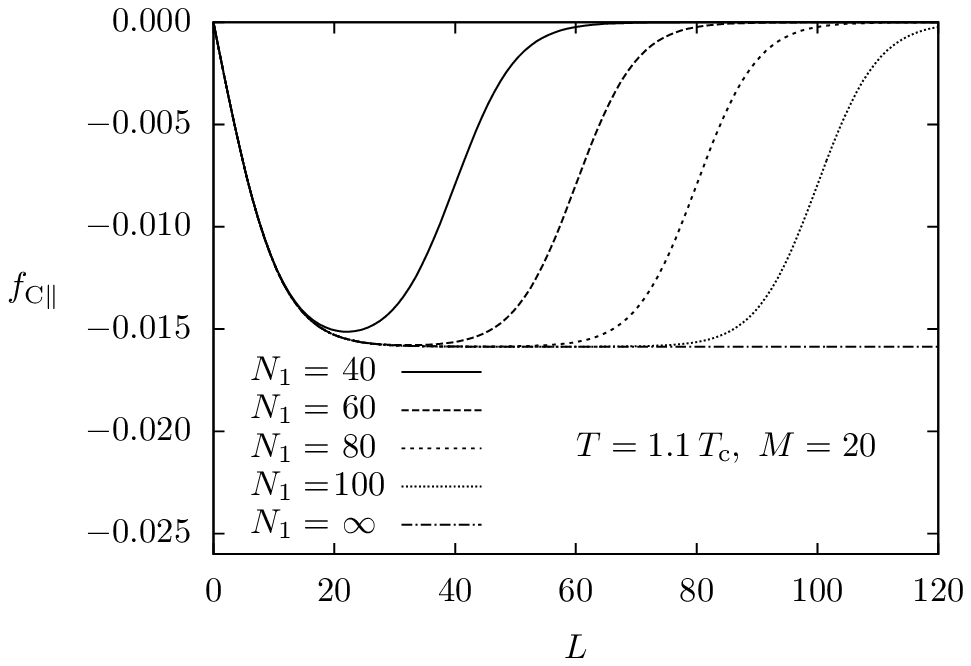} & & \includegraphics[width=0.45 \textwidth]{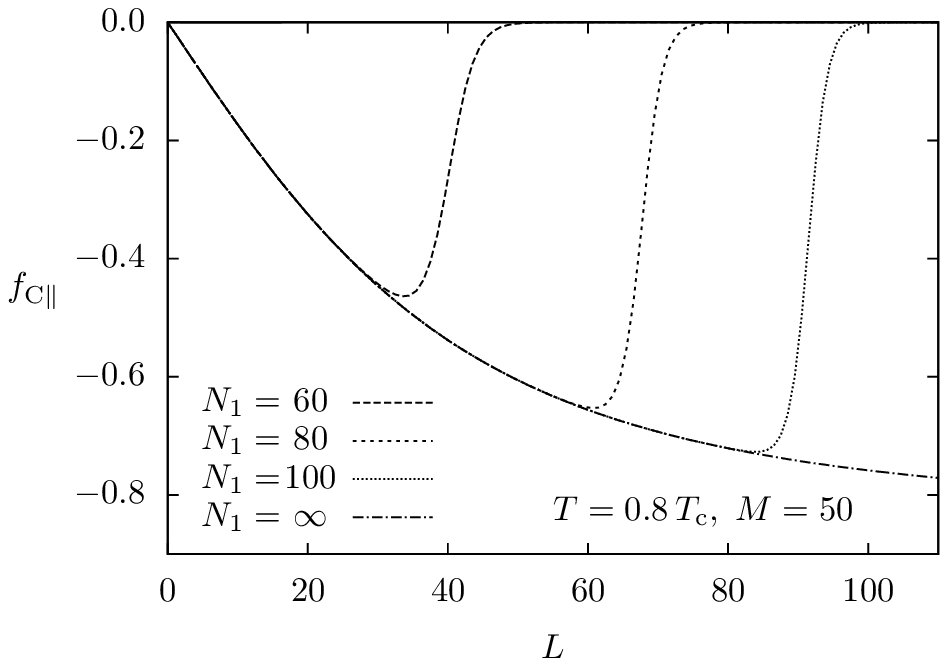}\\
\end{tabular}
\end{center}
\caption{\label{sec3:fig3} The lateral critical Casimir force as a function of $L$ for different values of $N_1$, $T$ and $M$. For all plots the surface field 
$h_1=0.8\,J$ ($\Tw\approx0.621$). Lines are drawn only to guide the eye. (a)~$T<\Tw$, (b)~and~(d)~$\Tw<T<\Tc$, and (c)~$T>\Tc$. The relevant values of  $T$ and $M$ are displayed on each plot.}
\end{figure*}

The results of numerical evaluations of the above formula are presented in Fig.~\ref{sec3:fig2}. For all temperatures the force is negative and for $P>0$ its absolute value is exponentially decreasing to zero upon increasing $P$. Thus the force acts on the upper inhomogeneity in the left direction and is large for $P\lesssim 0$. In Sec.~\ref{sec6} we investigate the relation between the decay of the force and the morphology of the capillary bridge spanning the inhomogeneities. In the limit $P\to\infty$ one obtains
\begin{multline}\label{sec3:flatPdecay}
\flateral\left(T,h_1,M,N_1=\infty,P\right)=\\
2\frac{t_1^{\left(2\right)}t_2^{\left(2\right)}}{t_1^{\left(1\right)}t_2^{\left(1\right)}}\sinh\left(2\xiS\right)^{-1} \ee^{-P/\xiS}+\mathrm{O}\left(\ee^{-P\left(\gamma_1+\gamma_3\right)}\right), 
\end{multline}
which allows to identify $\xiS$ as the scale of exponential decay of the force for large $P$.
Additionally, by direct calculations we check that
\begin{multline}\label{sec3:flatPsymmetry}
\flateral\left(T,h_1,M,N_1=\infty,P\right)\\
+\flateral\left(T,h_1,M,N_1=\infty,-P\right)=-2\sigma\left(T,h_1,M\right).
\end{multline}
Equation \eqref{sec3:flatPsymmetry} shows that for $N_1=\infty$ the force $\flateral$ has a point of symmetry $\left(P=0, \flateral=-\sigma\right)$. In App.~\ref{secC} we argue that this is quite a general property of the lateral critical Casimir force in systems with the strip geometry. The two formulas above show that the force  approaches $-2\sigma$ for $P\to-\infty$  and $0$ for $P\to\infty$. Both limiting values are approached exponentially with characteristic length--scale $\xiS$.

The above results are displayed in Fig.~\ref{sec3:fig2}. In Fig.~\ref{sec3:fig2}(a) the lateral force is plotted for $T=0.8\,\Tc$ and different values of $N_1$. The curves corresponding to increasing $N_1$--values approach the red line plotted for $N_1=\infty$, see \eqref{sec3:flatP}. In Fig.~\ref{sec3:fig2}(b) the same comparison is done for $T=0.5\,\Tc$, which is below $\Tw$. This time the curves converge to the $N_1=\infty$ line much slower and the differences are still visible for $N_1=200$. This is related to the fact that the difference between the plots for finite and infinite $N_1$ is of the order of $\exp\left(-N_1/\xiAS\right)$, and for $T=0.5\,\Tc<\Tw$ the length--scale $\xiAS$ is much larger than $200$, see \eqref{sec2:xiASasymp}. In Fig.~\ref{sec3:fig2}(c) the plots of the force for $N_1=\infty$ and different temperatures are displayed. Despite different limiting values for $P\to-\infty$ and different length--scales of the decay, all the functions have a point of symmetry. In Fig.~\ref{sec3:fig2}(d) we present the same functions in rescaled coordinates. This confirms that the limiting value for large negative $P$ is $-2\sigma$ and the scale of decay is given by $\xiS$ for all temperatures. In these new coordinates the plots do not coincide and thus one concludes that it is not possible to describe them with a single, temperature--independent scaling function.

\subsection{Finite inhomogeneities}

In this section, on the basis of the formula \eqref{secA:freeen}, we numerically evaluate the lateral force for finite values of $N_1$.

The behavior of the force for large values of $L$ can be investigated by expanding the formula for the force in the limit $L\to\infty$
\begin{multline}
\flateral\left(T,h_1,M,N_1,L\right)=8 t_1^{\left(1\right)}t_1^{\left(2\right)}t_2^{\left(1\right)}t_2^{\left(2\right)}\\
\times\left(\sum_{k=1}^{M+1}t_1^{\left(k\right)}t_3^{\left(k\right)}\ee^{-N_1\gamma_k}\right)^{-2}\cosh^2\left(\frac{N_1}{2\xiAS}\right)\\
\times\sinh\frac{1}{2\xiS}\ee^{-L/\xiS}+\mathrm{O}\left(\ee^{-L\left(\gamma_1+\gamma_3\right)}\right).
\end{multline}

Note that the characteristic length--scale is given by $\xiS$. The same type of decay has been observed in the limit $N_1\to\infty$ with $P=L-N_1$ fixed, see \eqref{sec3:flatPdecay}. 

The plots are presented in Fig.~\ref{sec3:fig3}. Generally, for small values of $L$ the force is almost the same as for $N_1=\infty$ and small $L$. When $L\approx N_1$, the decay is very similar to the one observed for $N_1=\infty$ and small $P$. The force is negative for $L>0$ and, upon increasing $L$, its absolute value increases from 0, reaches a maximum for $L=\Last$ and then decreases exponentially towards $0$. We study the behavior of $\Last$ in Sec.~\ref{sec6:comparison}.

Comparison of Fig.~\ref{sec3:fig3}(b) and Fig.~\ref{sec3:fig3}(d) shows how the force changes upon increasing $M$ --- the decay of the force becomes less rapid and takes place at smaller values of $L$. This observation is in agreement with the fact that for large $N_1$ the rapid decay of the lateral force occurs around $L=N_1$ with the corrections of the order of $\exp\left(-N_1/\xiAS\right)$, which follows from \eqref{sec3:flatPdecay}. Increasing $M$ increases $\xiAS$ (see \eqref{sec2:xiASasymp}) and thus makes the corrections more pronounced.

Note that all above plots correspond to $N_1>M$. When $N_1$ is small, the force behaves quite differently --- it is very small and locally symmetric around its minimum at $L=\Last$; the crossover is observed for $N_1\approx M$.

The properties of the force reported here can be related to known results for a colloidal particle immersed in the critical fluid above an inhomogeneous wall \cite{Troendle10, Troendle11}. In our model the role of the colloidal particle is played by the inhomogeneity on the top wall. Although the reported results are for a three--dimensional system close to the critical point, when the size of the particle is of the order of the width of the inhomogeneity on the wall, the observed lateral force is in qualitative agreement with our results.

\section{Normal force}\label{sec4}
We now move to the study of the force acting in the direction normal to the walls. Because the dependence of energy on $M$ in our formulas is implicit, our analysis necessarily has numerical character.

\subsection{Definition}

The leading contribution to the vertical component of the force in the strip is the same as in a homogeneous symmetric Ising strip. It is proportional to the length of the system $N$ and thus becomes infinite in the thermodynamic limit. This vertical force in a homogeneous system has already been studied \cite{Evans94, Nowakowski09} and therefore we do not repeat the discussion of this dominant term here. The next term stems from the interaction of the inhomogeneity on one wall with a homogeneous second wall --- it dominates for $L\gg N_1$. Although, to our knowledge, this term has not been studied in detail yet, we concentrate on the force due to the interaction between two inhomogeneities. Therefore, we focus on the excess free energy of the inhomogeneous strip
\begin{multline}\label{sec4:excess_en}
F_{\text{excess}}\left(T,h_1,M,N_1,L\right)=\\
F\left(T,h_1,M,N_1,L\right)-\lim_{L\to\infty}F\left(T,h_1,M,N_1,L\right),
\end{multline}
which is the free energy of our system after subtracting both above contributions. In this way one obtains the contribution to the free energy due to the interaction between the two inhomogeneities on the walls mediated by the Ising strip. The function $F$ in the above equation is the reduced free energy of the strip \eqref{sec2:red_energy}. Note that $F_\text{excess}$ is evaluated in the thermodynamic limit $N\to\infty$ and vanishes for $L\to\infty$. Since the difference between $F_\text{excess}$ and $F$ does not depend on $L$, this change does not modify the lateral component of the critical Casimir force. After a straightforward calculation, from \eqref{secA:freeen} we get
\begin{multline}
F_{\text{excess}}\left(T,h_1,M,N_1,L\right)=F\left(T,h_1,M,N_1,L\right)\\
+2\ln\left(\sum_{k=1}^{M+1}t_1^{\left(k\right)}t_3^{\left(k\right)}\ee^{-N_1 \gamma_k}\right).
\end{multline}

Similarly to \eqref{sec2:flateral},  the excess normal critical Casimir force is evaluated via the discrete version of the gradient as
\begin{multline}\label{sec4:fnormal}
\fnormal\left(T,h_1,M,N_1,L\right)=\\
-\Bigg[F_{\mathrm{excess}}\left(T,h_1,M+\frac{1}{2},N_1,L\right)\\
- F_{\mathrm{excess}}\left(T,h_1,M-\frac{1}{2},N_1,L\right)\Bigg],
\end{multline}
where $M\in\mathbb{Z}+\frac{1}{2}$ and $L,N_1\in\mathbb{Z}$. Note, that the allowed values of $M$ and $L$ are different than those in the case of $\flateral$.

\begin{figure*}
\begin{center}
\begin{tabular}{lcl}
(a) & & (b)\\
\includegraphics[width=\columnwidth]{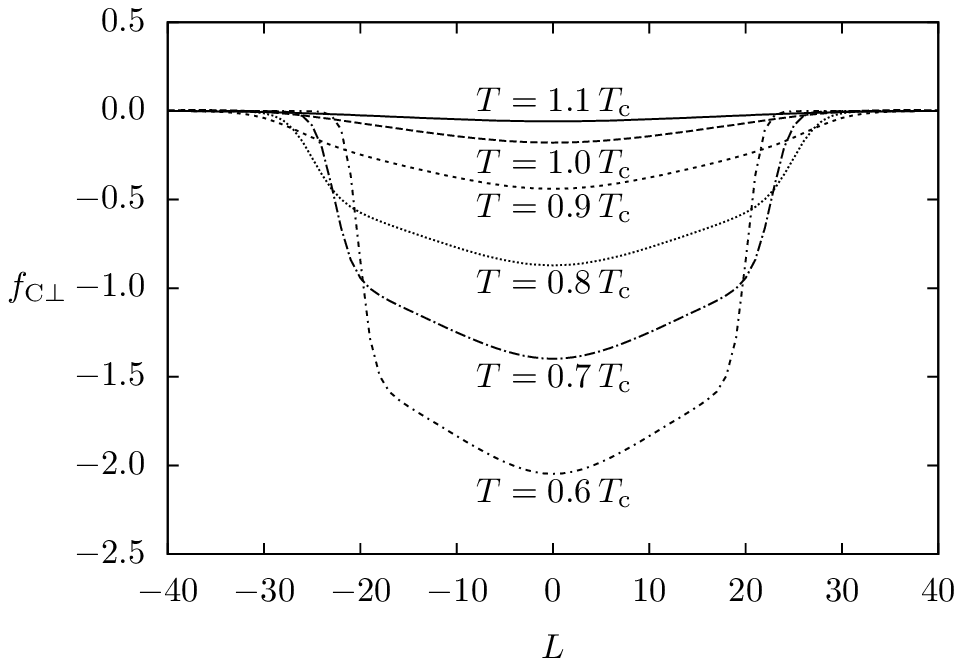} & & \includegraphics[width=\columnwidth]{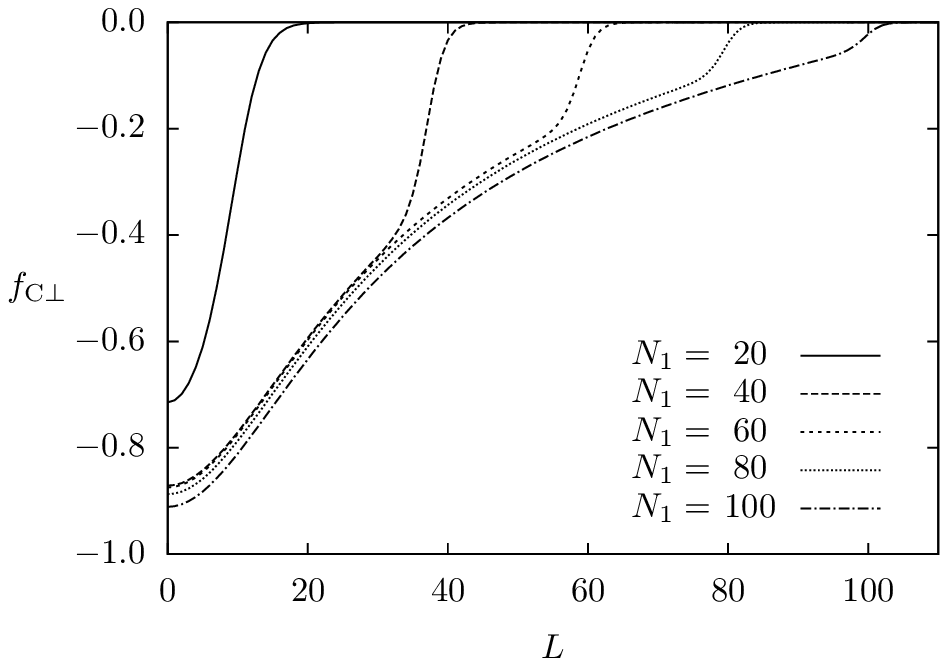}\\
\end{tabular}
\end{center}
\caption{\label{sec4:fig1} Normal critical Casimir force as a function of the shift $L$. $M=20.5$ and $h_1=0.8\,J$ on both plots. The discrete points are connected to guide the eye. (a) The force for different temperatures and $N_1=30$, (b) the force for $T=0.8\,\Tc$ and different values of $N_1$.}
\end{figure*}

\begin{figure*}
\begin{center}
\begin{tabular}{lcl}
(a) & & (b)\\
\includegraphics[width=\columnwidth]{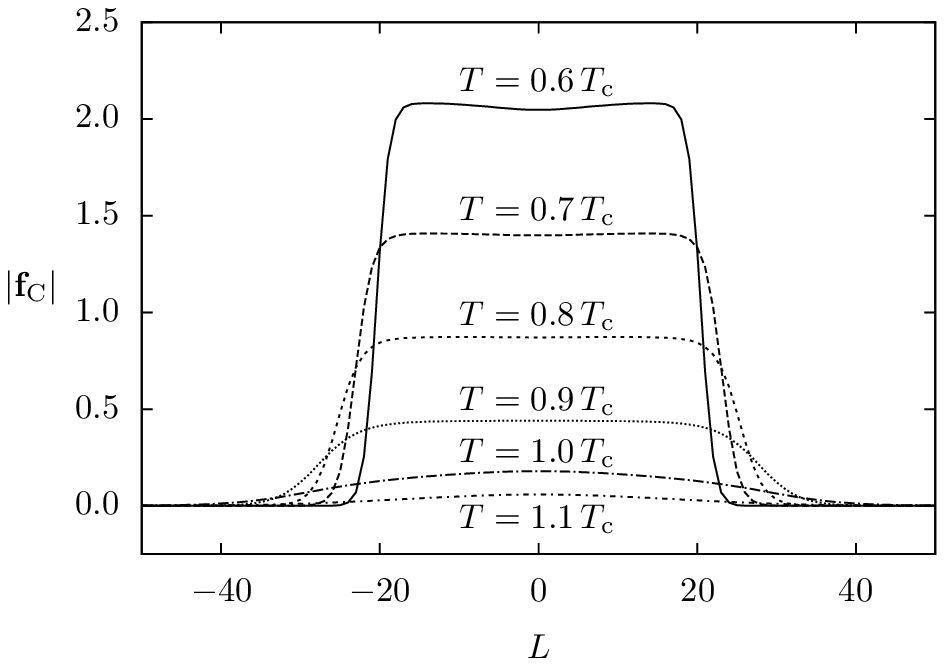} & & \includegraphics[width=\columnwidth]{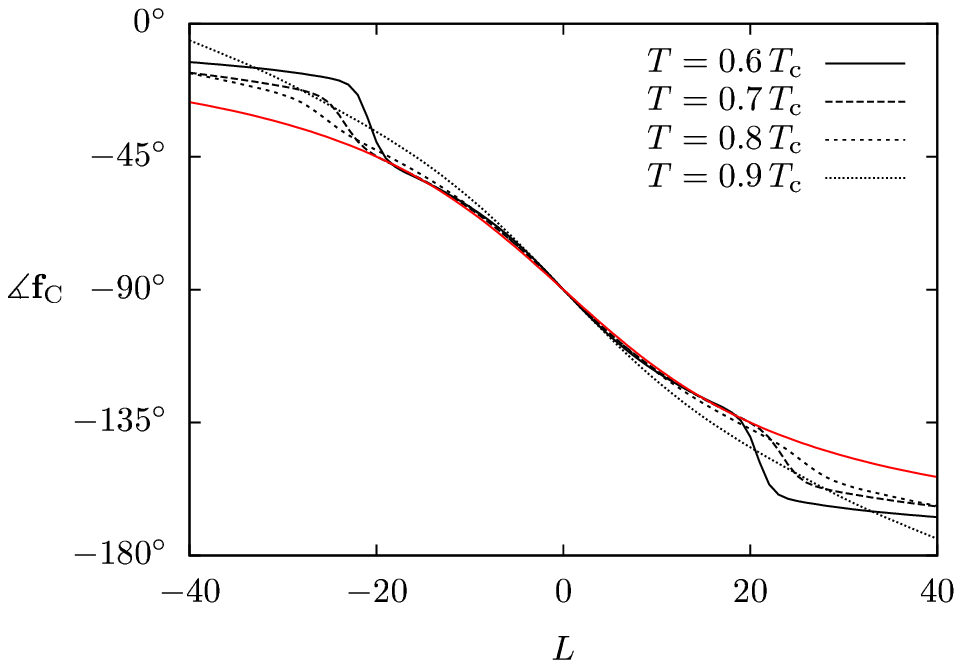}\\
\end{tabular}
\end{center}
\caption{\label{sec4:fig2} The excess total critical Casimir force which acts on the top wall as function of the shift $L$. For all plots $M=20$, $N_1=30$ and $h_1=0.8\,J$. Discrete points were connected to guide the eye. (a) The absolute value (length of the vector) of the force for different temperatures. (b) The angle at which the force acts. The angle $0^\circ$ means that the force acts to the right, $-90^\circ$ means that the force act to the bottom and $-180^\circ$ means that the force acts to the left. The red line shows the angle of the auxiliary line connecting the midpoints of the two inhomogeneities.}
\end{figure*}

\subsection{Excess normal force}

The excess normal critical Casimir force evaluated for different values of parameters is presented in Fig.~\ref{sec4:fig1}. The force is an even function of $L$ and is negative (attractive) for all values of $L$. The absolute value of the force is maximal for $L=0$ and is decreasing upon increasing $L$. For $L$ close to $N_1$ the absolute value of the force rapidly decreases to zero 

In Fig.~\ref{sec4:fig1}(a) we compare the excess normal force for different temperatures. The minimum at $L=0$ gets deeper upon decreasing $T$. This behavior can be explained by noting that for $L=0$ and $N_1>M$ there are two vertical interfaces connecting the edges of the inhomogeneities along the strip. In the first approximation, the energy of these interfaces is $2M\sigma_\infty$, and the leading corrections come from the finite length of interfaces and from the interaction between them. Indeed, the value of the force in the minimum differs a little from $-2\sigma_\infty$. This is especially visible for $T\geqslant\Tc$, where the minimum is still non--zero although $\sigma_\infty=0$. 

In Fig.~\ref{sec4:fig1}(b) we study the dependence of the force on $N_1$. For $L<N_1$, unlike for the lateral force, the excess normal force still decreases a little upon increasing $N_1$. This is probably caused by the interaction between two interfaces present in the system. For $L\approx N_1$ the force decays to zero quite rapidly. Finally, when $N_1<M$ the force rapidly decreases upon decreasing $N_1$. The plot of the force for $N_1=10$ is not included in Fig.~\ref{sec4:fig1}(b) because it would be indistinguishable from $\fnormal=0$. To our knowledge, the problem of  interaction between two interfaces in a two--dimensional Ising model has not been studied yet.

To our knowledge, the properties of the excess normal critical Casimir force reported in this paper have not been studied so far. The normal force between a wall with periodic pattern and a homogeneous wall discussed in \cite{Sprenger06, ParisenToldin10} should become similar to the normal force for $L\to \infty$; this case, however, was not considered here.

\subsection{Excess total force}\label{sec4:C}

In this section we analyze the absolute value and the direction of the excess total critical Casimir force which acts on the top wall.  We define the excess total force as a two--dimensional vector
\begin{widetext}
\begin{multline}\label{sec4:ftotal}
\ftotal\left(T,h_1,M,N_1,L\right)=\frac{1}{2}\Bigg[\flateral\left(T,h_1,M,N_1,L+\frac{1}{2}\right)+\flateral\left(T,h_1,M,N_1,L-\frac{1}{2}\right),\\
\fnormal\left(T,h_1,M+\frac{1}{2},N_1,L\right)+\fnormal\left(T,h_1,M-\frac{1}{2},N_1,L\right) \Bigg],
\end{multline}
\end{widetext}
where $M,L\in\mathbb{Z}$. The averaging of the forces inherent in \eqref{sec4:ftotal} was introduced to tackle the problem of different domains of $\flateral$ and $\fnormal$: the lateral force is defined for integer $M$ and half--integer $L$, while the normal force is defined for half--integer $M$ and integer $L$.

The results are presented in Fig.~\ref{sec4:fig2}. In Fig.~\ref{sec4:fig2}(a) the absolute value of the force is plotted for different temperatures. The force is an even function of the shift $L$. For small $L$ the force is almost constant and, upon increasing $L$, for $L\approx N_1$ it starts to decrease to $0$. A more detailed analysis shows that the plateau for small $L$ takes the form of a shallow minimum for $L=0$ surrounded by two maxima. This behavior, which is clearly visible on the plot for $T=0.6\,\Tc$, is also present for all temperatures below $\Tc$. The length--scale characterizing the decay for $L\approx N_1$ is growing upon increasing $T$. For $T>\Tc$ it so large that the plateau has disappeared. In Fig.~\ref{sec4:fig2}(b) we study the angle at which the force acts. It is decreasing upon increasing $L$ and has a point of symmetry at $\left(0,-90^\circ\right)$ --- this symmetry is a simple consequence of the symmetries of $\flateral$ and $\fnormal$. For small $L$ the angle is almost the same as the angle of the line connecting midpoints of the two inhomogeneities (the values of this angle are denoted by red line in Fig.\ref{sec4:fig2}(b)). For larger $L$, when the absolute value of the force starts to decrease, the angle decreases, which means that the horizontal component of the force decays slower than the vertical one.

From our numerical analysis it follows that for $N_1$ large enough, when $\left|L\right|$ is smaller than $N_1$, the force is almost constant and it is almost parallel to the direction of the line connecting midpoints of inhomogeneities. When $\left|L\right|$ is larger than $N_1$, the force is very small and almost horizontal. Note, however, that this result depends on the way in which the averaging inherent in the definition of the total force \eqref{sec4:ftotal} is implemented. In particular, a continuous model is free from this problem. There, one could compare the normal and lateral force evaluated at exactly the same point.

\section{Magnetization}\label{sec5}
Our numerical analysis of magnetization is based on the formulas derived in App.~\ref{secA}~and~\ref{secB}. The complexity of the algorithm used to calculate the relevant matrix elements grows fast with the width of the strip $M$, and allows us to determine the magnetization profiles in inhomogeneous system only for $M<24$.

\subsection{Homogeneous strip}

The magnetization in a homogeneous strip has already been studied in \cite{Stecki94,Maciolek96}; below, we  recall the major results relevant for our analysis of inhomogeneous case.

\begin{figure*}
\begin{center}
\begin{tabular}{lcl}
(a) & & (b)\\
\includegraphics[width=\columnwidth]{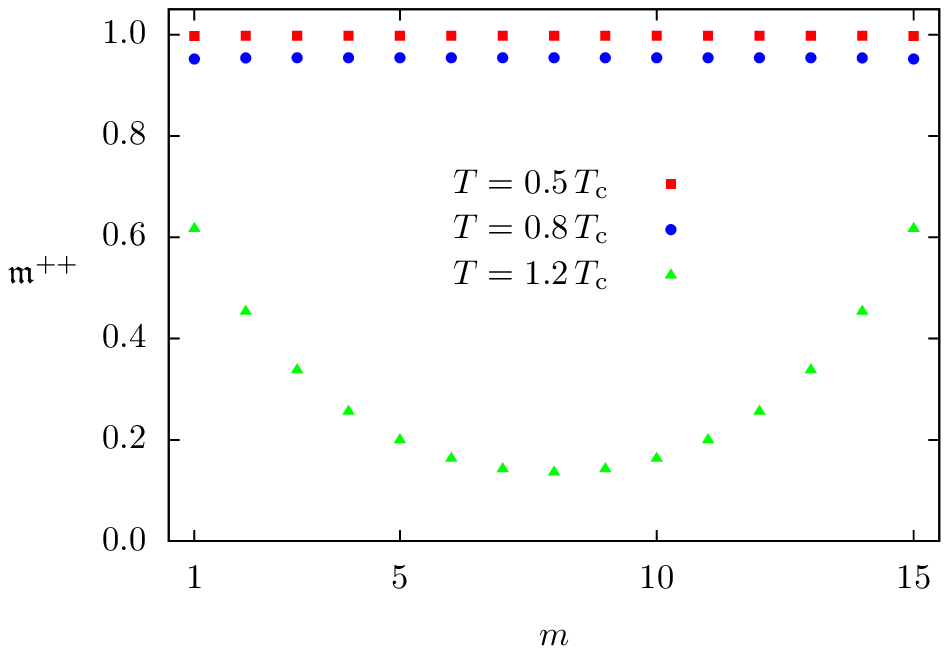} & & \includegraphics[width=\columnwidth]{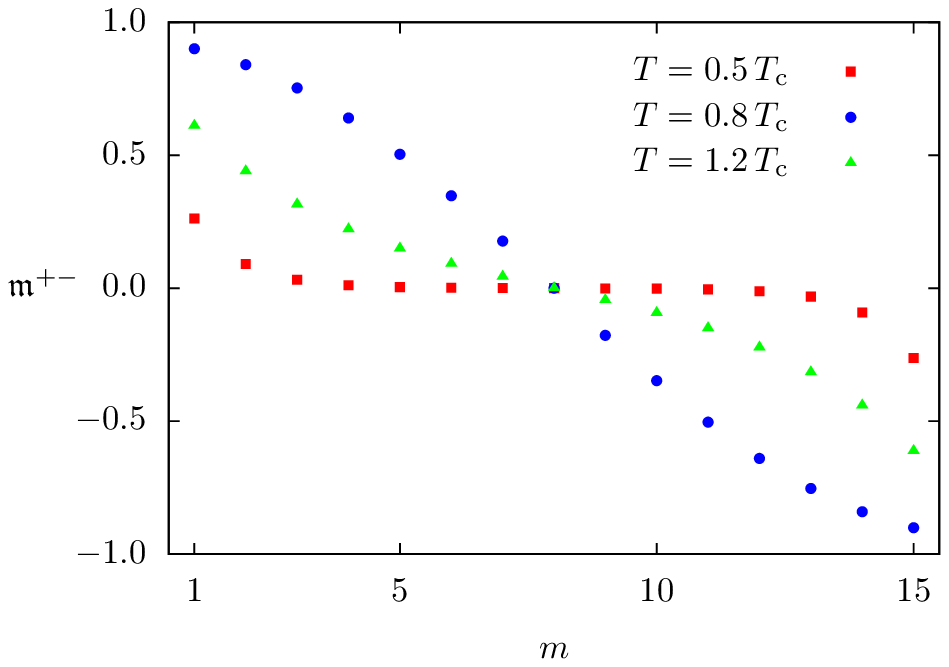}\\
\end{tabular}
\end{center}
\caption{\label{sec5:fig1} Magnetization profiles in homogeneous two--dimensional Ising strip for different temperatures. For both plots $M=15$ and $h_1=0.8\,J$ ($\Tw\approx0.621\,\Tc$). (a) Symmetric case (b) antisymmetric case. }
\end{figure*}

Above $\Tc$, for both symmetric and antisymmetric walls, the magnetization in the strip is almost equal to its supercritical bulk value $\mgn_0\left(T\right)=0$, except for the vicinity of the walls, where the non--zero magnetization is induced by the surface field. Below $\Tc$ the magnetization profile depends on the surface fields. For symmetric case, the magnetization in the middle of the strip practically attains the bulk value $+\mgn_0\left(T\right)$ (for $h_1>0$). Close to the walls this bulk--like value is modified, and, depending on the strength of the surface field, the magnetization is either decreased or increased. For antisymmetric case the situation is more complicated. There exists an interface separating regions with positive and negative magnetization which undergoes a delocalization transition \cite{Parry92} at an $M$--dependent temperature $T_{\text{w},M}\approx\Tw$.  Above this temperature, the interface is located in the middle of the system, while below it the interface is pinned to one of the walls. However,  due to the symmetry of the system, the probability of pinning is the same for both walls and the resulting equilibrium magnetization profile has almost zero value in the middle of the strip. Note that even though the interface can be in two different positions below $T_{\text{w},M}$, there is no phase transition in the system (for finite $M$ the system is quasi one--dimensional and cannot have a phase transition \cite{vanHove50}). Instead, the interface jumps from one wall to another on the length--scale of $\xiAS$ \cite{Parry92}. Typical magnetization profiles are presented in Fig.~\ref{sec5:fig1}.

The magnetization profiles for a homogeneous strip can be obtained from our results by an appropriate limiting procedure:
\begin{subequations}
\begin{align}
&\mgn^{++}_{m}\left(T,h_1,M\right)=\lim_{n\to\infty}\mgn_{m,n}\left(T,h_1,M,N_1,L\right)=\Xi^m_;,\\
&\nonumber\mgn^{+-}_{m}\left(T,h_1,M\right)=\\
&\lim_{n\to-\infty}\lim_{L\to-\infty}\lim_{N_1\to\infty}\mgn_{m,n}\left(T,h_1,M,N_1,L\right)=\Xi^m_{1;1}.
\end{align}
\end{subequations}
These expressions for $\mgn^{++}_{m}$ and   $\mgn^{+-}_{m}$ are identical with those discussed in \cite{Stecki94}.

\subsection{Inhomogeneity on one wall}

In this section we discuss an auxiliary inhomogeneous strip with an inhomogeneous bottom wall and a homogeneous upper wall. The Hamiltonian of this system is given by \eqref{sec1:Hamiltonian} with the surface field $h_n^\prime$ given by \eqref{sec1:hprime}, and $h_n^{\prime\prime}=+h_1$ for all $n$. The magnetization profile for such a system can be obtained by a direct calculation, but here we use our results derived in App.~\ref{secA} in the limit $L\to\infty$ instead. 

When the length of the bottom wall inhomogeneity is infinite ($N_1=\infty$), the magnetization takes the following form
\begin{multline}
\mgn_{m,n}^{1}\left(T,h_1,M,N_1=\infty\right)=\\
\begin{cases}\displaystyle \Xi^m_; +\sum_{k=1}^{M+1}t_3^{\left(k\right)}/t_1^{\left(1\right)} \ee^{n\left(\gamma_k+\gamma_1\right)}\Xi^m_{;k,1} & \text{for }n\leqslant 0,\\
\displaystyle-\sum_{k=1}^{M+1} t_1^{\left(k\right)}/t_1^{\left(1\right)}\ee^{-n\left(\gamma_k-\gamma_1\right)} \Xi^m_{k;1} & \text{for }n>0. \end{cases}
\end{multline}
For the definitions of coefficients $\gamma_k$, $t_i^{\left(k\right)}$, and matrix elements $\Xi^m$ see \eqref{secA:gamma}, \eqref{secA:t}, and \eqref{secA:mel}, respectively. In the limit $\left|n\right|\to\infty$ this formula simplifies to
\begin{subequations}
\begin{align}
\nonumber&\mgn_{m,n}^{1}\left(T,h_1,M,N_1=\infty\right)=\Xi^m_;+\mathrm{O}\left(\ee^{-\left|n\right|/\xiS}\right)\\
&\hspace{5.3cm}\text{for }n\to-\infty,\\
\nonumber&\mgn_{m,n}^{1}\left(T,h_1,M,N_1=\infty\right)=-\Xi^m_{1,1}\\
\label{sec5:mgn1infasymp}&-t_1^{\left(2\right)}/t_1^{\left(1\right)}\ee^{-n/\xiAS}\Xi^{m}_{2;1}+\mathrm{O}\left(\ee^{-n/\xiASp}\right) \text{ for }n\to\infty,
\end{align}
\end{subequations}
which shows that away from the points of discontinuity of the surface field the magnetization profile along the column is the same as in the homogeneous strip (for column number $n\to\infty$ the surface field is negative at the bottom wall and positive at top wall, thus $\mgn^1_{m,n}\to \mgn_m^{-+} = -\mgn_m^{+-}$). The perturbation of magnetization caused by the change of the sign of the surface field on one wall extends on the length--scale $\xiS$ towards the symmetric side and on length--scale $\xiAS$ towards the antisymmetric side. Since for all temperatures $\xiS<\xiAS$, the perturbation is always asymmetric.

\begin{figure*}
\begin{center}
\begin{tabular}{lcl}
(a) & & (b)\\
\includegraphics[width=\columnwidth]{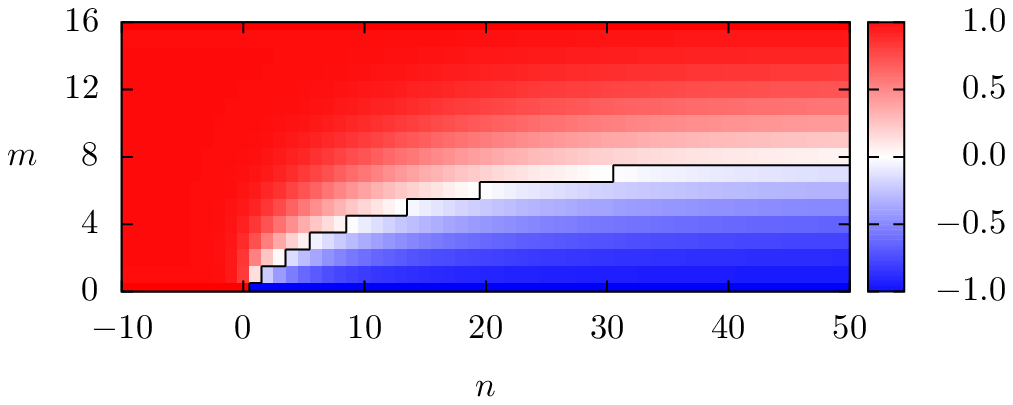} & & \includegraphics[width=\columnwidth]{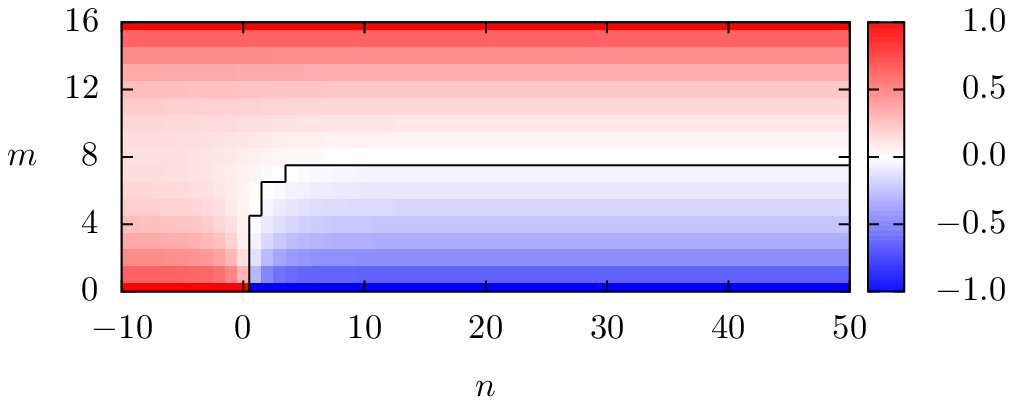}\\
(c) & & (d)\\
\includegraphics[width=\columnwidth]{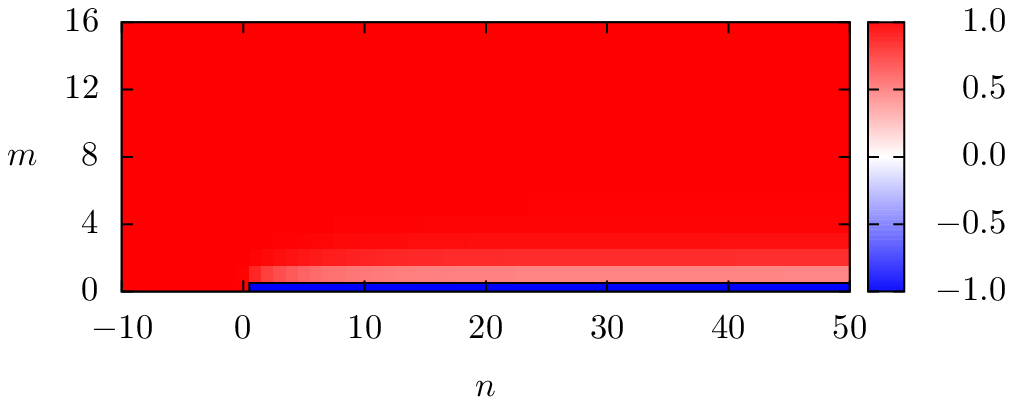} & & \includegraphics[width=\columnwidth]{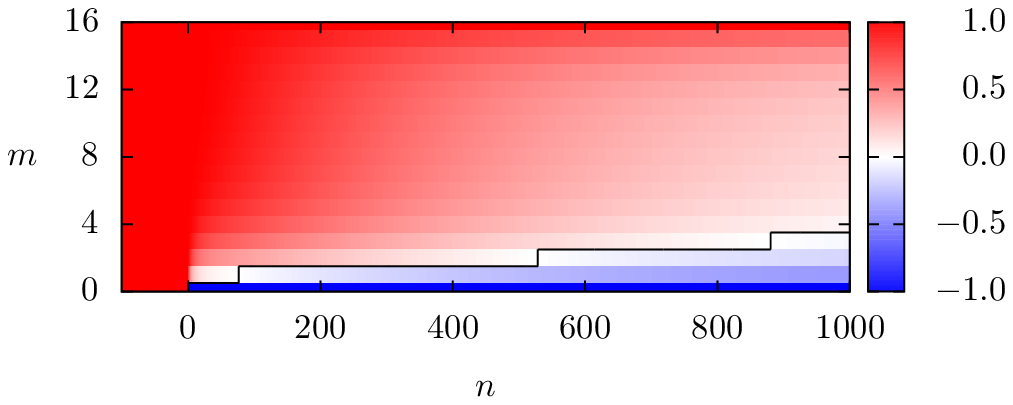}\\
\end{tabular}
\end{center}
\caption{\label{sec5:fig2} Plots of magnetization profiles in a strip with only one (bottom) inhomogeneous wall; the inhomogeneity has infinite extent. For every plot $M=15$ and $h_1=0.8\,J$ ($\Tw\approx0.621\,\Tc$). Each site of the lattice is represented by one rectangle. The color represents the value of magnetization and changes continuously from red for $\mgn=+1$, via white for $\mgn=0$, to blue for $\mgn=-1$. In rows $m=0$ and $m=16$ the spins are fixed and they represent the sign of the surface field. On all plots the black line representing the interface was plotted in such a way that it separates regions with positive and negative magnetization. (a) $T=0.8\,\Tc$, (b) $T=1.2\,\Tc$, (c) $T=0.5\,\Tc$, (d) $T=0.55\,\Tc$.}
\end{figure*}

The plots of magnetization profiles for $N_1=\infty$ and different temperatures are shown in Fig.~\ref{sec5:fig2}. On all plots each lattice site is represented by a rectangle whose color indicates the value of magnetization: the red color corresponds to $\mgn=+1$, white to $\mgn=0$, and blue to $\mgn=-1$. Additionally, we place the interface as the line that separates regions with different sign of magnetization.

For $T>\Tw$ (Fig.~\ref{sec5:fig2}(a) and (b)) the magnetization profiles approach exponentially those in a homogeneous strip. For $T<\Tw$ the situation is slightly more complicated. Because $\xiAS\gg 1$, there exist a range of $n>0$ for which the magnetization stays almost constant upon changing column number $n$ but it is, nevertheless, different from the magnetization in homogeneous antisymmetric strip. This is illustrated in Fig.~\ref{sec5:fig2}(c), where $\xiAS\approx9770$. Typically, in this regime the strength of the surface field is not big enough to make the magnetization negative, even in the surface layer next to the inhomogeneity on the wall. In Fig.~\ref{sec5:fig2}(d) the temperature is slightly increased to $T=0.55\,\Tc$ ($\xiAS\approx526$) and the displayed range of $n$ is expanded, so that the slow approach of the column magnetization profile to the homogeneous case can be observed. These magnetization profiles are in a qualitative agreement with the order parameter profiles in the scaling limit calculated in a three--dimensional system using Monte Carlo simulations \cite{ParisenToldin10}.

\begin{figure}
\begin{center}
\includegraphics[width=0.9\columnwidth]{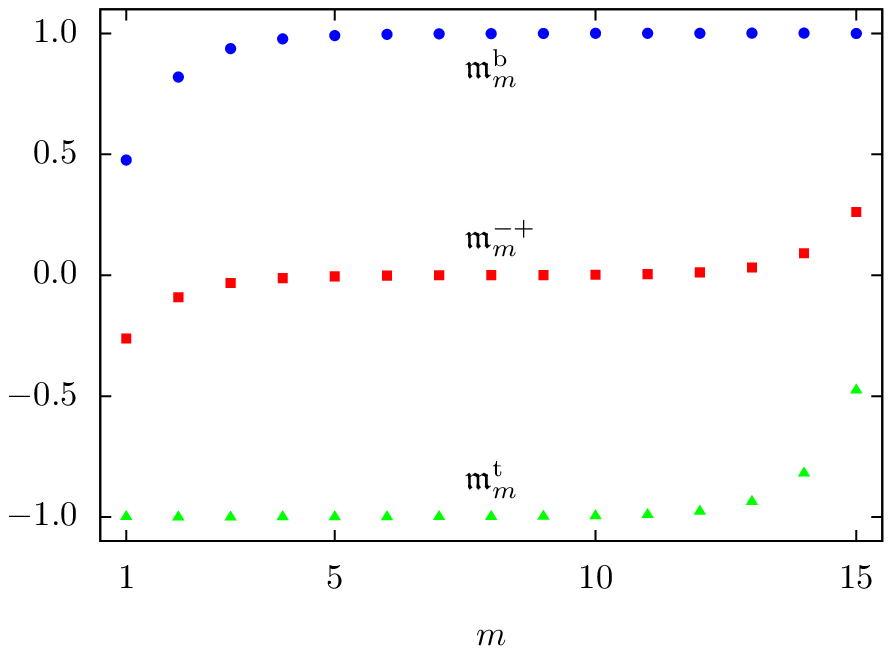}
\end{center}
\caption{\label{sec5:fig3} The decomposition of a column magnetization profile $\mgn_m^{-+}$ of a homogeneous antisymmetric Ising strip into two profiles $\mgn_m^{\mathrm{b}}$ and 
$\mgn_m^{\mathrm{t}}$ for the system with an interface located close to the bottom and the top wall, respectively. Because of the symmetry with respect to reversing all spins, both locations of the interface have the same probability and $\mgn_m^{-+}=\left(\mgn_m^{\mathrm{b}}+\mgn_m^{\mathrm{t}}\right)/2$. The plots are done for $T=0.5\,\Tc$, $h_1=0.8\,J$ and $M=15$.}
\end{figure}

From \eqref{sec5:mgn1infasymp} it follows that, when $1\ll n\ll \xiAS$ the magnetization profile is given by
\begin{subequations}\label{sec5:mgnsep}
\begin{equation}
\mgn_{m}^{\mathrm{b}}\left(T,h_1,M\right)=-\Xi^{m}_{1;1}-t_1^{\left(2\right)}/t_1^{\left(1\right)}\Xi^m_{2;1}.
\end{equation}
This is the magnetization profile in a homogeneous antisymmetric Ising strip with the interface pinned to the bottom wall. By taking the limit $N_1\to\infty$ and $L\to-\infty$, and reversing all spins in the system one can determine the column magnetization profile for the interface pinned to the top wall
\begin{equation}
\mgn_{m}^{\mathrm{t}}\left(T,h_1,M\right)=-\Xi^{m}_{1;1}+t_1^{\left(2\right)}/t_1^{\left(1\right)}\Xi^m_{2;1}.
\end{equation} 
\end{subequations}
Because of the symmetry, in the homogeneous antisymmetric system both configurations occur with the same probability and 
$\mgn^{-+}_m=-\mgn^{+-}_m=\left(\mgn_m^{\mathrm{b}}+\mgn_m^{\mathrm{t}}\right)/2$. An example of the decomposition of the equilibrium magnetization profile for $T<\Tw$ is presented in Fig.~\ref{sec5:fig3}.

To our knowledge, the formulas \eqref{sec5:mgnsep} have not been derived before. Because, for finite $M$, the system is quasi one--dimensional, no phase transition is possible, and any infinitesimal perturbation produces only infinitesimal change. Thus, it is not possible to pin the interface to one of the walls in the usual way by breaking the symmetry by a small perturbation to the homogeneous strip.

\begin{figure*}
\begin{center}
\begin{tabular}{lcl}
(a) & & (b)\\
\includegraphics[width=0.45 \textwidth]{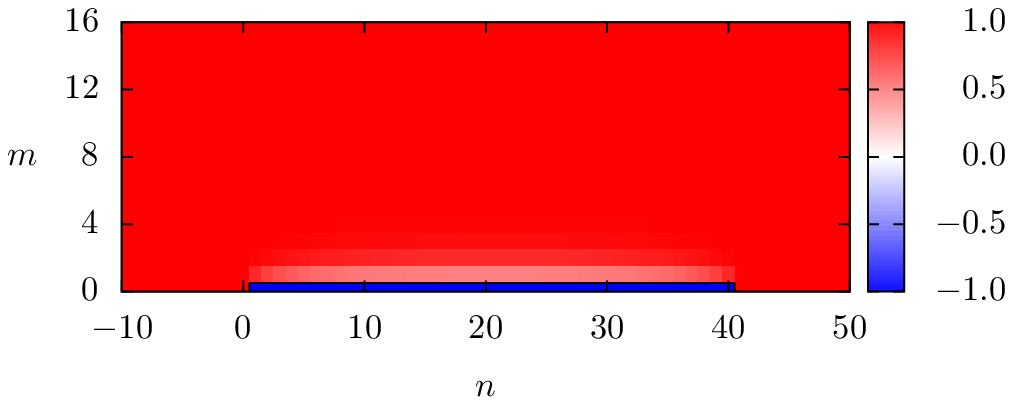} & & \includegraphics[width=0.45 \textwidth]{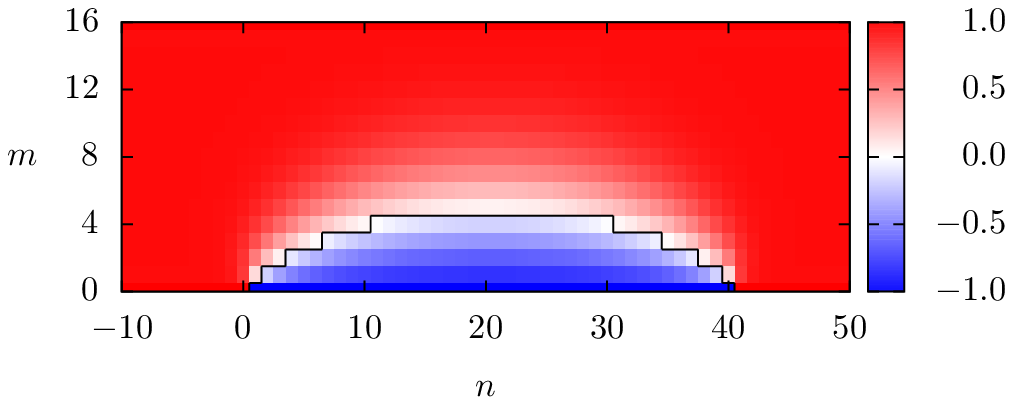}\\
(c) & & (d)\\
\includegraphics[width=0.45 \textwidth]{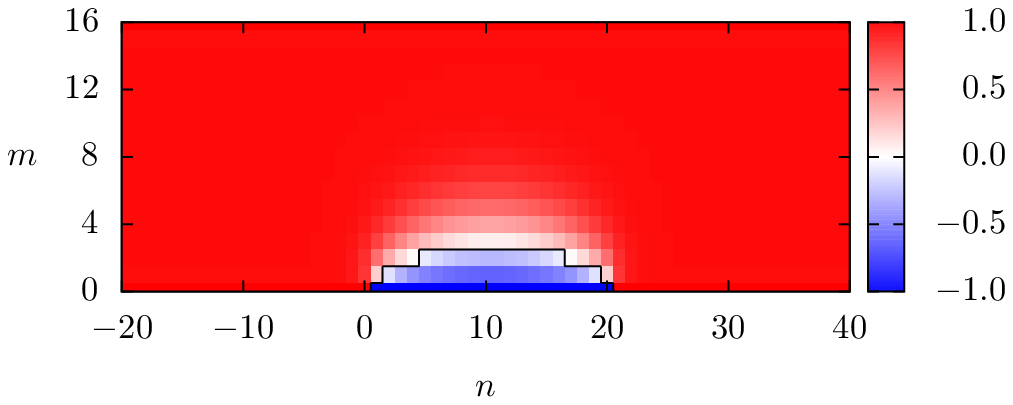} & & \includegraphics[width=0.45 \textwidth]{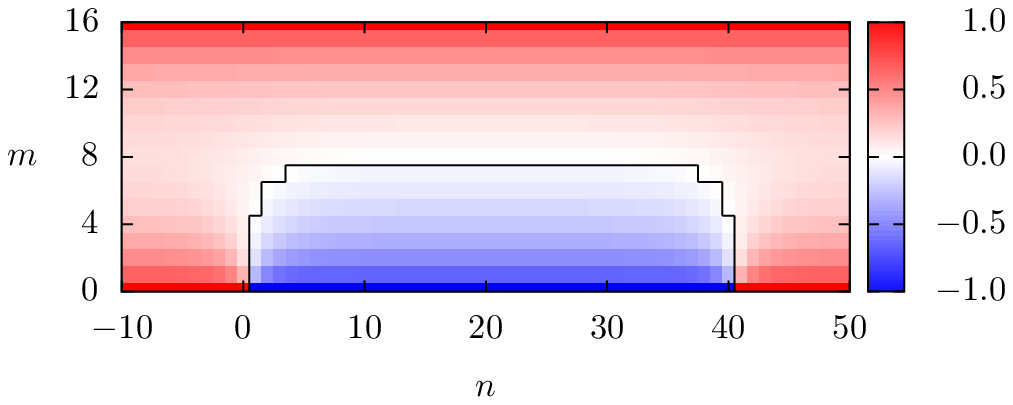}\\
\end{tabular}
\end{center}
\caption{\label{sec5:fig4} The magnetization profiles in the strip with one  finite--size inhomogeneity located at the bottom wall. For every plot $M=15$, $h_1=0.8\,J$ ($\Tw\approx0.621\,\Tc$). The color coding is the same as in Fig.~\ref{sec5:fig2}. In the rows $m=0$ and $m=16$ the spins are fixed and they represent the sign of the surface field. (a)~$T=0.5\,\Tc$ and $N_1=40$, (b)~$T=0.8\,\Tc$ and $N_1=40$, (c)~$T=0.8\,\Tc$ and $N_1=20$, (d)~$T=1.2\,\Tc$ and $N_1=40$.}
\end{figure*}

In the next step we again consider a strip with one inhomogeneity but now of finite length $N_1$.  The analytical results are rather complicated and we do not present them here. In Fig.~\ref{sec5:fig4}  the plots of magnetization profiles for different temperatures and lengths of inhomogeneities are presented.

For all temperatures and positions away from the inhomogeneity, the column magnetization profile approaches  $\mgn_m^{++}$ exponentially with the length--scale $\xiS$. For $T<\Tw$ the magnetization profile in columns with antisymmetric surface field is equal to $\mgn_{m}^\mathrm{b}$ with minor modifications close to the edges of the inhomogeneity; these modifications decay exponentially with a length--scale $\xiASp$, see Fig.~\ref{sec5:fig4}(a). This picture is correct when $N_1\ll\xiAS$; if this is not the case, the magnetization profile looks similarly to the one presented in Fig.~\ref{sec5:fig2}(d). For $T>\Tw$ a droplet of negative magnetization is formed above the inhomogeneity. When $\xiAS$ is of the order of $N_1$, the height of the droplet depends on $N_1$ (in Fig.~\ref{sec5:fig4}(b) and (c), $\xiAS\approx 14.8$), and when $\xiAS\ll N_1$ the droplet reaches the middle of the strip (in Fig.~\ref{sec5:fig4}(d), $\xiAS\approx 2.7$).

For $T>\Tc$ the bulk magnetization is zero and regions of positive and negative magnetization are induced by the surface field. In this case, our definition of the position of the interface may be misleading. In particular, in the middle of the strip the magnetization is very close to zero and the interface separates regions with almost the same magnetization. 

The results for magnetization in the case $N_1<M$ are in a qualitative agreement with the order parameter profile obtained within the mean field theory in the scaling limit for a semi--infinite system with one inhomogeneity on the wall \cite{Sprenger05}.

\subsection{Inhomogeneities on both walls}

\begin{figure*}
\begin{center}
\begin{tabular}{l}
(a)\\
\includegraphics[width=0.9\textwidth]{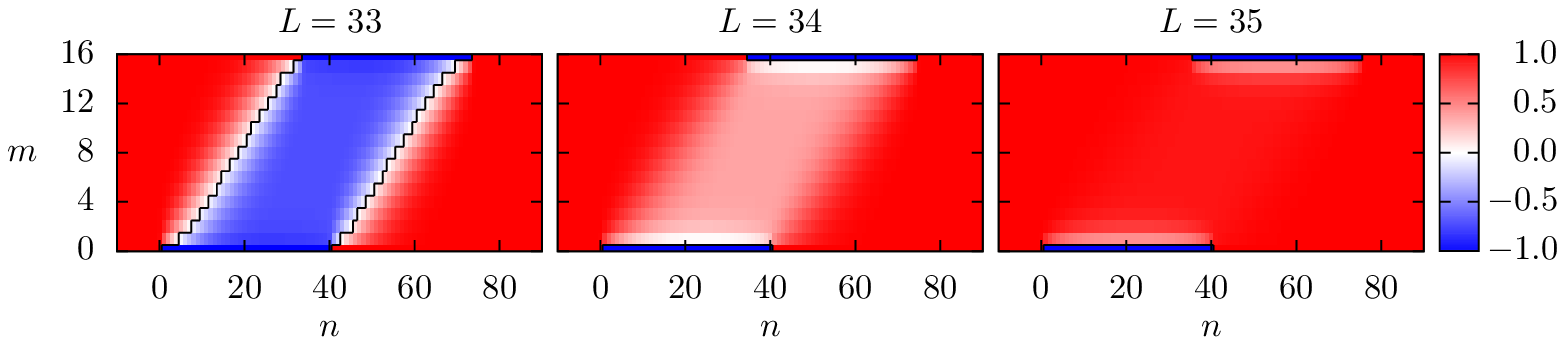}\\
(b)\\
\includegraphics[width=0.9\textwidth]{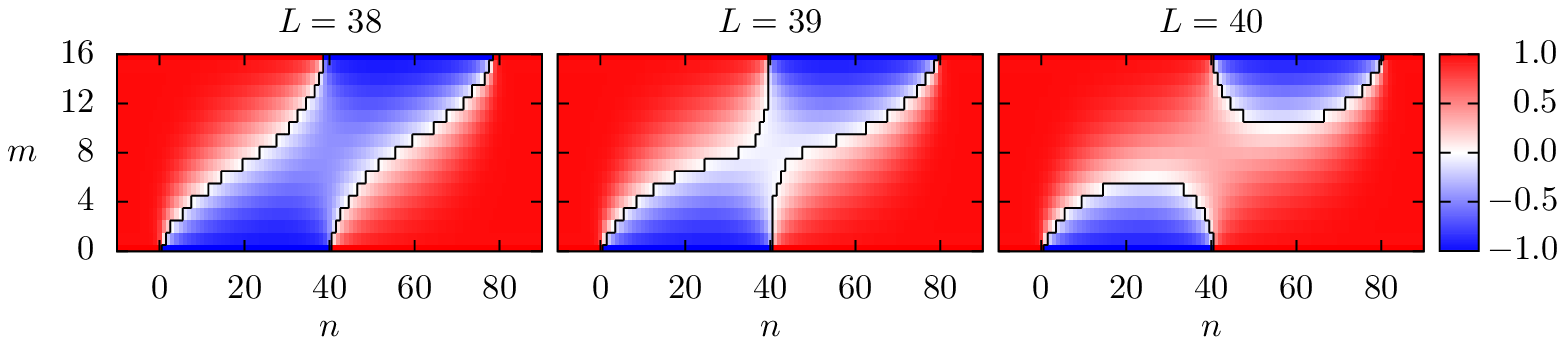}\\
(c)\\
\includegraphics[width=0.9\textwidth]{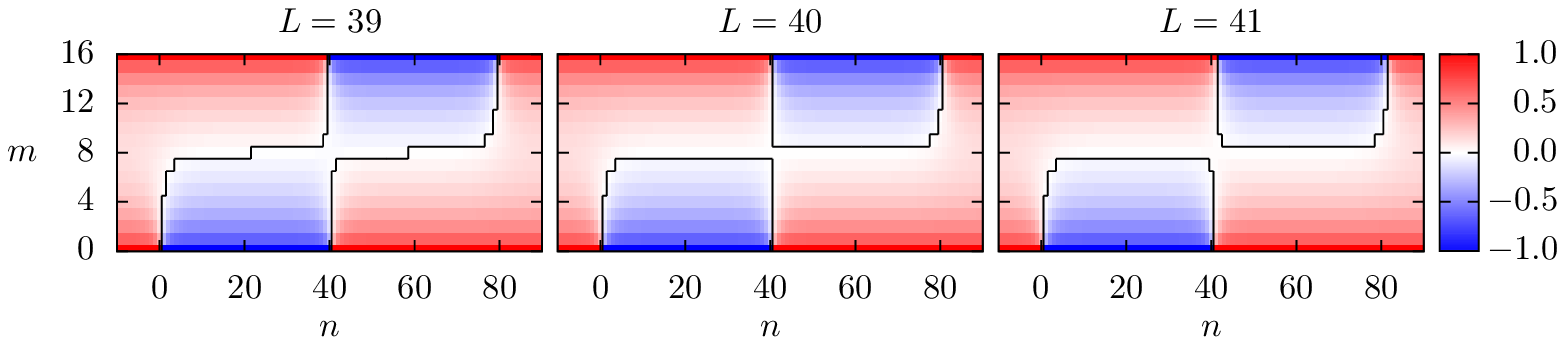}
\end{tabular}
\end{center}
\caption{\label{sec5:fig5} Magnetization profiles of the system with two finite inhomogeneities, displayed for different shifts $L$, close to the point of breaking of the bridge. For every plot $M=15$, $h_1=0.8\,J$ ($\Tw\approx0.621\,\Tc$) and $N_1=40$. The color coding is like in the previous figure. (a)~$T=0.5\,\Tc$, (b)~$T=0.8\,\Tc$, (c)~$T=1.2\,\Tc$.}
\end{figure*}

When both walls contain inhomogeneities and the absolute value of the shift $\left|L\right|$ is large, there exist two regions of negative magnetization close to each inhomogeneity (see Fig.~\ref{sec5:fig5}). If  $\left|L\right|$ becomes small then it may happen that these two regions coalesce and form a capillary bridge, i.e., a region of negative magnetization connecting the walls (for precise definition of a capillary bridge see Sec.~\ref{sec6:A}). At the same time, upon increasing $L$ from $0$, we observe breaking of the capillary bridge. In Fig.~\ref{sec5:fig5} a typical scenario of this phenomenon, taking place at different temperatures, is presented.

Breaking of the capillary bridge for $T<\Tw$ is presented in Fig.~\ref{sec5:fig5}(a). In this case, the interfaces are almost straight lines and, inside the bridge, the magnetization is almost constant --- it varies only close to the inhomogeneities and the interfaces. The value of the magnetization inside the bridge is not equal to its bulk value $-\mgn_0\left(T\right)$, and its value increases upon increasing $L$. The breaking of the capillary bridge takes place when the magnetization between inhomogeneities exceeds 0. This behavior can be explained by assuming that (similarly to the case of the homogeneous antisymmetric strip) there are two different configurations --- one with a capillary bridge  and one without --- and the resulting magnetization profile is an average of these two. Upon increasing $L$ the probability of a system with the bridge decreases.  In this case, we do not see any way to determine magnetization profiles of these two configurations. However, in higher--dimensional systems, where phase transitions are possible, one expects a first--order bridging transition. Note that breaking of the capillary bridge happens for $L<N_1$.

Typical scenario of breaking of the capillary bridge for $\Tw<T<\Tc$ is presented in Fig.~\ref{sec5:fig5}(b). Here, the interfaces are curved and the width  of the bridge is the smallest in the middle of the strip. The magnetization is also varying --- its absolute value is the largest close to the inhomogeneities and the smallest in the middle of the bridge. Upon increasing $L$, after breaking of the bridge took place, the droplets located next to the inhomogeneities remain deformed for some range of $L$. This might suggest that in a system in which a phase transition can take place, one could observe a continuous bridging transition. In this case, breaking of capillary bridge happens for $L$ close to $N_1$. 

Above $\Tc$ the magnetization is induced only by the surface fields and large regions of almost zero magnetization surround the interface. As can be seen in Fig.~\ref{sec5:fig5}(c), the bridge is formed in the process of coalescence of two droplets of negative magnetization.  Even close to the point of coalescence, the shape of the two separate droplets undergoes only minor changes. Like in the previous case, the breaking of the capillary bridge occurs for $L$ close to $N_1$

\begin{figure*}
\begin{center}
\begin{tabular}{l}
(a)\\
\includegraphics[width=0.9\textwidth]{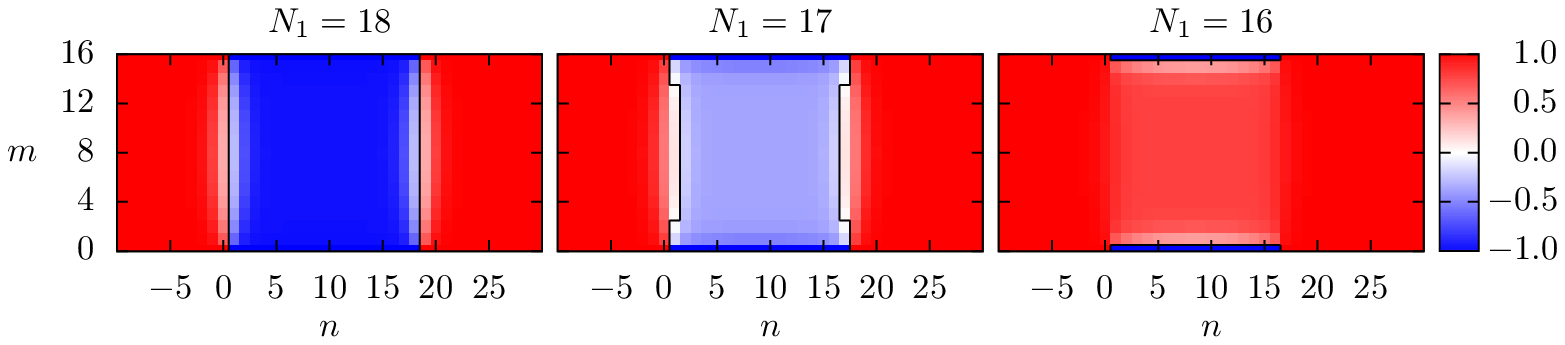}\\
(b)\\
\includegraphics[width=0.9\textwidth]{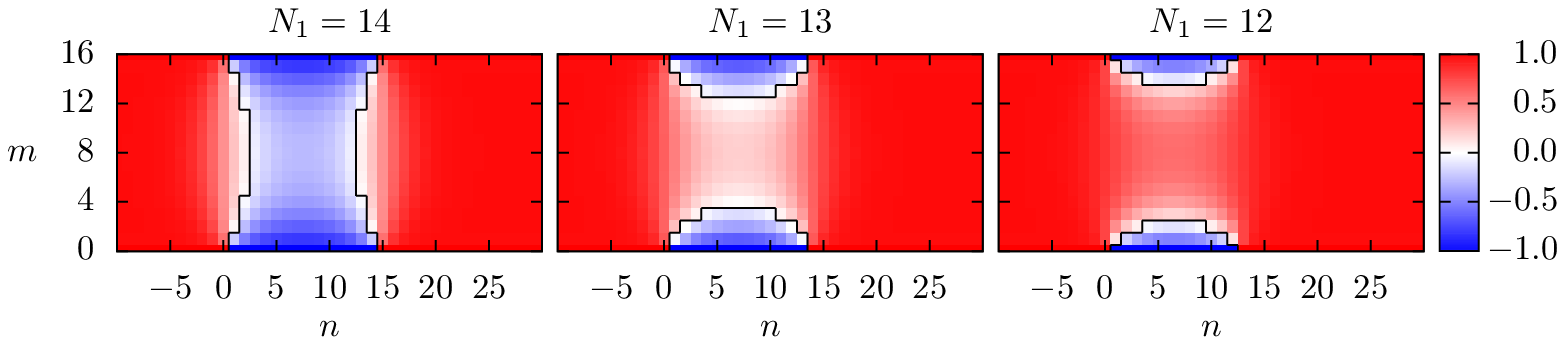}\\
(c)\\
\includegraphics[width=0.9\textwidth]{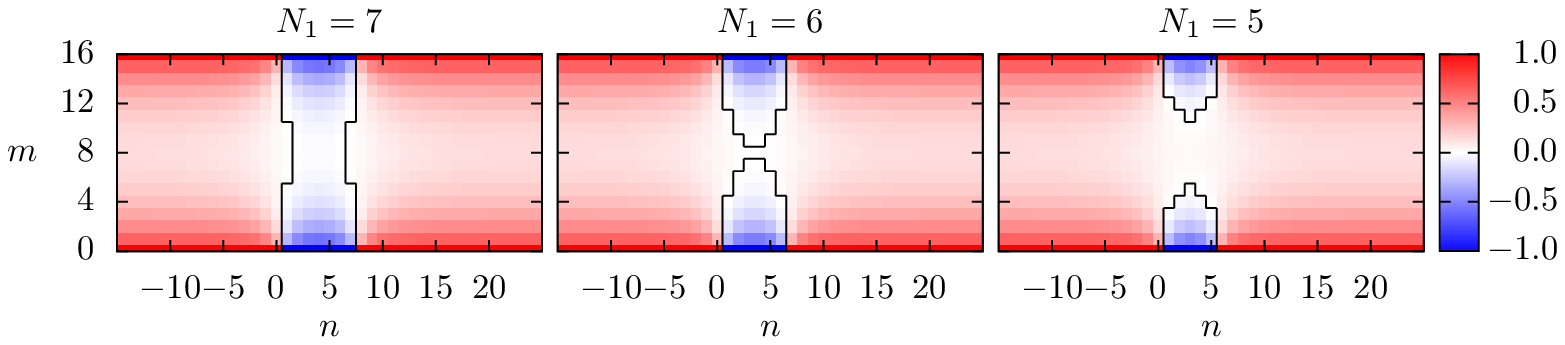}
\end{tabular}
\end{center}
\caption{\label{sec5:fig6} Magnetization profiles of the system with two finite inhomogeneities for different values of $N_1$ and fixed $L=0$, close to the point of breaking of the bridge. For every plot $M=15$ and $h_1=0.8\,J$ ($\Tw\approx0.621\,\Tc$). The color coding is the same as in Fig.~\ref{sec5:fig2}.  (a)~$T=0.5\,\Tc$, (b)~$T=0.8\,\Tc$, (c)~$T=1.2\,\Tc$.}
\end{figure*}

The scenario described above takes place only when the length of inhomogeneities $N_1$ is large enough. When $N_1$ is small, there can be no capillary bridge in the system, even for a zero shift. To study this phenomenon, we fix $L=0$ and decrease  $N_1$ towards zero. As expected, upon decreasing $N_1$ we observe breaking of the capillary bridge. This situation is illustrated in Fig.~\ref{sec5:fig6}.

When $T<\Tw$ the situation is almost the same as in the previous scenario --- the magnetization is almost independent on row number $m$ and it changes from negative to positive values upon decreasing $N_1$. This confirms our hypothesis about the existence of two magnetization configurations, each  with a different probability. Breaking of the bridge takes place for $N_1$ slightly higher than $M$.

When $\Tw<T<\Tc$ the situation is again similar to breaking of the capillary bridge for fixed $N_1$ upon increasing $L$. The absolute value of magnetization is the largest close to inhomogeneities and the smallest in the middle of the system where the bridge is the thinnest. In this case the breaking of the bridge occurs for $N_1$ slightly smaller than $M$.

Finally, when $T>\Tc$ the bridge gets broken for $N_1\approx 2\xiAS\approx 2\xiS$, when the size of the inhomogeneity is too small to induce the region of negative magnetization that reaches the middle of the system. This is in agreement with the picture of two coalescing droplets of negative magnetization presented above.

We note that upon increasing temperature there are no sharp transitions between different scenarios of breaking of the capillary bridge described above. Instead, for $T\approx \Tw$ and $T\approx \Tc$ one type of behavior of magnetization changes smoothly into the other. This situation reflects the lack of phase transitions in our 2D strip \cite{vanHove50}. We cannot exclude the possibility that the transition between scenarios happens in slightly different temperatures. To study this phenomenon in detail one would need to consider higher dimensional system in which real phase transitions take place or at least a 2D Ising strip for higher widths $M$ where pseudotransitions become sharper.

\section{Capillary bridge}\label{sec6}
In this section we study the phenomenon of breaking of the capillary bridge quantitatively and we look for the relation between the breaking of the bridge and the properties of the lateral critical Casimir.

\subsection{Point of breaking of capillary bridge}\label{sec6:A}

We start this section with a precise definition of a capillary bridge used in this paper. The condition for the existence of a capillary bridge in a strip is the presence of a path, going through pairs of neighboring spins, connecting a surface spin on the bottom wall with a surface spin on the top wall, such that at all lattice sites along this path the magnetization is negative. This definition implies that a bridge exists when interfaces connect the edges of inhomogeneities across the strip and that there is no bridge when each interface connects the edges of the same inhomogeneity.

Other reasonable definitions of the capillary bridge can differ from ours only for some special magnetization profiles like two interfaces touching each other at a single point or exactly zero magnetization in the middle of the system --- according to the above definition there is no capillary bridge in both of these cases.

The point of breaking of the capillary bridge $L_1\left(T,h_1,M,N_1\right)$ is defined such that for $L=L_1-1/2$ there is a capillary bridge while for $L=L_1+1/2$ there is no bridge. (Here we consider only $L\geqslant 0$.) The value of $L_1$ is a half--integer.

\begin{figure}
\begin{center}
\begin{tabular}{l}
(a) \\[1mm]
 \includegraphics[width=0.9\columnwidth]{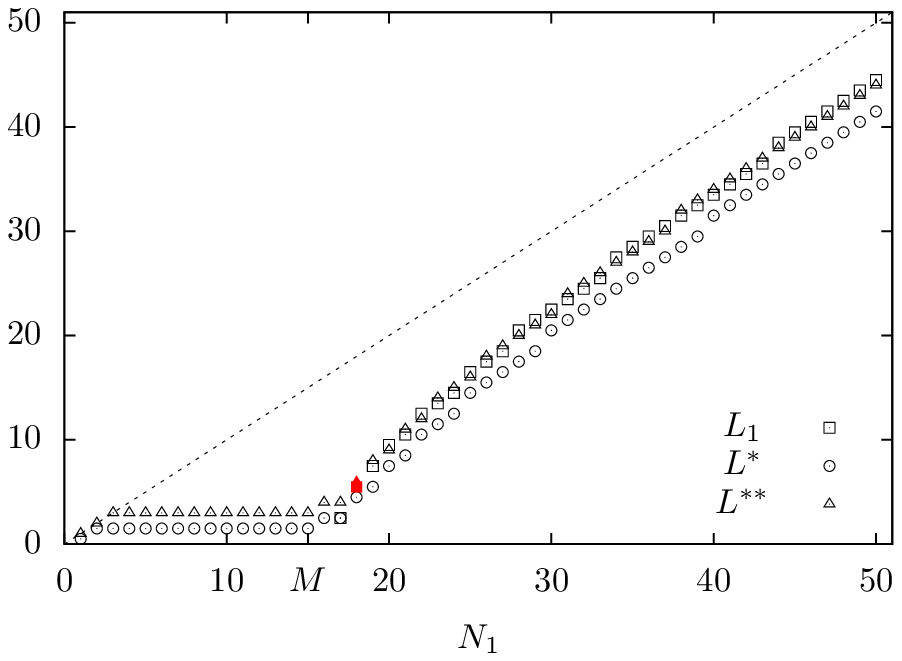}\\
(b) \\[1mm]
 \includegraphics[width=0.9\columnwidth]{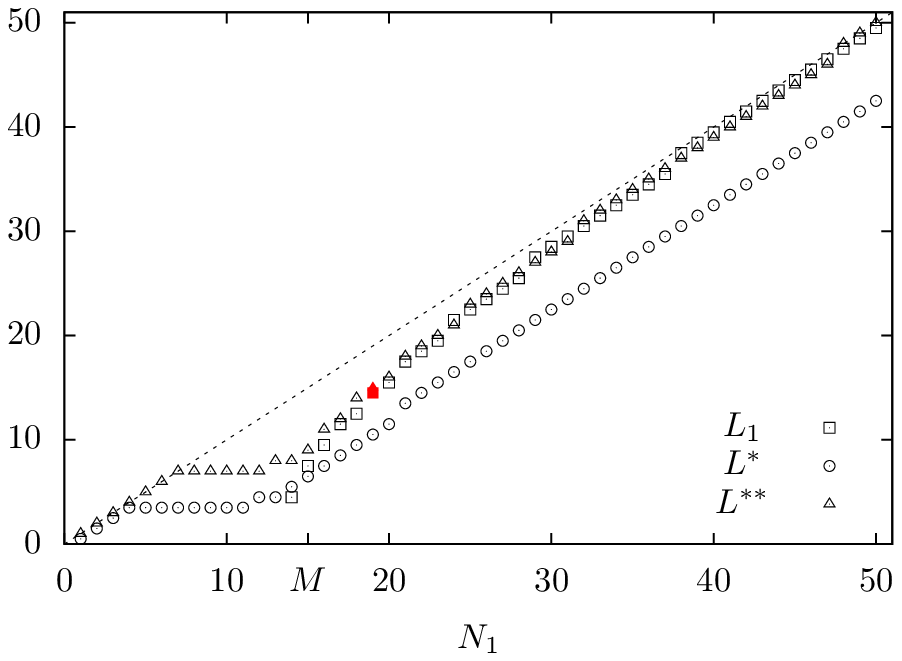}\\
(c) \\[1mm]
 \includegraphics[width=0.9\columnwidth]{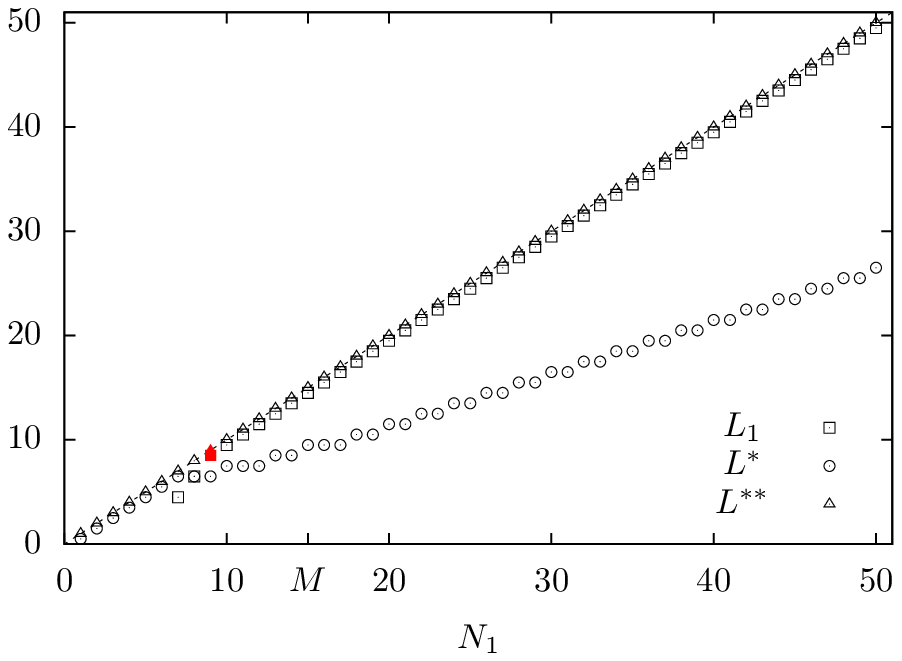}
\end{tabular}
\end{center}
\caption{\label{sec6:fig1} The comparison of the point of breaking of the capillary bridge $L_1$ as function of $N_{1}$ with the minimum $\Last$ and the inflection point $\Laast$ of the lateral critical Casimir force. On all plots $h_1=0.8\,J$ and $M=15$. (a) $T=0.5\,\Tc$, (b) $T=0.8\,\Tc$, (c) $T=1.2\,\Tc$. The red triangle and the red square denote the value $N_1=\bar{N}_1$ starting from which the difference between the point of breaking of the bridge $L_1$ and the inflection point $\Laast$ is minimal (see \eqref{sec6:N1bar}).}
\end{figure}

\begin{figure}
\begin{center}
\includegraphics[width=0.9\columnwidth]{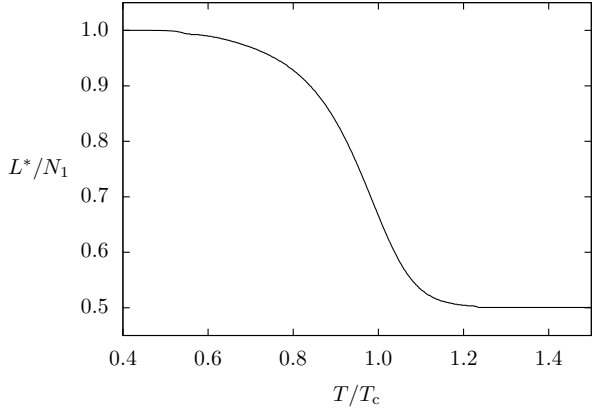}
\end{center}
\caption{\label{sec6:fig2} The minimum of the lateral critical Casimir force $\Last$ divided by the length of inhomogeneities $N_1$ in the limit $N_1\to\infty$. The ratio was calculated for $M=15$ and $h_1=0.8\,J$. The small irregularities at $T\approx 0.6\,\Tc$ and $T\approx 1.2\,\Tc$ are artifacts of the numerical procedure of calculating the limit $N_1\to\infty$.}
\end{figure}

In Fig.~\ref{sec6:fig1} the value of $L_1$ is  plotted as a function of $N_1$ for different temperatures. From the previous section we know that when $N_1$ is small there is no capillary bridge and thus $L_1$ is defined only for $N_1$ above certain threshold. The value $L_1$ grows upon increasing $N_1$. When $N_1$ is large enough, we have $L_1=N_1-1/2$.

\subsection{Comparison with the lateral force}\label{sec6:comparison}

To study, how the breaking of the capillary bridge influences the lateral force, we define two special points:  the point of the minimum of the force $\Last\left(T,h_1,M,N_1\right)$ which satisfies
\begin{multline}
\flateral\left(T,h_1,M,N_1,\Last-1\right)\geqslant\flateral\left(T,h_1,M,N_1,\Last\right)<\\
\flateral\left(T,h_1,M,N_1,\Last+1\right),
\end{multline}
and the inflection point of the force $\Laast\left(T,h_1,M,N_1\right)$ which is defined via
\begin{multline}
\flateral^{\prime\prime}\left(T,h_1,M,N_1,\Laast-1/2\right)>0>\\
\flateral^{\prime\prime}\left(T,h_1,M,N_1,\Laast+1/2\right),\quad \Laast>0,
\end{multline}
where the discrete version of the derivative $f^\prime\left(L\right)=f(L+1/2)-f(L-1/2)$ has been used, see \eqref{sec2:flateral}. Because $\flateral$ and $\flateral^{\prime\prime}$ are defined only for half--integer values of $L$, $\Last$ is a half--integer while $\Laast$ is an integer.

In Fig.~\ref{sec6:fig1} these two quantities are compared with $L_1$. Unlike the point of breaking of the bridge, $\Last$ and $\Laast$ are defined for all values of $N_1$. For fixed $M$ and $T<\Tc$, for the smallest values of $N_1$ we observe $\Last=N_1-1/2$ and $\Laast=N_1$. Upon increasing $N_1$ both functions have a plateau whose value depends on temperature. This plateau ends at $N_1\approx M$ and, upon further increase of $N_1$, both $\Last$ and $\Laast$ grow. When $N_1$ is of the order of $\xiAS$, the relation $\Laast=N_1$ is recovered and it remains fulfilled as $N_1$ is increased to infinity. For $N_1$ larger than $\xiAS$ the location of the minimum of the force $\Last$ grows (approximately) linearly with $N_1$ but the proportionality coefficient is smaller than 1. When $T>\Tc$, the behavior of $\Last$ and $\Laast$ is very similar, except that we do not observe the plateau for $N_1<M$.

In Fig.~\ref{sec6:fig2} we investigate the behavior of the ratio $\Last/N_1$ for large $N_1$. For small temperatures it is almost equal to 1. Upon increasing the temperature it decreases and for large temperatures tends to $1/2$.

Our comparison shows that the inflection point of the lateral force $\Laast$ is a better approximation of the point of breaking of the capillary bridge $L_1$ than the minimum of the force $\Last$. Taking into account the fact that $L_1$ is a half--integer and $\Laast$ is an integer, the agreement between these two quantities is very good. We define $\bar{N}_1\left(T,h_1,M\right)$  to be the smallest integer for which
\begin{multline}\label{sec6:N1bar}
\left|L_1\left(T,h_1,M,N_1\right)-\Laast\left(T,h_1,M,N_1\right)\right|=1/2 \\
\quad \text{for all } N_1\geqslant \bar{N}_1.
\end{multline}
The value of $\bar{N}_1$ is for all temperatures significantly smaller than the value of $N_1$ for which asymptotic relations $L_1=N_1-1/2$ and $\Laast=N_1$ are reached.

\subsection{Continuous analogs of $L_1$ and $\Laast$}

\begin{figure}
\begin{center}
\begin{tabular}{l}
(a) \\[1mm]
\includegraphics[width=0.9\columnwidth]{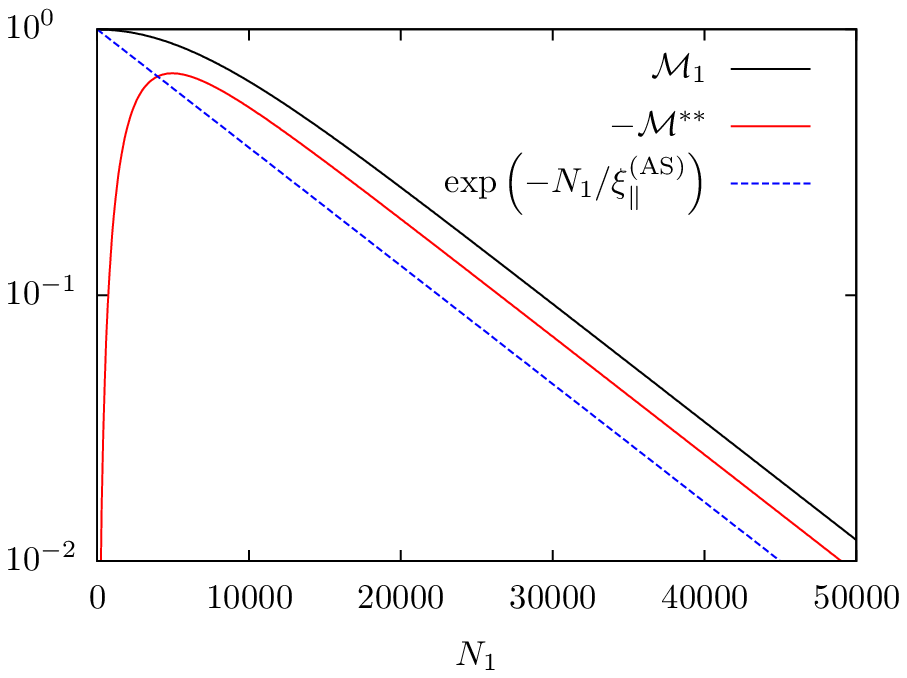}\\
(b) \\[1mm]
\includegraphics[width=0.9\columnwidth]{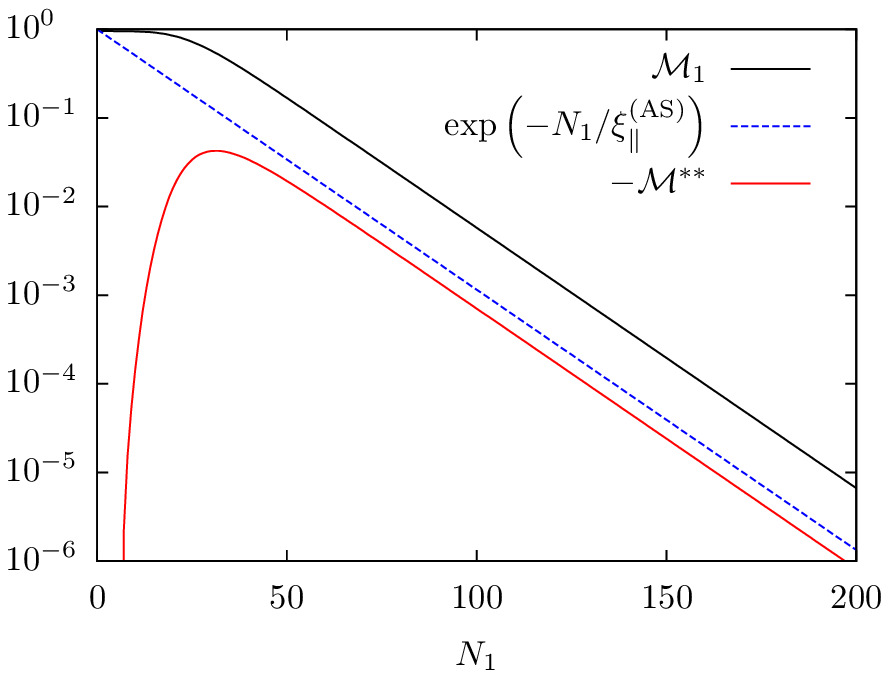}\\
(c) \\[1mm]
\includegraphics[width=0.9\columnwidth]{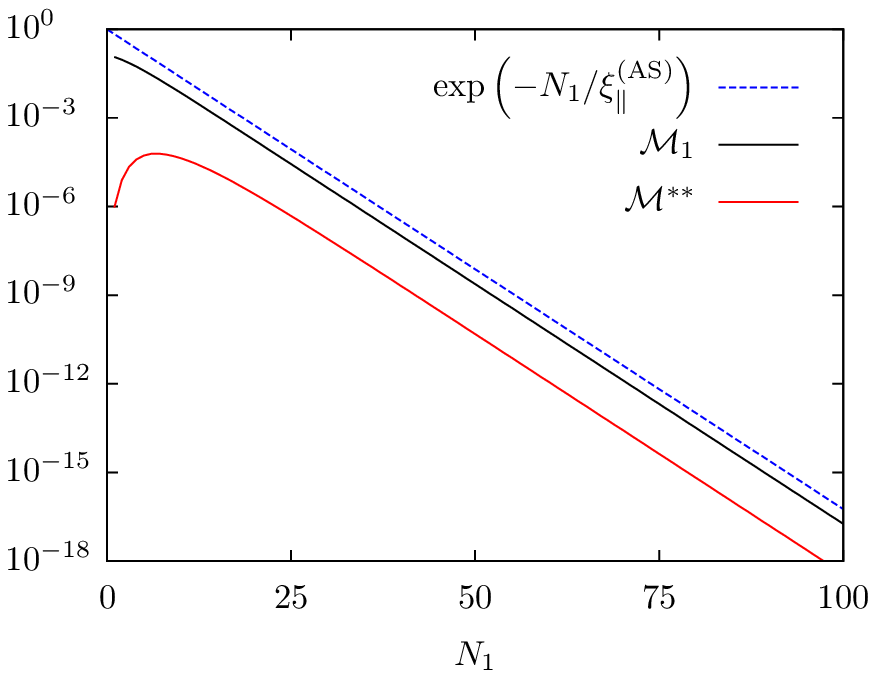}
\end{tabular}
\end{center}
\caption{\label{sec6:fig3} Comparison of the functions $\mathcal{M}_1$ and $\mathcal{M}^{\ast\ast}$ that measure how close to $N_1$ is the point of breaking of capillary bridge and inflection point of the lateral critical Casimir force, respectively. Points for integer values of $N_1$ are connected by lines to guide the eye. For all considered temperatures, both of these quantities are proportional to $\exp\left(-N_1/\xiAS\right)$ for large $N_1$. $M=15$ and $h_1=0.8\,J$ for all plots. (a) $T=0.5\,\Tc$, (b) $T=0.8\,\Tc$, (c) $T=1.2\,\Tc$. Note that on the first two graphs the function $-\mathcal{M}^{\ast\ast}$ is plotted instead of $\mathcal{M}^{\ast\ast}$.}
\end{figure}

The agreement between $L_1$ and $\Laast$, though the best possible for functions with a discrete domain and image, can be still tested in a model (from the same universality class) with a continuous order parameter, for which $L$  takes real values. As  shown above, for all temperatures, both the point of breaking of capillary bridge and the inflection point of the lateral force approach $N_1$ upon increasing $N_1$. To study this in more detail we introduce two real functions which measure how close these quantities are to $N_1$
\begin{align}
\nonumber&\mathcal{M}_1\left(T,h_1,M,N_1\right)=\mgn_{\left(M-1\right)/2,N_1}\left(T,h_1,M,N_1,N_1\right)=\\
&\hspace{2.4cm}\mgn_{\left(M-1\right)/2,N_1+1}\left(T,h_1,M,N_1,N_1\right),\\
\nonumber&\mathcal{M}^{\ast\ast}\left(T,h_1,M,N_1\right)=\frac{1}{2}\Big[\flateral^{\prime\prime}\left(T,h_1,M,N_1,N_1-1/2\right)+\\
&\hspace{3.0cm}\flateral^{\prime\prime}\left(T,h_1,M,N_1,N_1+1/2\right)\Big].
\end{align}
Both of these functions are defined for integer values of $N_1$ and take real values. The function $\mathcal{M}_1$ represents the magnetization in the center of the system ($m=\left(M-1\right)/2$ and $n=N_1$) with the shift $L=N_1$. (This definition is correct only for odd $M$.) When for $L=N_1$ there is no capillary bridge in the system  this magnetization is positive, and upon increasing $N_1$ the value of $\mathcal{M}_1$ should approach  $0$. The value of $\mathcal{M}^{\ast\ast}$ represents the average of second (discrete) derivatives of the lateral force calculated at two points around $L=N_1$. This is an interpolated value of this derivative at $L=N_1$ assuming that it is locally linear. When, upon increasing $N_1$, the inflection point approaches $N_1$ this value should go to $0$.

We have checked numerically that for large $N_1$ both functions approach 0 exponentially with the same length--scale $\xiAS$. Plots of them for three different temperatures are presented in Fig.~\ref{sec6:fig3}. We note that although the length--scales are the same, the amplitudes multiplying exponential decay are different. We also note that, depending on temperature and $N_1$, value of $\mathcal{M}^{\ast\ast}$ can be positive or negative.

This agreement between continuous measures of the point of breaking of the capillary bridge and the inflection point suggests that the observed relation between $L_1$ and $\Laast$ is not coincidental. In App.~\ref{secC} we provide an argument that this relation is quite general and should hold for a wide range of systems.

\section{Summary and conclusions}\label{sec7}
In this paper we have studied a two--dimensional Ising strip of width $M$ bounded by  inhomogeneous planar walls. The influence of the walls on the system was represented by a short--range surface field acting on spins located next to the corresponding walls. This surface field is equal to $+h_1$, except for a group of $N_1$ surface sites, where the field is $-h_1$. The inhomogeneities on the walls are shifted laterally by a distance $L$ (see Fig.~\ref{sec1:fig1}). Using a specially adapted method based on the exact diagonalization of the transfer matrix, we have derived formulas for the free energy and magnetization profile of the strip. The formulas depend on some coefficients and matrix elements that can only be calculated numerically. We have studied the critical Casimir force acting between the walls and its relation to the magnetization profile. Because the walls (and thus the surface field) are inhomogeneous, the force has both normal and lateral component.

We have found that for all temperatures the lateral component of the critical Casimir force always acts in the direction opposite to the shift. For small shifts ($L\approx 0$) the force grows linearly and, upon increasing $L$ (for $1\ll L\ll N_1$) it saturates at $-2\sigma$, where $\sigma\left(T,h_1,M\right)$ is a surface tension in the strip (see \eqref{sec2:sigma}). If $L$ is further increased ($L>N_1$), the force decays exponentially with the characteristic length--scale $\xiS$ (see \eqref{sec2:xiAS}).

Next, we have investigated the excess normal critical Casimir force (i.e., the part of the normal force that comes from the interaction of two inhomogeneities). This force is always negative (attractive) and is the strongest for $L=0$. Similarly to the lateral component, for $\left|L\right|>N_1$, upon increasing $\left|L\right|$ the excess normal force decays exponentially with the length--scale $\xiS$. We have also discovered that for $\left|L\right|<N_1$ the vector of the excess total force has almost constant length and only its direction is changing upon changing $L$.

To explain the observed properties of the critical Casimir force, we have calculated magnetization profiles in the strip. We have discovered that the rapid decay of the critical Casimir force observed for $\left|L\right|>N_1$ is correlated with the disappearance of the capillary bridge. We have considered two different scenarios of breaking of the bridge: increasing $L$ for fixed $N_1$ and decreasing $N_1$ for $L=0$. In both cases, below the wetting temperature $\Tw$, this morphological transition looks to be of the first order while for $T>\Tw$ it turns continuous. Because in the Ising strip with finite width there cannot be a real phase transition, this observation is only a suggestion of possible behavior of higher--dimensional systems. Unfortunately, numerical complications accompanying the evaluation of magnetization profiles were too substantial to consider systems large enough to study precisely the nature of observed transitions. 

Additionally, by introducing an inhomogeneity to the Ising strip, we were able to extract two different magnetization profiles (see \eqref{sec5:mgnsep}) of the homogeneous strip with opposite surface fields for temperatures below the wetting temperature $\Tw$. Due to the nearly spontaneously broken symmetry, these two profiles have always the same probability and standard calculation cannot discern them. 

Finally, we have studied the relation between the properties of the critical Casimir force and the breaking of the capillary bridge. It turns out that the point of breaking of the bridge is rather related to the inflection point than to the minimum of the force. To support this observation, we have given some general arguments, which suggest that the link between the inflection point of the critical Casimir force and breaking of the capillary bridge should be present for a wide range of systems. The existence of such a link would be very useful because typically the calculation of the magnetization profile is much more expensive numerically than determining the critical Casimir force. 

\begin{acknowledgments}
The authors would like to thank Siegfried Dietrich, Anna Macio{\l}ek and Matthias Tr\"ondle for helpful and inspiring discussions. The support from the Polish National Science Center (Narodowe Centrum Nauki) via the grant 2011/03/B/ST3/02638 is also gratefully acknowledged.
\end{acknowledgments}
\appendix
\section{Calculation of free energy for inhomogeneous Ising strip}\label{secA}
In this section we present detailed calculations of the free energy of the inhomogeneous Ising strip using the column transfer matrix. We concentrate on the case $0< L< N_1$. Other possibilities require only minor modifications of the calculations presented below.

\subsection{Modification of the original system} 

\begin{figure}[t]
\begin{center}
\includegraphics[width=0.9\columnwidth]{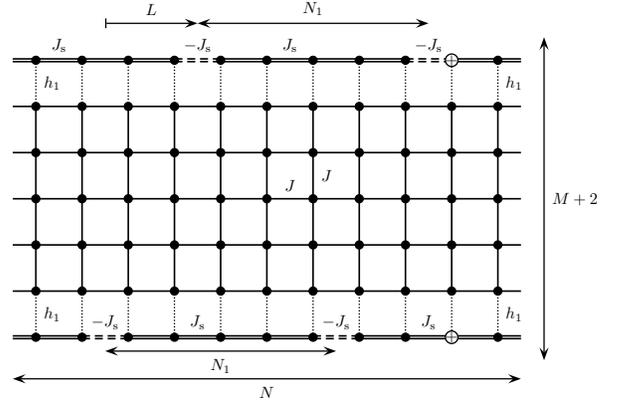}
\end{center}
\caption{\label{secA:fig1} Schematic plot of the system after modification. Two rows of spins ($m=0$ and $m=M+1$) are added. Spins in added rows interact with their neighbors from the same row with coupling $J_\mathrm{s}$ or $-J_\mathrm{s}$ and with the spins in rows $m=1$ and $m=M$ with a coupling $h_1$. Two of the added spins are in fixed state $+1$. Geometrical parameters $N_1$, $L$ and $N$, and couplings $J$ between neighboring spins inside the strip are the same as in the original model (see Fig.~\ref{sec1:fig1}).}
\end{figure}

Instead of our original model, following \cite{Abraham73}, we consider a system presented schematically in Fig.~\ref{secA:fig1} where two rows of spins ($m=0$ and $m=M+1$) are added. These extra spins are coupled with their nearest neighbors in the same row via coupling $J_\mathrm{s}$ or $-J_\mathrm{s}$. In the calculation we will apply the limit $J_\mathrm{s}\to\infty$, which puts all pairs of spins coupled with the energy $J_\mathrm{s}$ or $-J_\mathrm{s}$ in the same or the opposite state, respectively. Because one spin in each of the added rows is already fixed, the above limit fixes all additional spins. They are in the state $+1$ with an exception of spins added above and below inhomogeneities --- they are fixed to $-1$. This procedure changes the total energy of the system by $-2NJ_\mathrm{s}$. The spins in rows $m=1$ and $m=M$ interact with added spins from the same column with a coupling $h_1$. Because the added spins are fixed, the interaction is equivalent to the influence of a surface magnetic field $\pm h_1$, and the sign depends on the state of the additional spin in the same column.

This proves that the system considered here is equivalent to the model presented in Fig.~\ref{sec1:fig1}, provided that the energy of each configuration is increased by $2NJ_\mathrm{s}$ and the limit $J_\mathrm{s}\to\infty$ is taken.

\subsection{Partition function}

The first step of the transfer matrix technique is to assign vectors to states of each spin
\begin{equation}
s_{m,n}=+1\mapsto v_{m,n}=\begin{pmatrix} 1 \\ 0\end{pmatrix}, \  s_{m,n}=-1\mapsto v_{m,n}=
\begin{pmatrix} 0 \\ 1\end{pmatrix}.
\end{equation} 
We denote by $\sigma_n$ a sequence of all spins in $n$-th column
\begin{equation}
\sigma_n=\left(s_{0,n},s_{1,n},s_{2,n},\ldots,s_{M,n},s_{M+1,n}\right),
\end{equation}
and assign to it a vector from $2^{M+2}$--dimensional complex vector space $\spX$
\begin{equation}\label{secA:std_basis}
\sigma_n\mapsto\left|\sigma_n\right>=v_{0,n}\otimes v_{1,n}\otimes v_{2,n}\otimes\ldots\otimes v_{M,n}\otimes v_{M+1,n}.
\end{equation}
Vectors assigned to all possible configurations of spins in one column form an orthonormal basis of $\spX$.
We define two operators  $ \opV_1$ and $\opV_2$:
\begin{subequations}\label{secA:v1v2}
\begin{align}
\nonumber&\left<\sigma_n\middle|\opV_1\left(K,K_\mathrm{0},K_\mathrm{M+1}\right)\middle| \sigma_{n^\prime}\right>=\exp\Bigg(K_0 s_{0,n}s_{0,n^\prime}\\
&\hspace{4mm}+K_{M+1}s_{M+1,n}s_{M+1,n^\prime}+K\sum_{m=1}^M s_{m,n}s_{m,n^\prime} \Bigg),\\
\nonumber&\left<\sigma_n\middle|\opV_2\left(K,H_1\right)\middle|\sigma_{n^\prime}\right>=\left<\sigma_n\middle|\sigma_{n^\prime}\right>\exp\Bigg(H_1s_{0,n}s_{1,n}\\
&\hspace{10mm}+H_1s_{M,n}s_{M+1,n}+K\sum_{m=1}^{M-1}s_{m,n}s_{m+1,n}\Bigg).
\end{align}
\end{subequations}
Operator $\opV_1$ represents the Boltzmann factor stemming from interaction of spins in two neighboring columns; spins in the $m=0$ row interact via coupling $\kB T K_0$, in row $M+1$ via coupling $\kB T K_{M+1}$, and via $k_\mathrm{B} T K=J$ in all other rows. Operator $\opV_2$ is diagonal in the basis defined above and represents the Boltzmann factor stemming from the interaction of spins within one column --- the top and bottom pair of spins interact via couplings $\kB T H_1=h_1$, and all other pairs of nearest--neighbor spins via coupling $J$.

Using these two operators we can rewrite the formula for the partition function (for $0<L<N_1$) by combining the Boltzmann factors for all interactions
\begin{multline}
\partf=\lim_{\Ks\to\infty} \ee^{-2N\Ks}\Tr\Big[\opV_1\left(K,-\Ks,\Ks\right)\\
\times\left[\opV_2\left(K,H_1\right)\opV_1\left(K,\Ks,\Ks\right)\right]^{L-1}\opV_2\left(K,H_1\right)\\
\times\opV_1\left(K,\Ks,-\Ks\right)\left[\opV_2\left(K,H_1\right)\opV_1\left(K,\Ks,\Ks\right)\right]^{N_1-L-1}\\
\times\opV_2\left(K,H_1\right)\opV_1\left(K,-\Ks,\Ks\right)\\
\times\left[\opV_2\left(K,H_1\right)\opV_1\left(K,\Ks,\Ks\right)\right]^{L-1}\\
\times\opV_2\left(K,H_1\right)\opV_1\left(K,\Ks,-\Ks\right)\opR\\
\times\left[\opV_2\left(K,H_1\right)\opV_1\left(K,\Ks,\Ks\right)\right]^{N-L-N_1-1}\opV_2\left(K,H_1\right)\Big],
\end{multline}
where $\Ks=J_\mathrm{s}/\left(\kB T\right)$ and $\opR$ is a projection operator which fixes spins in rows $m=0$ and $m=M+1$. This formula was obtained by multiplying Boltzmann factors for all columns (see Fig.~\ref{secA:fig1}). Additional factor $\exp\left(-2N \Ks\right)$ was included to compensate the change of energy caused by the two extra rows of spins as compared to the original system. Using the cyclic property of the trace the above formula can be simplified
\begin{equation}\label{secA:partfunc1}
\partf=\Tr\left[\opF_{\downarrow}{\opV}^{L}\opF_{\uparrow}{\opV}^{N_1-L}\opF_{\downarrow}{\opV}^{L}\opF_{\uparrow}\opR{\opV}^{N-N_1-L}\right],
\end{equation}
where
\begin{multline}\label{secA:Vprime1}
\opV=\lim_{\Ks\to\infty}\ee^{-2\Ks}\left[\opV_2\left(K,H_1\right)\right]^{1/2}\\
\times\opV_1\left(K,\Ks,\Ks\right) \left[\opV_2\left(K,H_1\right)\right]^{1/2}
\end{multline}
is a symmetrized transfer matrix for a homogeneous two--dimensional Ising strip \cite{Kaufman49}, and
\begin{subequations}\label{secA:opF1}
\begin{align}
\nonumber&\opF_{\downarrow}=\lim_{\Ks\to\infty}\left[\opV_2\left(K,H_1\right)\right]^{1/2}\opV_1\left(K,-\Ks,\Ks\right)\\
&\hspace{15mm}\times\left[\opV_1\left(K,\Ks,\Ks\right)\right]^{-1}\left[\opV_2\left(K,H_1\right)\right]^{-1/2},\\
\nonumber&\opF_{\uparrow}=\lim_{\Ks\to\infty}\left[\opV_2\left(K,H_1\right)\right]^{1/2}\opV_1\left(K,\Ks,-\Ks\right)\\
&\hspace{16mm}\times\left[\opV_1\left(K,\Ks,\Ks\right)\right]^{-1}\left[\opV_2\left(K,H_1\right)\right]^{-1/2}
\end{align}
\end{subequations}
are the ``spin flip'' operators that change the sign of the surface field at the bottom and top wall, respectively. Similar operator which changes the sign of the surface field has been introduced in \cite{Abraham06}. Here, we have also used the fact that operator $\opR$ commutes with $\opV_2$.

\subsection{Derivation of transfer matrix}

To simplify the formula for operators $\opV_1$ and $\opV_2$ \eqref{secA:v1v2} we introduce a set of operators on $\spX$
\begin{equation}
\tau^{\alpha}_m=\underbrace{\tau^0\otimes\ldots\otimes\tau^0}_{\text{$m$ factors}}\otimes\tau^\alpha\otimes\underbrace{\tau^0\otimes\ldots\otimes\tau^0}_{\text{$M+1-m$ factors}},
\end{equation}
where $m=0,1,2,\ldots,M+1$, $\alpha=x,y,z$, $\tau^0$ is the two--dimensional identity matrix and $\tau^\alpha$ are the three Pauli matrices
\begin{equation}
\tau^x=\begin{pmatrix} 1 &0\\ 0&-1 \end{pmatrix},\quad \tau^y=\begin{pmatrix} 0&-\iu\\ \iu&0 \end{pmatrix}, \quad \tau^z=\begin{pmatrix} 0&-1\\ -1&0 \end{pmatrix}.
\end{equation}
With the help of the relation $\tau^x_m \left|\sigma_n\right>=s_{m,n} \left|\sigma_n\right>$ we get
\begin{subequations}
\begin{align}
\nonumber&\opV_1\left(K,K_0,K_{M+1}\right)=\left(\ee^{K_0}\opI-\ee^{-K_0}\tau^z_0\right)\\
&\hspace{4mm}\times\left(\ee^{K_{M+1}}\opI-\ee^{-K_{M+1}}\tau^z_{M+1}\right)\prod_{m=1}^M\left(\ee^{K}\opI-\ee^{-K}\tau^z_{m}\right),\\
\nonumber&\opV_2\left(K,H_1\right)=\\
&\hspace{6mm}\exp\left(H_1 \tau^x_0\tau^x_1+H_1\tau^x_M\tau^x_{M+1}+K\sum_{m=1}^{M-1}\tau^x_m\tau^x_{m+1}\right),
\end{align}
where $\opI=\tau^0\otimes\ldots\otimes\tau^0$ is the identity operator on $\spX$. The formula for $\opV_1$ can be further simplified using $\left(\ee^{K}\opI-\ee^{-K}\tau^z\right)=\left(2\sinh 2K\right)^{1/2}\exp\left(-\Kast \tau^z\right)$:
\begin{multline}
\opV_1\left(K,K_0,K_{M+1}\right)=\\
\mathcal{A}\exp\left(-\Kast_0 \tau^z_0-\Kast_{M+1}\tau^z_{M+1}-\Kast\sum_{m=1}^{M} \tau^z_m\right),
\end{multline}
\end{subequations}
where $\mathcal{A}=\left(2\sinh 2K\right)^{M/2}\left(2\sinh 2K_0\right)^{1/2}\linebreak\left(2\sinh 2K_{M+1}\right)^{1/2}$ and the dual coupling $\Kast$ is defined via $\exp\left(2\Kast\right)=\coth K$, which for $K>0$ simplifies to \eqref{sec2:dual}.

Following \cite{Kaufman49}, we introduce spinors $\Gamma_k$, $k=0,1,2,\ldots,2M+3$
\begin{equation}
\Gamma_{2j}=\left(\prod_{i=0}^{j-1}-\tau_i^z\right)\tau_j^x,\quad \Gamma_{2j+1}=\left(\prod_{i=0}^{j-1}-\tau_i^z\right)\tau_j^y.
\end{equation}
These operators on $\spX$ satisfy the anticommutation relation $\Gamma_{i}\Gamma_j+\Gamma_j\Gamma_i=
2\,\opI\, \delta_{i,j}$, and allow to transform the transfer operators
\begin{subequations}
\begin{align}
\nonumber&\opV_1\left(K,K_0,K_{M+1}\right)=\mathcal{A}\exp\Bigg(\iu\Kast\sum_{m=1}^{M}\Gamma_{2m}\Gamma_{2m+1}\\
&\hspace{15mm}+\iu \Kast_0\Gamma_0\Gamma_1+\iu \Kast_{M+1}\Gamma_{2M+2}\Gamma_{2M+3}\Bigg),\\
\nonumber&\opV_2\left(K,H_1\right)=\exp\Bigg(\iu K \sum_{m=1}^{M-1}\Gamma_{2m+1}\Gamma_{2m+2}+\iu H_1\Gamma_1\Gamma_2\\
&\hspace{38mm}+\iu H_1\Gamma_{2M+1}\Gamma_{2M+2}\Bigg).
\end{align}
\end{subequations}
Using the above formula in \eqref{secA:Vprime1} we get
\begin{multline}
\opV=\left(2\sinh 2 K\right)^{M/2}\exp\Bigg(\iu \frac{K}{2} \sum_{m=1}^{M-1}\Gamma_{2m+1}\Gamma_{2m+2}\\
+\iu \frac{H_1}{2}\Gamma_1\Gamma_2+\iu \frac{H_1}{2}\Gamma_{2M+1}\Gamma_{2M+2}\Bigg)\\
\times\exp\left(\iu \Kast\sum_{m=1}^M \Gamma_{2m}\Gamma_{2m+1}\right)\\
\times\exp\Bigg(\iu \frac{K}{2} \sum_{m=1}^{M-1}\Gamma_{2m+1}\Gamma_{2m+2}+\iu \frac{H_1}{2}\Gamma_1\Gamma_2\\
+\iu \frac{H_1}{2}\Gamma_{2M+1}\Gamma_{2M+2}\Bigg),
\end{multline}
where we have used the fact that in the limit $\Ks\to\infty$ we have $\Ks^\ast\to 0$ and $\ee^{-\Ks}\left(2\sinh 2\Ks\right)^{1/2}\to 1$. Similarly, from \eqref{secA:opF1} we get
\begin{subequations}\label{secA:opF2}
\begin{align}
\opF_{\downarrow}&=\iu\Gamma_0\Gamma_1\,\cosh H_1+ \Gamma_0\Gamma_2\,\sinh H_1, \\
\opF_{\uparrow}&=\iu \Gamma_{2M+2}\Gamma_{2M+3}\,\cosh H_1- \Gamma_{2M+1}\Gamma_{2M+3}\,\sinh H_1.
\end{align}
\end{subequations}
Operator $\opR$ is the projection operator that sets to $+1$ the values of spins in the top and bottom rows
\begin{multline}\label{secA:R}
\opR=\frac{1}{2}\left(\tau^0+\tau^x\right)\otimes\tau^0\otimes\ldots\otimes\tau^0\otimes \frac{1}{2}\left(\tau^0+\tau^x\right)\\
=\frac{1}{4}\left(\opI+\Gamma_0\right)\left(\opI+\iu\,\Gamma_{2M+3}\opU\right),
\end{multline}
where we have introduced the operator $\opU=\iu^{M+2}\Gamma_0\Gamma_1\Gamma_2\ldots\Gamma_{2M+3}$, which is a symmetry, i.e., $\opU^2=\opI$.

\subsection{Diagonalization of transfer matrix}

The diagonalization of the transfer matrix $\opV$ has already been done in the context of a homogeneous Ising strip \cite{Abraham73, Maciolek96}. Here we only sketch the procedure and recall formulas needed in our calculation.
 
A new set of spinors $\opG_k$, $k=0,1,2,\ldots,2M+3$ is generated by a linear transformation of $\Gamma$ spinors:
\begin{equation}
\label{secA:Gamma2G}
\opG_i=\sum_{j=0}^{2M+3} \mS_{j,i}\Gamma_j,
\end{equation}
where $\mS$ is a real, orthogonal matrix chosen such that
\begin{equation}\label{secA:opvG}
\opV=\exp\left(\frac{\iu}{2}\sum_{k=1}^{M+1}\gamma_k \opG_{2k-1}\opG_{2k}+\frac{\iu}{2}\gamma_0\opG_{2M+3}\opG_0\right),
\end{equation}
and the coefficients $\gamma_k$, $k=0,1,2,\ldots,M+1$ are yet to be determined. We assume that $0\leqslant\gamma_0<\gamma_1<\gamma_2<\ldots<\gamma_{M+1}$. 

The transfer matrix $\opV$ can be rewritten with the help of fermionic operators
\begin{equation}\label{secA:f_from_G}
f_0=\frac{1}{2}\left(\opG_0+\iu\opG_{2M-1}\right), \quad f_k=\frac{1}{2}\left(\opG_{2k}+\iu \opG_{2k-1}\right), 
\end{equation}
where $k=1,2,\ldots,M+1$, to the following simple form 
\begin{equation}
\opV=\exp\left[-\sum_{k=0}^{M+1}\gamma_k\left(\fd_k f_k-\frac{1}{2}\opI\right)\right].
\end{equation}
The above formula shows that the transfer matrix $\opV$ is diagonal in the occupation number basis defined by the above fermionic operators $f_k$. The ``vacuum'' eigenvector $\left|0\right>\in\spX$ satisfies 
$f_k\left|0\right>=0 \quad\text{for } k=0,1,2,\dots,M+1,$
and the corresponding eigenvalue of $\opV$ is 
\begin{equation}
\Lambda_0=\exp\left(\frac{1}{2}\sum_{k=0}^{M+1}\gamma_k\right).
\end{equation}
All other eigenvectors can be obtained by acting with creation operators $\fd_k$ on $\left|0\right>$. In our case, $\gamma_0=0$ and $\gamma_k>0$ for $k=1,2,\ldots,M+1$ (see below), and thus $\Lambda_0$ is the largest eigenvalue and is associated with two vectors $\left|0\right>$ and $\fd_0\left|0\right>$. 

The formulas for coefficients $\gamma_k$ and matrix $\mS$ are rather complicated although well known \cite{Maciolek96}; here we recall only the final results. The smallest coefficient is $\gamma_0=0$; all other coefficients can be determined numerically from equation
\begin{equation}\label{secA:gamma}
\cosh\gamma_k=\cosh 2\Kast\cosh 2K-\cos\omega_k, \quad \gamma_k>0,
\end{equation}
where $k=1,2,3,\ldots,M+1$, parameters $\omega_k$ are the solutions of
\begin{equation}\label{secA:omega}
M \omega-\delta^\prime\left(\omega\right)-\phi\left(\omega\right)=l\pi, \quad l\in\mathbb{Z},
\end{equation}
and
\begin{subequations}
\begin{align}
\ee^{\iu\phi\left(\omega\right)}&=\frac{W\ee^{\iu\omega}-1}{\ee^{\iu\omega}-W},\quad \phi\left(\pi\right)=0,\\
\ee^{\iu\delta^\prime\left(\omega\right)}&=\left[\frac{\left(\ee^{\iu\omega}-C\right)\left(\ee^{\iu\omega}-D\right)}{\left(C\ee^{\iu\omega}-1\right)\left(D\ee^{\iu\omega}-1\right)}\right]^{1/2}, \quad \delta^\prime\left(0\right)=\pi,\\
W&=\left(\cosh 2\Kast+1\right)\left(\cosh 2K-\cosh 2H_1\right),\\
C&=1/\left(\tanh K\tanh \Kast\right),\\
D&=\tanh K/\tanh \Kast.
\end{align}
\end{subequations}
The solutions of \eqref{secA:omega} satisfy $0<\omega_k<\pi$. Below $\Tw$ and above $\Tc$ one or two additional solutions for $\omega=\iu u$, $u>0$, can exist, see \cite{Maciolek96}.

After determining values of the coefficients $\gamma_k$, the elements of the matrix $\mS$ can be calculated according to the following procedure: first the components of $2M+4$--dimensional vectors $y^{\left(k\right)}$ for each $k=1,2,3,\ldots,M+1$ are calculated
\begin{subequations}
\begin{align}
&y_0^{\left(k\right)}=0,\\
\nonumber&y_1^{\left(k\right)}=\iu E_k A\Big\{\left[\sinh K \ee^{\iu \delta^\ast\left(\omega_k\right)}+\cosh K\right]\ee^{\iu \omega_k}\\
&\hspace{6mm}+\left[\sinh K+\cosh K \ee^{\iu \delta^\ast\left(\omega_k\right)}\right]R\left(\omega_k\right)\ee^{2\iu\omega_k}\Big\},\\
\nonumber&y_2^{\left(k\right)}=E_k B \Big\{\left[\sinh K \ee^{\iu \delta^\ast\left(\omega_k\right)}+\cosh K\right]\ee^{\iu \omega_k}\\
&\hspace{7mm}+\left[\sinh K+\cosh K \ee^{\iu \delta^\ast\left(\omega_k\right)}\right]R\left(\omega_k\right)\ee^{2\iu\omega_k}\Big\},\\
&y_{2j+1}^{\left(k\right)}=-\iu E_k \left[\ee^{\iu\delta^\ast\left(\omega_k\right)}\ee^{\iu\omega_k\left(j+1\right)}+R\left(\omega_k\right)\ee^{-\iu\omega_k\left(j-2\right)}\right],\\
&y_{2j+2}^{\left(k\right)}=E_k \left[ \ee^{\iu\omega_k\left(j+1\right)}+R\left(\omega_k\right)\ee^{\iu \delta^\ast\left(\omega_k\right)}\ee^{-\iu\omega_k\left(j-2\right)}\right],\\
\nonumber&y_{2M+1}^{\left(k\right)}=-\iu E_k B\Big\{\left[\sinh K+\cosh K \ee^{\iu \delta^\ast\left(\omega_k\right)}\right]\ee^{\iu \omega_k \left(M+1\right)}\\
&\hspace{8mm}+\left[\sinh K\ee^{\iu\delta^\ast\left(\omega_k\right)}+\cosh K\right]R\left(\omega_k\right)\ee^{-\iu \omega_k\left(M-2\right)}\Big\},\\
\nonumber&y_{2M+2}^{\left(k\right)}=-E_k A\Big\{\left[ \sinh K+\cosh K\ee^{\iu\delta^\ast\left(\omega_k\right)}\right]\ee^{\iu \omega_k\left(M+1\right)}\\
&\hspace{8mm}+\left[\sinh K\ee^{\iu \delta^\ast\left(\omega_k\right)}+\cosh K\right]R\left(\omega_k\right)\ee^{-\iu\omega_k\left(M-2\right)}\Big\},\\
&y_{2M+3}^{\left(k\right)}=0,
\end{align}
\end{subequations}
where $j=1,2,3,\ldots,M$. The $E_k$ are normalization constants, and
\begin{subequations}
\begin{align}
A&=\cosh H_1 \frac{\sinh 2H_1}{\ee^{-\gamma_k}-\cosh 2H_1}+\sinh H_1,\\
B&=\sinh H_1 \frac{\sinh 2H_1}{\ee^{-\gamma_k}-\cosh 2H_1}+\cosh H_1,\\
\nonumber\ee^{\iu\delta^\ast\left(\omega_k\right)}&=\left[\frac{\left(\ee^{\iu\omega_k}-C\right)\left(D\ee^{\iu\omega_k}-1\right)}{\left(C\ee^{\iu\omega_k}-1\right)\left(\ee^{\iu\omega_k}-D\right)}\right]^{1/2},\\
&\hspace{27mm}\delta^\ast\left(0\right)=\pi\, \Theta\left(T-\Tc\right),\\
R\left(\omega_k\right)&=-\ee^{\iu \delta^\prime\left(\omega_k\right)}\ee^{-\iu\omega_k}\frac{\ee^{\iu\omega_k}W-1}{\ee^{\iu\omega_k}-W},
\end{align}
\end{subequations}
where $\Theta(x)$ is the Heaviside step function. Then, the absolute values of the normalization constants $E_k$ are determined by the requirement 
$\left(y^{\left(k\right)}\right)^\dagger y^{\left(k\right)}=1$ and their phases are chosen such that $y_{2j+1}^{\left(k\right)}$ are real numbers and 
$y_{2j}^{\left(k\right)}$ are imaginary for all $k=1,2,3,\ldots,M+1$ and $j=0,1,2,\ldots,M+1$.
Finally, the elements of matrix $\mS$ are
\begin{subequations}
\begin{align}
\mS_{j,2k-1}&=\sqrt{2}\operatorname{Re}\left(y_{j}^{\left(k\right)}\right), &\mS_{j,2k}&=\sqrt{2}\operatorname{Im}\left(y_{j}^{\left(k\right)}\right),\\
\mS_{j,0}&=\delta_{j,0}, & \mS_{j,2M+3}&=\delta_{j,2M+3}.
\end{align}
\end{subequations}
We note that the matrix $\mS$ is not unique. For the above particular choice of  normalization constants $E_k$
\begin{subequations}\label{secA:Sproperties}
\begin{align}
\label{secA:Szeros}\mS_{2l,2k+1}&=\mS_{2l+1,2k}=0,\\
\label{secA:Ssym}\mS_{2a,2k}&=s_k\mS_{2M+3-2a,2k-1}, &s_k=\pm 1,\\
\label{secA:Uvac}\opU\left|0\right>&=+\left|0\right>.
\end{align}
\end{subequations}

\subsection{Free energy}

In order to calculate the free energy of the system \eqref{secA:partfunc1}, we first note that
\begin{subequations}\label{secA:VandG}
\begin{align}
\opV^n\opG_{2k}\opV^{-n}&=\fd_k \ee^{-n\gamma_k}+f_k \ee^{n\gamma_k},\\
\opV^n\opG_{0}\opV^{-n}&=\opG_{0}=\fd_0+f_0,\\
\opV^n\opG_{2k-1}\opV^{-n}&=\iu\left(\fd_k \ee^{-n\gamma_k}-f_k \ee^{n\gamma_k} \right), \\
\opV^n\opG_{2M+3}\opV^{-n}&=\opG_{2M+3}=\iu\left(\fd_0-f_0\right),
\end{align}
\end{subequations}
where $k=1,2,3,\ldots,M+1$, which follows directly from \eqref{secA:opvG} and \eqref{secA:f_from_G}, and
\begin{multline}\label{secA:trace}
\Tr\left(\opA \opV^N \opR \right)=\frac{1}{2}\Lambda_0^N\Big[\left<0\middle| \opA\middle| 0\right>+\left<0\middle|f_0\opA\middle|0\right>+\left<0\middle|\opA\fd_0\middle|0\right>\\
+\left<0\middle|f_0\opA\fd_0\middle|0\right>+\mathrm{O}\left(\ee^{-N\gamma_1}\right) \Big],
\end{multline} 
where $\opA$ is an arbitrary operator independent of the length of the system $N$. The second formula was proved in \cite{Stecki94}; it follows from \eqref{secA:R}, the commutation of operators $\opR$ and $\opV$, and from the fact that $\left|0\right>$ and $\fd_0\left|0\right>$ are two eigenvectors of the transfer matrix with the same, largest eigenvalue $\Lambda_0$.

Using the cyclic property of the trace, we transform \eqref{secA:partfunc1} to the following form
\begin{multline}
\partf=\Tr\Big[\opF_\downarrow \left(\opV^L \opF_\uparrow \opV^{-L} \right)\left(\opV^{N_1}\opF_\downarrow\opV^{-N_1}\right)\\
\times\left(\opV^{N_1+L}\opF_\uparrow\opV^{-\left(N_1+L\right)}\right)\opV^N \opR \Big].
\end{multline}
From \eqref{secA:opF2}, \eqref{secA:Gamma2G} and \eqref{secA:VandG} it follows that
\begin{subequations}\label{secA:FVcommutator}
\begin{align}
\opV^n\opF_\downarrow\opV^{-n}&=\left(\fd_0+f_0\right)\sum_{k=1}^{M+1}\left(t_1^{\left(k\right)}f_k\ee^{n\gamma_k}+t_3^{\left(k\right)}\fd_k\ee^{-n\gamma_k}\right),\\
\nonumber\opV^n\opF_\uparrow\opV^{-n}&=\sum_{k=1}^{M+1}\left(-t_4^{\left(k\right)}f_k\ee^{n\gamma_k}-t_2^{\left(k\right)}\fd_k\ee^{-n\gamma_k}\right)\\
&\hspace{32mm}\times\left(\fd_0-f_0\right),
\end{align}
\end{subequations}
where we have introduced
\begin{subequations}\label{secA:t}
\begin{align}
t_1^{\left(k\right)}&=\mS_{2,2k}    \sinh H_1 + \mS_{1,2k-1}    \cosh H_1,\\
t_2^{\left(k\right)}&=\mS_{2M+2,2k} \cosh H_1      - \mS_{2M+1,2k-1} \sinh H_1,\\
t_3^{\left(k\right)}&=\mS_{2,2k}    \sinh H_1      - \mS_{1,2k-1}    \cosh H_1,\\
t_4^{\left(k\right)}&=\mS_{2M+2,2k} \cosh H_1      + \mS_{2M+1,2k-1} \sinh H_1.
\end{align}
\end{subequations}
We note that, from \eqref{secA:Ssym} it follows that 
\begin{equation}\label{secA:tsymmetry}
t_4^{\left(k\right)}=s_k t_1^{\left(k\right)}, \qquad t_2^{\left(k\right)}=-s_k t_3^{\left(k\right)}.
\end{equation}
Using the above formulas in \eqref{secA:trace} with $\opA=\opF_\downarrow \left(\opV^L \opF_\uparrow \opV^{-L} \right)\left(\opV^{N_1}\opF_\downarrow\opV^{-N_1}\right)\left(\opV^{N_1+L}\opF_\uparrow\opV^{-\left(N_1+L\right)}\right)$~we get
\begin{multline}
\partf=\Lambda_0^N\Bigg[\left(\sum_{k=1}^{M+1}t_1^{\left(k\right)}t_2^{\left(k\right)}\ee^{-L\gamma_k}\right)^2\\
-\left(\sum_{k=1}^{M+1} t_1^{\left(k\right)}t_2^{\left(k\right)}\ee^{-\left(N_1+L\right)\gamma_k}\right)\left(\sum_{k=1}^{M+1}t_1^{\left(k\right)}t_2^{\left(k\right)}\ee^{-\left(N_1-L\right)\gamma_k}\right)\\
+\left(\sum_{k=1}^{M+1}t_1^{\left(k\right)}t_3^{\left(k\right)}\ee^{-N_1\gamma_k}\right)^2+\mathrm{O}\left(\ee^{-N\gamma_1}\right)\Bigg].
\end{multline}
To calculate the contribution to the free energy associated with two inhomogeneities, we subtract from the (dimensionless) full free energy 
$F_{\text{full}}=-\ln \partf$ the (dimensionless) free energy of the system of the same size but with homogeneous, symmetric surface fields 
$-N\ln\Lambda_0+\mathrm{O}\left(\ee^{-N\gamma_1}\right)$. After taking the limit $N\to\infty$ we get the (dimensionless) free energy
\begin{multline}
F\left(T,h_1,M,N_1,L\right)=-\ln\Bigg[\left(\sum_{k=1}^{M+1}t_1^{\left(k\right)}t_2^{\left(k\right)}\ee^{-L\gamma_k}\right)^2\\
-\left(\sum_{k=1}^{M+1} t_1^{\left(k\right)}t_2^{\left(k\right)}\ee^{-\left(N_1+L\right)\gamma_k}\right)\left(\sum_{k=1}^{M+1}t_1^{\left(k\right)}t_2^{\left(k\right)}\ee^{-\left(N_1-L\right)\gamma_k}\right)\\
+\left(\sum_{k=1}^{M+1}t_1^{\left(k\right)}t_3^{\left(k\right)}\ee^{-N_1\gamma_k}\right)^2\Bigg].
\end{multline}

The above formula for the reduced free energy is valid for $0\leqslant L\leqslant N_1$. For all other possible shifts of the inhomogeneities  similar calculations were performed. Because they differ only by minor details, we skip the derivations and give the final formula that includes all possible cases
\begin{multline}\label{secA:freeen}
F\left(T,h_1,M,N_1,L\right)=-\ln\Bigg[\left(\sum_{k=1}^{M+1}t_1^{\left(k\right)}t_2^{\left(k\right)}\ee^{-\left|L\right|\gamma_k}\right)^2\\
-\left(\sum_{k=1}^{M+1} t_1^{\left(k\right)}t_2^{\left(k\right)}\ee^{-\left(N_1+\left|L\right|\right)\gamma_k}\right)\left(\sum_{k=1}^{M+1}t_1^{\left(k\right)}t_2^{\left(k\right)}\ee^{-\left|N_1-\left|L\right|\right|\gamma_k}\right)\\
+\left(\sum_{k=1}^{M+1}t_1^{\left(k\right)}t_3^{\left(k\right)}\ee^{-N_1\gamma_k}\right)^2\Bigg].
\end{multline}

\subsection{Magnetization profiles}

In this section we present our method of calculation of the magnetization profile. We concentrate on columns with indices $n$ such that $1\leqslant n\leqslant L\leqslant N_1$; the calculation for other $n$--values  requires only minor changes.

We start from noting that
\begin{equation}
\tau^x_m \left|\sigma_n\right>=s_{m,n}\left|\sigma_n\right>,
\end{equation}
where $s_{m,n}=\pm 1$ denotes a state of a spin located in the $m$-th row and $n$-th column and the vector $\left|\sigma_n\right>$ is defined in \eqref{secA:std_basis}. Thus, to derive the magnetization at a particular lattice site we need to calculate a matrix element similar to the one needed for evaluation of the partition function but now with the operator $\tau^z_m$  inserted in the appropriate place. This operator commutes with $\opV_2$, and for $1\leqslant n\leqslant L\leqslant N_1$ we have
\begin{multline}
\mgn_{m,n}=\left<s_{m,n}\right>=\frac{1}{\partf}\Tr\Big[\opF_{\downarrow}{\opV}^{n}\tau^x_m\opV^{L-n}\opF_{\uparrow}\\
\times{\opV}^{N_1-L}\opF_{\downarrow}{\opV}^{L}\opF_{\uparrow}\opR{\opV}^{N-N_1-L}\Big].
\end{multline}
To simplify the formula we note that
\begin{equation}
\tau^x_m=\iu^m \Gamma_0 \Gamma_1\ldots \Gamma_{2m},
\end{equation}
and, using \eqref{secA:trace} and \eqref{secA:FVcommutator}, after some algebra we get
%\begin{linenomath}
%\begin{multline}
%\mgn_{m,n}=-\frac{\Lambda_0^N}{\partf}\Bigg[\sum_{a,b,c,d=1}^{M+1} t_1^{\left(a\right)}t_2^{\left(b\right)}t_3^{\left(c\right)}t_2^{\left(d\right)}\exp[-n\gamma_a\\
%-\left(L-n\right)\gamma_b-\left(N_1-n\right)\gamma_c-\left(N_1+L-n\right)\gamma_d]\Xi^m_{a;b,c,d}\\
%+\left(\sum_{k=1}^{M+1} t_1^{\left(k\right)}t_2^{\left(k\right)}\ee^{-L\gamma_k}\right)\sum_{a,b=1}^{M+1}t_1^{\left(a\right)}t_2^{\left(b\right)} \exp[-n\gamma_a\\
%-\left(L-n\right)\gamma_b]\Xi^m_{a;b}-\left(\sum_{k=1}^{M+1} t_1^{\left(k\right)}t_2^{\left(k\right)}\ee^{-\left(N_1-L\right)\gamma_k}\right)\\
%\times\sum_{a,b=1}^{M+1}t_1^{\left(a\right)}t_2^{\left(b\right)} \exp\left[-n\gamma_a-\left(L+N_1-n\right)\gamma_b\right]\Xi^m_{a;b}\\
%+\left(\sum_{k=1}^{M+1} t_1^{\left(k\right)}t_3^{\left(k\right)}\ee^{-N_1\gamma_k}\right)\\
%\times\sum_{a,b=1}^{M+1}t_1^{\left(a\right)}t_3^{\left(b\right)} \exp\left[-n\gamma_a-\left(N_1-n\right)\gamma_b\right]\Xi^m_{a;b}\Bigg],
%\end{multline}
%\end{linenomath}
\begin{widetext}
\begin{multline}
\mgn_{m,n}=-\frac{\Lambda_0^N}{\partf}\Bigg[\sum_{a,b,c,d=1}^{M+1} t_1^{\left(a\right)}t_2^{\left(b\right)}t_3^{\left(c\right)}t_2^{\left(d\right)}\exp\left[{-n\gamma_a-\left(L-n\right)\gamma_b-\left(N_1-n\right)\gamma_c-\left(N_1+L-n\right)\gamma_d}\right]\Xi^m_{a;b,c,d}\\
+\left(\sum_{k=1}^{M+1} t_1^{\left(k\right)}t_2^{\left(k\right)}\ee^{-L\gamma_k}\right)\sum_{a,b=1}^{M+1}t_1^{\left(a\right)}t_2^{\left(b\right)} \exp\left[-n\gamma_a-\left(L-n\right)\gamma_b\right]\Xi^m_{a;b}\\
-\left(\sum_{k=1}^{M+1} t_1^{\left(k\right)}t_2^{\left(k\right)}\ee^{-\left(N_1-L\right)\gamma_k}\right)\sum_{a,b=1}^{M+1}t_1^{\left(a\right)}t_2^{\left(b\right)} \exp\left[-n\gamma_a-\left(L+N_1-n\right)\gamma_b\right]\Xi^m_{a;b}\\
+\left(\sum_{k=1}^{M+1} t_1^{\left(k\right)}t_3^{\left(k\right)}\ee^{-N_1\gamma_k}\right)\sum_{a,b=1}^{M+1}t_1^{\left(a\right)}t_3^{\left(b\right)} \exp\left[-n\gamma_a-\left(N_1-n\right)\gamma_b\right]\Xi^m_{a;b}\Bigg],
\end{multline}
\end{widetext}
where we have introduced the following notation for matrix elements:
\begin{multline}\label{secA:mel}
\Xi^m_{a,b,\ldots,c;j,k,\ldots,l}=\\
\iu^m \left<0\middle| f_a f_b \ldots f_c \Gamma_1\Gamma_2\ldots \Gamma_{2m} \fd_j \fd_k\ldots \fd_l\middle|0\right>,
\end{multline}
where ``$a,b,\ldots,c$'' and ``$j,k,\ldots,l$'' stand for an arbitrary number of indices, possibly zero. The algorithm used to calculate matrix elements of the above type is described in Appendix~\ref{secB}.

Similar formulas for magnetization can be derived for all possible values of $n$ and $L$.

%For $n\leqslant 0\leqslant L<N_1$
%\begin{linenomath}
%\begin{multline}
%\mgn_{m,n}=\Xi^m+\frac{\Lambda_0^N}{\partf}\Bigg[\sum_{a,b,c,d=1}^{M+1} t_3^{\left(a\right)} t_2^{\left(b\right)} t_3^{\left(c\right)} t_2^{\left(d\right)}\exp\left[n\gamma_a+\left(n-L\right)\gamma_b+\left(n-N_1\right)\gamma_c+\left(n-N_1-L\right)\gamma_d\right]\Xi^m_{;a,b,c,d}\\
%+\left(\sum_{k=1}^{M+1}t_1^{\left(k\right)}t_2^{\left(k\right)}\ee^{-L\gamma_k}\right)\sum_{a,b=1}^{M+1}t_3^{\left(a\right)}t_2^{\left(b\right)}\left(\ee^{n\gamma_a+\left(n-L\right)\gamma_b}+\ee^{\left(n-N_1\right)\gamma_a+\left(n-N_1-L\right)\gamma_b}\right)\Xi^m_{;a,b}\\
%+\left(\sum_{k=1}^{M+1}t_1^{\left(k\right)}t_3^{\left(k\right)}\ee^{-N_1\gamma_k}\right)\sum_{a,b=1}^{M+1} \left(t_3^{\left(a\right)}t_3^{\left(b\right)}\ee^{n\gamma_a+\left(n-N_1\right)\gamma_b}-t_2^{\left(a\right)}t_2^{\left(b\right)}\ee^{\left(n-L\right)\gamma_a+\left(n-N_1-L\right)\gamma_b}\right)\Xi^m_{;a,b} \\
%-\left(\sum_{k=1}^{M+1}t_1^{\left(k\right)}t_2^{\left(k\right)}\ee^{-\left(N_1-L\right)\gamma_k}\right)\sum_{a,b=1}^{M+1} t_3^{\left(a\right)}t_2^{\left(b\right)}\ee^{n\gamma_a+\left(n-N_1-L\right)\gamma_b}\Xi^m_{;a,b}\\
%+\left(\sum_{k=1}^{M+1}t_1^{\left(k\right)}t_2^{\left(k\right)}\ee^{-\left(N_1+L\right)\gamma_k}\right)\sum_{a,b=1}^{M+1} t_2^{\left(a\right)}t_3^{\left(b\right)}\ee^{\left(n-L\right)\gamma_a+\left(n-N_1\right)\gamma_b}\Xi^m_{;a,b}\Bigg]
%\end{multline}
%\end{linenomath}

\section{Calculation of matrix elements}\label{secB}
In this Appendix we describe the method of numerical calculation of matrix elements $\Xi$ given by \eqref{secA:mel}. For simplicity, we are considering only the matrix element
\begin{equation}\label{secB:mel1}
\Xi_{a;b}^m=\iu^m \left<0\middle|f_a \Gamma_1 \Gamma_2 \Gamma_3\ldots \Gamma_{2m} \fd_b\middle| 0\right>
\end{equation}
but our method can be easily adapted to any matrix element $\Xi$.
\subsection{Transformation into Pfaffian}

It follows from \eqref{secA:Gamma2G} and \eqref{secA:f_from_G}  that every spinor $\Gamma_k$ is a linear combination 
of creation and annihilation operators $\fd_k$ and $f_k$.  Therefore the Wick's theorem \cite{Wick50} can be applied with the purpose of transforming matrix element \eqref{secB:mel1} into a Pfaffian:
\begin{equation}
\Xi_{a;b}^m=\iu^m\Pf A_m,
\end{equation}
where $A_m$ is a skew--symmetric matrix with elements built from appropriate contractions between operators in the matrix element \eqref{secB:mel1}:
\begin{widetext}
\begin{equation}
A_m=\begin{pmatrix}
0                  & \iu \mS_{1,2a-1} & \mS_{2,2a} & \iu\mS_{3,2a-1} & \mS_{4,2a} & \cdots & \iu\mS_{2m-1,2a-1} & \mS_{2m,2a} & \delta_{a,b} \\
-\iu \mS_{1,2a-1}  & 0      & -\iu K_{1,2} & 0 & -\iu K_{1,4} & \cdots & 0 &-\iu K_{1,2m} & -\iu\mS_{1,2b-1}\\
-\mS_{2,2a}        & \iu K_{1,2}& 0       &-\iu K_{2,3}& 0 & \cdots & -\iu K_{2,2m-1} & 0 & \mS_{2,2b}\\
-\iu\mS_{3,2a-1}   & 0      & \iu K_{2,3}& 0 & -\iu K_{3,4}& \cdots & 0& -\iu K_{3,2m}& -\iu\mS_{3,2b-1}\\
-\mS_{4,2a}        & \iu K_{1,4}& 0 & \iu K_{3,4}& 0 & \cdots & -\iu K_{4,2m-1} & 0 & \mS_{4,2b}\\
\vdots	& \vdots   & \vdots & \vdots &\vdots & \ddots & \vdots & \vdots & \vdots \\
-\iu\mS_{2m-1,2a-1}& 0      & \iu K_{2,2m-1}& 0& \iu K_{4,2m-1}& \cdots & 0 & -\iu K_{2m-1,2m}& -\iu\mS_{2m-1,2b-1} \\
-\mS_{2m,2a}       &\iu K_{1,2m}& 0& \iu K_{3,2m}& 0& \cdots & \iu K_{2m,2m-1}& 0 & \mS_{2m,2b}\\
-\delta_{a,b}      &\iu \mS_{1,2b-1} & -\mS_{2,2b} & \iu \mS_{3,2b-1} & -\mS_{4,2b} & \cdots & \iu \mS_{2m-1,2b-1} & -\mS_{2m,2b} & 0
\end{pmatrix}.
\end{equation}
\end{widetext}
In the above matrix we have introduced the elements $K_{k,l}$ of another antisymmetric matrix
\begin{multline}
K_{k,l}=\frac{1}{2}\left(\left<0\middle| \Gamma_k \Gamma_l\middle| 0\right>-\left<0\middle| \Gamma_l \Gamma_k\middle|0\right>\right)=\\
\sum_{j=0}^{M+1}\left(\mS_{k,2j-1}\mS_{l,2j}-\mS_{k,2j}\mS_{l,2j-1}\right), 
\end{multline}
where $k,l=1,2,\ldots,2M+2$, and we have used the fact that $K_{k,l}$ is non--zero only when $k$ and $l$ have a different parity, which follows from \eqref{secA:Szeros}.

\subsection{Elementary operations on matrix}

A particular method of calculation of the Pfaffian of a matrix similar to ours was described in \cite{Stecki94}. There, the determinant of the matrix was calculated using Gauss elimination method and the Pfaffian was obtained from the identity $\left(\Pf A\right)^2=\det A$. In our case this method cannot be applied directly, because it does not allow to determine the sign of the Pfaffian. In \cite{Stecki94} there was only a single Pfaffian and its sign was easy to guess. In this case we deal with a sum of Pfaffians and their signs are relevant. Therefore, we have developed a method of the direct calculation of the Pfaffian using Gauss elimination (a similar method was described in \cite{Gonzalez11}).

To calculate the Pfaffian of matrix $A_m$ we apply a sequence of three elementary operations. These operations must preserve the skew--symmetry of the matrix and therefore they must be done simultaneously on rows and columns.  These operations are: swapping the positions of rows $i$ and $j$ followed by swapping positions of columns $i$ and $j$; multiplying row $i$ by a non--zero number followed by multiplying column $i$ by the same number; and adding row $j$ multiplied by a number to row $i$  followed by adding column $j$ multiplied by the same number to column $i$. Note  that: swapping two rows and columns changes the sign of the Pfaffian (while the sign of determinant remains unchanged); multiplying a row and column by a number  multiplies the Pfaffian by the same number; and adding two rows and columns leaves the Pfaffian unchanged.

\subsection{Transformation of matrix $A_m$}

We start by simplifying the form of matrix $A_m$. We multiply all even rows and columns except for the last one by imaginary unit $\iu$. This changes Pfaffian by a factor $\iu^m$. Next, we move the last row and column to the second row and column. This requires $2m$ swappings and therefore does not change the Pfaffian. The result is
\begin{equation}
\Xi^m_{a,b}=\Pf A_m^\prime,
\end{equation}
where
\begin{widetext}
\begin{equation}
A_m^\prime=\begin{pmatrix}
0                   & \delta_{a,b} & -\mS_{1,2a-1} & \mS_{2,2a} & -\mS_{3,2a-1} & \mS_{4,2a} & \cdots & -\mS_{2m-1,2a-1} & \mS_{2m,2a}\\
-\delta_{a,b}      & 0 &-\mS_{1,2b-1} & -\mS_{2,2b} & - \mS_{3,2b-1} & -\mS_{4,2b} & \cdots & - \mS_{2m-1,2b-1} & -\mS_{2m,2b}\\
\mS_{1,2a-1}  & \mS_{1,2b-1} & 0                & K_{1,2} & 0 &  K_{1,4} & \cdots & 0 & K_{1,2m}\\
-\mS_{2,2a}        & \mS_{2,2b} & -K_{1,2}      & 0       & K_{2,3}& 0 & \cdots &  K_{2,2m-1} & 0\\
\mS_{3,2a-1}   & \mS_{3,2b-1}& 0                & - K_{2,3}& 0 &  K_{3,4}& \cdots & 0&  K_{3,2m}\\
-\mS_{4,2a}        & \mS_{4,2b} & -K_{1,4}& 0   & -K_{3,4}& 0 & \cdots &  K_{4,2m-1} & 0\\
\vdots	 & \vdots & \vdots   & \vdots & \vdots &\vdots & \ddots & \vdots & \vdots \\
\mS_{2m-1,2a-1}& \mS_{2m-1,2b-1} & 0      & - K_{2,2m-1}& 0& - K_{4,2m-1}& \cdots & 0 &  K_{2m-1,2m}\\
-\mS_{2m,2a}       & \mS_{2m,2b} &-K_{1,2m}& 0& - K_{3,2m}& 0& \cdots & - K_{2m,2m-1}& 0
\end{pmatrix}.
\end{equation}
\end{widetext}
The above formula proves that the matrix element $\Xi_{a;b}^m$ is real. We also note that increasing $m$ adds columns and rows to matrix $A_m^\prime$ without modifying the already existing elements.

\subsection{Reduction of matrix $A_m^\prime$}

Further simplification of matrix $A_m^\prime$ depends on values of its particular elements and therefore it is done numerically. We denote by $a_{i,j}$ the element of the matrix under reduction in $i$-th row and $j$-th column.

The reduction is done only in odd rows, starting from the first row and going down the matrix. For each row $k$ under reduction we first look at the element $a_{k,k+1}$, which is just above the diagonal. If it is zero, we look for non--zero elements in this row among $a_{k,k+2}, a_{k,k+3}, \ldots, a_{k,2m+2}$. If all of them are zero the Pfaffian of the whole matrix is zero and the reduction process is halted. If $a_{k,l}$ is the first non--zero element in the sequence, we switch row and column $k$ with $l$ and multiply row and column $k$ by $-1$. Each of these elementary operations changes sign of the Pfaffian and thus performing both of them leaves the Pfaffian unchanged. After this operation $a_{k,k+1}\neq 0$. To each row and column $l>k+1$ we now add row and column $k+1$ multiplied by $-a_{k,l}/a_{k,k+1}$. This operation does not change the Pfaffian and leads to $a_{k,l}=0$ for all $l>k+1$. This finishes the reduction process of $k$-th row. Afterwards the reduction is continued in the next odd row.

We denote by $A_m^{\prime\prime}$ the matrix after the reduction. Because the whole process is preserving the Pfaffian, $\Pf A_m=\Pf A_m^{\prime\prime}$. In all odd rows of the matrix $A_m^{\prime\prime}$, above the diagonal there is only one non--zero element. Straightforward calculation shows that the Pfaffian of this matrix is 
\begin{equation}
\Pf A_m=\Pf A_m^\prime =\Pf A_m^{\prime\prime}=\prod_{k=1}^{m+1}a^{\prime\prime}_{2k-1,2k},
\end{equation}
where $a^{\prime\prime}_{i,j}$ denotes an element of matrix $A_m^{\prime\prime}$ in $i$-th row and $j$-th column.

\subsection{Calculation of $\Pf A_m$ for different $m$}

As noted above, increasing $m$ adds two rows and columns to the matrix $A_m$, leaving all other elements unchanged. This property allows to speed up the calculation of Pfaffians for different values of $m$ in a way similar to the one described in \cite{Stecki94}. We perform the reduction for matrix $A_{M+1}$ and Pfaffians of all matrices $A_m$ can be calculated from the elements of matrix $A_{M+1}^{\prime\prime}$. Note that special care must be taken when an element of the reduced matrix above the diagonal is zero which results in $\Pf A_m=0$ for some $m$; we omit the details of dealing with this problem here.

\subsection{Numerical stability}

It is well known \cite{Trefethen97} that methods based on Gaussian reduction of the matrix are numerically unstable. Therefore a special care was taken to provide and check the stability of our calculation.

During the reduction process we treat all elements of the matrix with absolute value smaller than $10^{-10}$ as equal to zero. Additionally, using the properties of operator $\opU$ (see \eqref{secA:Uvac}) it can be shown that
\begin{equation}
\Xi^{M+1}_{a;b}=-\delta_{a,b},
\end{equation}
which provides an independent check of our results for Pfaffians. For all matrix elements calculated numerically the above formula was satisfied with an absolute numerical error not exceeding $10^{-5}$.

\section{General relation between the lateral critical Casimir force and breaking of the capillary bridge}\label{secC}
In this Appendix we provide an argument that some of the properties derived in this paper for a two--dimensional Ising model via exact analysis should also hold for a much wider class of systems. For this purpose we compare the free energy of a system with a periodic pattern of inhomogeneities on both walls with a system with only a single inhomogeneity on each wall, i.e., the case studied in detail in this paper.

\subsection{Class of systems}

We consider quite a general system which fulfills the following conditions: (I) The system is a 2D strip limited by two flat walls. (II) The walls are build from two different types of surfaces, denoted by ``$+$'' and ``$-$''; each wall is build from the ``$+$'' surface except for one inhomogeneity of length $N_1$, where the surface is of the ``$-$'' type. The shift between inhomogeneities on the walls is measured by the distance $P$ between the right end of the bottom inhomogeneity and the left end of top inhomogeneity  (see Fig.~\ref{secC:fig2}). The interaction between the walls and the system is short--ranged. (III) There exists a symmetry of the order parameter which, together with switching of ``$+$'' and ``$-$'' surfaces, leaves the free energy of the system unchanged. (IV) The free energy per unit length of a homogeneous system with the same type of surfaces is smaller than the free energy of a homogeneous system with different surfaces on each wall. (V) At the point of change of the type of surface the system is perturbed in comparison with the homogeneous one; this perturbation decays exponentially with the length--scales $\xiS$ or $\xiAS$, depending on the types of surfaces in 
the strip.

Additionally, we assume certain standard properties of thermodynamical systems like, for example, the invariance of the free energy under rotations and reflections. For simplicity we will also assume that (VI) the order parameter of the system is a continuous real function. This is not necessary for the argument to hold but it allows us to avoid the usage of discrete versions of derivatives. The assumption of real values of the order parameter simplifies the discussion of the capillary bridging in Sec.~\ref{secC:d}, but for a vector order parameter one can introduce a projection, such that the order parameter preferred by ``$+$'' (``$-$'') type of surface is projected to positive (negative) values. Additionally, the assumption (I) can be weakened --- the dimension of the system can be arbitrary, as long as the pattern on the walls changes only in one direction. Note that (V) is not satisfied at $T=\Tc$ and therefore the argument does not work at the critical point.

We believe that the conditions stated in this section are not difficult to satisfy and therefore our argument works for a wide range of systems.

In this section we do not display the dependence of discussed quantities on temperature $T$ and the width of the strip $M$.

\subsection{System with periodic inhomogeneities}

\begin{figure}
\begin{center}
\includegraphics[width=0.9\columnwidth]{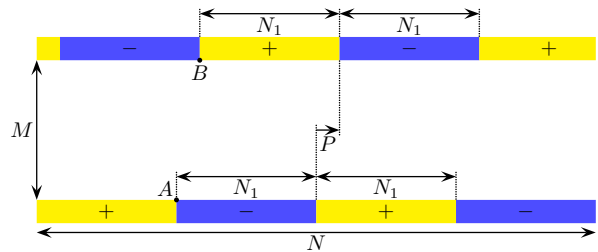}
\end{center}
\caption{\label{secC:fig1} Schematic plot of the system with periodic pattern on each wall. Regions of the wall with the surface of the type ``$+$'' (``$-$'') are marked yellow (blue).}
\end{figure}

We first consider a system with a periodic pattern of inhomogeneities on both walls. Each wall consists of segments of length $N_1$ with alternating type of surface. The segments are shifted by $P$ from the antisymmetric position. The total length of the system is $N=2n N_1$, where $n$ is the number of periods. We assume periodic boundary conditions in the horizontal direction. A schematic plot of the system is presented in Fig.~\ref{secC:fig1}.

The periodicity of the system implies that its free energy $\Fper$ is a periodic function of $P$ with a period $2N_1$. The symmetry of switching the types of surface (III) implies that $\Fper$ is symmetric around $P=0$:
\begin{equation}\label{secC:Fper_symmetry}
\Fper\left(P, N_1, N\right)=\Fper\left(-P, N_1, N\right).
\end{equation}
From (IV) and (V) it follows that for $N_1$ large enough the function $\Fper$ has a maximum for $P=0$ (opposite surface field in every column) and a minimum for $P=N_1$ (identical  surface field in every column). Around $P=0$ the properties of $\Fper$ depend on the type of bridging transition present in the system. If there is no transition (like in a two--dimensional Ising strip with finite $M$), $\Fper$ is analytic around $P=0$ and one has $\frac{\partial \Fper}{\partial P}\left(P=0\right)=0$ and $\frac{\partial^2 \Fper}{\partial P^2}<0$. If it is a first--order transition, there is a jump in first derivative: $\lim_{P\to 0^+}\frac{\partial \Fper}{\partial P}=-\lim_{P\to 0^-}\frac{\partial \Fper}{\partial P}< 0$. If the transition at $P=0$ is continuous, one again has $\frac{\partial \Fper}{\partial P}\left(P=0\right)=0$ and $\frac{\partial^2 \Fper}{\partial P^2}<0$ for $P$ around $0$, but at $P=0$ the second derivative may not exist.

The free energy of the system in the limit $N_1\gg \left|P\right|$ can be expressed as
\begin{multline} \label{secC:fe_periodic}
\Fper\left(P,N_1,N=2nN_1\right)=N f_{++}+2n\left|P\right|\sigma\\
+2n\,\delta f\left(P\right)+\mathrm{O}\left(\ee^{-N_1/\xiAS}\right),
\end{multline}
where $f_{++}$ is the free energy per unit length of a homogeneous strip with identical walls and $f_{++}+\sigma$ is the free energy per column in a homogeneous strip with opposite types of surfaces on the walls. The first two terms of \eqref{secC:fe_periodic} describe the free energy of the system without taking into account perturbations of the order parameter around the points where the sign of surface field changes (this corresponds to Deriaguin approximation \cite{Derjaguin34}). The function $\delta f\left(P\right)$ is the correction to the free energy stemming from the changes of the surface field which take place at points that are close to each other (for example point $A$ and $B$ in Fig.~\ref{secC:fig1}). There are $2n$ pairs of such points in the strip, and the pairs are separated by a distance of the order of $N_1$ (for $N_1\gg\left|P\right|$). Therefore, all other corrections to the free energy are of the order of $\exp\left(-N_1/\xiAS\right)$ (or even smaller) and are negligible in the limit $N_1\gg \left|P\right|$.

From \eqref{secC:fe_periodic} one can derive the derivatives of $\delta f\left(P\right)$ for $P\neq 0$:
\begin{subequations}\label{secC:derivatives}
\begin{align}
\delta f^\prime \left(P\right)&=\frac{1}{2} \frac{\partial \fper}{\partial P}+\sigma \sign\left(P\right)+\mathrm{O}\left(\ee^{-N_1/\xiAS}\right),\\
\delta f^{\prime\prime\prime} \left(P\right)&=\frac{1}{2}\frac{\partial^3 \fper}{\partial P^3}+\mathrm{O}\left(\ee^{-N_1/\xiAS}\right),
\end{align}
\end{subequations}
where we have introduced $\fper=\Fper/n$ the free energy of the system per one period. The term $\mathrm{O}\left(\ee^{-N_1/\xiAS}\right)$ present in \eqref{secC:fe_periodic} has the form of a function of $P$ multiplied by the exponent and therefore the corrections to the derivative are of the same order.

\subsection{System with two inhomogeneities}

\begin{figure}
\begin{center}
\includegraphics[width=0.9\columnwidth]{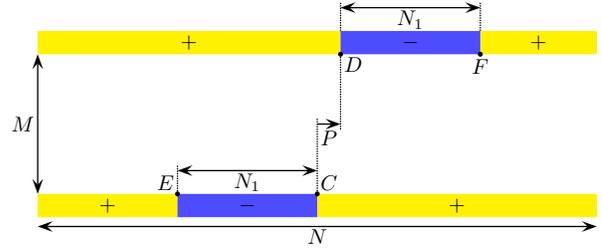}
\end{center}
\caption{\label{secC:fig2} Schematic plot of the system with two defects. Regions of wall with the surface of the type ``$+$'' (``$-$'') are marked yellow (blue).}
\end{figure}

In this section we return to the strip with two inhomogeneities, one on each wall. Total length of the system is $N$ and we assume periodic boundary conditions in the horizontal direction. A schematic plot of the system is presented in Fig.~\ref{secC:fig2}.

The free energy $\Fsingle$ of this system for $N_1\gg \left|P\right|$ can be written as
\begin{multline}\label{secC:fe_single}
\Fsingle\left(P,N_1,N\right)=N f_{++}+2N_1\sigma\\
-P \left[1-\sign\left(P\right)-\delta_{P,0}\right] \sigma+\delta f\left(P\right)\\
+2 f_0+\mathrm{O}\left(\ee^{-N_1/\xiAS}\right),
\end{multline}
where the first three terms represent the free energy without taking into account the perturbations of the order parameter close to the points of change of the surface field, $\delta f\left(P\right)$ is a contribution from two points of change of the field separated by a distance $P$ (points $C$ and $D$ in Fig.~\ref{secC:fig2}) and $f_0$ is an excess free energy of a single point of change of sign of the surface field (points $E$ and $F$ in Fig.~\ref{secC:fig2}). The additional terms coming from the overlap of perturbations of the order parameter due to distant points of change of the surface field are of the order of $\exp\left(-N_1/\xiAS\right)$ (or even smaller). It is crucial that the free energy of interaction $\delta f \left(P\right)$ in \eqref{secC:fe_single} is exactly the same function as in \eqref{secC:fe_periodic}. This is justified by the fact that the region around points $A$ and $B$  in Fig.~\ref{secC:fig1}  is exactly the same as the one around points $C$ and $D$ in Fig.~\ref{secC:fig2}. The differences between these two strips pop up at distances of the order of $N_1$ and thus the corresponding difference in free energies is of the order of $\exp\left(-N_1/\xiAS\right)$.

The lateral force $\flateral\left(P,N_1\right)=-\partial \Fsingle/\partial P$ can now be calculated form \eqref{secC:fe_single}. 
Using \eqref{secC:derivatives} we obtain for $P\neq 0$
\begin{subequations}
\begin{align}
\flateral\left(P,N_1\right)&=-\sigma-\frac{1}{2}\frac{\partial \fper}{\partial P}+\mathrm{O}\left(\ee^{-N_1/\xiAS}\right),\\
\frac{\partial^2\flateral}{\partial P^2}&=-\frac{1}{2}\frac{\partial^3 \fper}{\partial P^3}+\mathrm{O}\left(\ee^{-N_1/\xiAS}\right)
\end{align}
\end{subequations}
From the symmetry of $\Fper=n \fper$ (see \eqref{secC:Fper_symmetry}) and the above equations it follows that the lateral critical Casimir force in the limit $N_1\to\infty$ is symmetric around $P=0$, and if there is no first--order transition then $\flateral\left(P=0,N_1=\infty\right)=-\sigma$. For a two--dimensional Ising strip we have derived this expression rigorously (see \eqref{sec3:flatPsymmetry}). These formulas also show, that the inflection point of the force for $N_1=\infty$ is at $P=0$ and for finite $P$ it is shifted by the term of the order of $\exp\left(-N_1/\xiAS\right)$.

\subsection{Order parameter and capillary bridge}\label{secC:d}

We now consider the order parameter profile and the existence of a capillary bridge. In a strip with a periodic pattern on the walls, when there is no first--order bridging transition, it follows from the symmetry (III) that for $P=0$  the order parameter is zero right in the middle between points $A$ and $B$ in Fig.~\ref{secC:fig1}, and there is neither negative nor positive order parameter bridge. In the region between points $A$ and $B$, when $P<0$ there is a bridge of negative order parameter, and there is no such bridge for $P>0$. Thus this bridge breaks exactly at $P=0$.

If the walls are changed such that there is only one inhomogeneity with surface of the type ``$-$'' on each wall, the overall order parameter will increase, which means that $P$ must be slightly decreased from 0 to form a bridge of negative order parameter in the system in Fig.~\ref{secC:fig2}. Because for small $P$ the surface field remains unchanged in the region with size of the order of $N_1$ around the bridge, then from (V) it follows that the breaking of the capillary bridge will occur for negative $P$ of the order of $\exp\left(-N_1/\xiAS\right)$. This agrees with our results for the two--dimensional Ising strip.

We note that the above argument does not work if there is a first--order bridging transition at $P=0$. In that case even a small change of the chemical structure of the wall may cause a substantial change of the equilibrium order parameter. A trace of this effect is visible in the Ising strip for $T<\Tw$: although for finite $M$ there is no transition, one needs to consider very large $N_1$ to see the exponential approach of the point of breaking of the capillary bridge to $L_1=N_1$ (see Fig.~\ref{sec6:fig3}(a)).

%\bibliography{is}
\end{document}